\documentclass{article}

\usepackage[usenames,dvipsnames,svgnames,table]{xcolor}
\usepackage{cite}
\usepackage{url}
\usepackage[normalem]{ulem}
\usepackage[T1]{fontenc}
\usepackage{bold-extra}
\usepackage{multirow}
\usepackage{pifont} 
\usepackage{placeins}
\usepackage{graphicx}
\usepackage{geometry}
\usepackage{inconsolata} 

\usepackage{amsmath}
\usepackage{latexsym}
\usepackage{amssymb}

\usepackage{tcolorbox}

\newcommand{\algname}[1]{{\sf {#1}}}
\newcommand{\salgname}[1]{{\sf {#1}}}

\newcommand{\pipe}{{\algname{\algname{PicNic}}}}
\newcommand{\spipe}{{\salgname{\algname{PicNic}}}}

\newcommand{\bpipe}{{\algname{\bf \pipe{}}}}
\newcommand{\acronym}{\algname{Pipeline for Cancer Inference}}

\newcommand{\xx}{\mathbf{x}}
\newcommand{\Probab}[1]{\mathcal{P}({#1})}

\newcommand{\g}[1]{\textsc{{#1}}}
\newcommand{\sg}[1]{{\small \g{{#1}}}}

\definecolor{fgreen}{rgb}{0.13,0.55,0.13}
\newcommand{\added}[1]{{\textcolor{fgreen}{{#1}}}} 

\renewcommand{\added}[1]{{#1}} 
\newcommand{\addedtwo}[1]{#1}  
\newcommand{\addedthree}[1]{#1}  
\newcommand{\deleted}[1]{}

\title{Algorithmic methods to infer the evolutionary trajectories in cancer progression}

\author{Giulio Caravagna$^{1,2,\star}$
 \and 
Alex Graudenzi$^{1,3}$
 \and
Daniele Ramazzotti$^1$ \and 
Rebeca Sanz-Pamplona$^4$\and
Luca De Sano$^1$\and
 Giancarlo Mauri$^{1,5}$\and
 Victor Moreno$^{4,6}$\and
Marco Antoniotti$^{1,7}$ \and 
Bud Mishra$^8$
}

\begin{document}

\maketitle
\footnotesize
\noindent
$^{1}\,${Department of Informatics, Systems and Communication, University of Milan-Bicocca, Milan, Italy.}\\
$^{2}\,${School of Informatics, University of Edinburgh, Edinburgh, UK.}\\
$^{3}\,${Institute of Molecular Bioimaging and Physiology of the Italian National Research Council (IBFM-CNR), Milan, Italy.}\\
$^{4}\,${Unit of Biomarkers and Susceptibility, Cancer Prevention and Control Program, Catalan Institute of Oncology (ICO), IDIBELL \& CIBERESP. Hospitalet de Llobregat, Barcelona, Spain.}\\
$^{5}\,${SYSBIO Centre of Systems Biology, Milano, Italy.}\\
$^{6}\,${Department of Clinical Sciences, Faculty of Medicine, University of Barcelona, Barcelona, Spain.}\\
$^{7}\,${Milan Center for Neuroscience, University of Milan-Bicocca, Milan, Italy.}\\
$^{8}\,${Courant Institute of Mathematical Sciences, New York University, New York, USA.}\\
$\quad$\\
$^{\star}\,${Corresponding author, {\tt giulio.caravagna@ed.ac.uk}.}\\

\tableofcontents
\listoffigures

\newpage
\normalsize

\begin{abstract}
The genomic evolution inherent to cancer relates directly to a renewed focus on the voluminous next generation sequencing (NGS) data, 
and machine learning for the inference of explanatory models of how the (epi)genomic events are choreographed in
cancer initiation and development. However, despite the increasing availability of multiple additional -omics data,
this quest has been frustrated by various theoretical and technical hurdles, mostly stemming from the dramatic heterogeneity of the disease.  
In this paper, we build on  our recent works on ``selective advantage'' relation among driver mutations in cancer progression and investigate its applicability to the modeling problem at the population level. Here, we introduce PiCnIc (Pipeline for Cancer Inference), a versatile, modular and customizable pipeline to extract ensemble-level  progression models  from cross-sectional sequenced cancer genomes.  The pipeline has many translational implications as it combines  state-of-the-art techniques for  sample stratification,  driver  selection,  identification of fitness-equivalent exclusive alterations and progression model inference.  We demonstrate PiCnIc's ability to reproduce much of the  current knowledge on colorectal cancer progression, as well as to suggest novel experimentally verifiable hypotheses. 
\end{abstract}


\noindent{\sc Statement of significance:} {\em A causality based new machine learning  Pipeline for Cancer Inference (\algname{PicNic}) is introduced to infer the underlying somatic evolution of ensembles of tumors from   next generation sequencing data. \algname{PicNic} combines  techniques for  sample stratification,  driver  selection and  identification of fitness-equivalent exclusive alterations to exploit a novel  algorithm based on Suppes' probabilistic causation. The accuracy and translational significance of the results are studied in details, with  an application to colorectal  cancer. \algname{PicNic} pipeline has been made publicly accessible for reproducibility, interoperability and for future enhancements. }

\section{Introduction}

{S}ince the late seventies evolutionary dynamics, with its interplay between variation and selection, has progressively provided the widely-accepted paradigm for the interpretation of cancer emergence and development \cite{nowell1976clonal,fidler1978tumor,dexter1978heterogeneity}. Random alterations of an organism's (epi)genome can sometimes confer a functional \added{\emph{selective advantage}}\footnote{\added{For this and other technical terms commonly used in the statistics and cancer biology communities we provide a Glossary in the Supplementary Material.}}
 to certain cells, in terms of adaptability and ability to survive and proliferate. Since the consequent \emph{clonal expansions} are naturally constrained by the availability of resources (metabolites, oxygen, etc.), further mutations in the emerging heterogeneous tumor populations are necessary to provide additional \emph{fitness} of different kinds that allow survival and proliferation in the unstable micro environment. Such further advantageous mutations will eventually allow some of their sub-clones to outgrow the competing cells, thus enhancing tumor's heterogeneity as well as its ability to overcome future limitations imposed by the rapidly exhausting resources.  \added{Competition, predation, parasitism and cooperation have been in fact theorized as co-present among cancer clones \cite{merlo2006cancer}.}

In the well-known vision of Hanahan and Weinberg \cite{hanahan2000hallmarks,hanahan2011hallmarks}, the phenotypic stages that characterize this multistep evolutionary process are called \emph{hallmarks}. These can be acquired by cancer cells in many possible alternative ways, as a result of a complex biological interplay at several spatio-temporal scales that is still only partially deciphered \cite{huang2009cancer}. In this framework, we distinguish ``alterations'' {driving the hallmark acquisition process} (i.e., {\em drivers}) by activating  \emph{oncogenes} or inactivating \emph{tumor suppressor genes},  from  those that are transferred to sub-clones  without  increasing their fitness (i.e., {\em passengers}) \cite{futreal2004census}. 
Driver identification is a modern challenge of cancer biology, as distinct cancer types exhibit very different combinations of drivers, some cancers display mutations in hundreds of genes \cite{vogelstein2013cancer}, and the majority of drivers is mutated at  low frequencies (``long tail'' distribution), hindering their detection only from the statistics of the recurrence at the population-level  \cite{garraway2013lessons}.  
 
Cancer  clones  harbour distinct types of alterations. The \emph{somatic} (or \emph{genetic}) ones involve either few nucleotides or larger chromosomal regions. They  are usually catalogued as \emph{mutations} - i.e.,  single nucleotide or  structural variants at multiple scales (insertions, deletions, inversions, translocations)  --  of which only some are detectable as {\em Copy Number Alterations} (CNAs),  most prevalent in many tumor types  \cite{zack2013pan}.  Also \emph{epigenetic} alterations, such as \emph{DNA methylation} and \emph{chromatin reorganization}, play a key role in the process \cite{baylin2011decade}.  The overall picture is confounded by  factors such as  \emph{genetic instability} \cite{weinberg2013biology}, \emph{tumor-microenvironment} interplay \cite{albini2007tumour,greaves2012clonal}, and by the influence of \emph{spatial organization} and \emph{tissue specificity} on tumor development \cite{nowak2003linear}\footnote{We mention that much attention has been recently casted on newly discovered cancer genes affecting global processes that are apparently not directly related to cancer development, such as cell signaling, chromatin and epigenomic regulation, RNA splicing, protein homeostasis, metabolism  and lineage maturation \cite{garraway2013lessons}.}.

\added{Significantly, in many cases, distinct driver alterations can damage in a similar way the same \emph{functional pathway}, leading to the acquisition of new hallmarks \cite{vogelstein2004cancer,nowak2006evolutionary,wood2007genomic,jones2008core,parsons2008integrated}. Such alterations individually provide an equivalent \emph{fitness gain} to cancer cells, as any additional alteration hitting the same pathway would provide no further selective advantage. This dynamic results in groups of driver alterations that form \emph{mutually exclusive} patterns across tumor samples from different patients (i.e., the sets of alterations that are involved in the same pathways tend not to occur mutated together). This phenomenon has significant translational consequences.}

An immediate challenge posed by this state of affairs is the dramatic {\em  heterogeneity} of cancer, both at the \emph{inter-tumor} and at the \emph{intra-tumor} levels \cite{fisher2013cancer}. The former manifests as different patients with the same cancer type can display few common alterations. This obsersvation led to the development of techniques to stratify tumors into {\em subtypes} with different genomic signatures, prognoses and response to  therapy  \cite{curtis2012genomic}. The latter form of heterogeneity refers to the observed genotypic and phenotypic variability among the cancer cells within a single neoplastic lesion, characterized by the coexistence of more than one cancer clones with distinct evolutionary histories \cite{gerlinger2012intratumor}.

Cancer heterogeneity poses a serious problem from the diagnostic and therapeutic perspective as, for instance, it is now acknowledged that a single biopsy  might  not be representative of other parts of the tumor, hindering the problem of devising effective treatment strategies  \cite{merlo2006cancer}.  Therefore, presently the quest for an extensive etiology of cancer heterogeneity and for the identification of cancer evolutionary trajectories is  central to cancer research, which attempts to exploit the  massive amount of  sequencing data  available through public projects such as \algname{The Cancer Genome Atlas} (\algname{TCGA}) \cite{tcga}. 

Such projects involve an increasing number of \emph{cross-sectional}  (epi)genomic profiles collected via single biopsies of patients with various cancer types, which might be used to extract trends of cancer evolution across a population of samples\footnote{\added{At the time of this writing, in {TCGA},  sample sizes per cancer type  are in the order of a few hundreds. Such numbers are expected to increase in the near future, with a clear benefit for all the statistical approaches to analyze cancer data which currently lack a proper background of data.}}. Higher resolution data such as \emph{multiple samples} collected from the same tumor \cite{gerlinger2012intratumor}, as well as  \emph{single-cell} sequencing data \cite{navin2011tumour}, might be complementarily used to face the same problem within a specific patient. However,  the lack of public data coupled to the problems of accuracy and reliability, currently prevents a straightforward application\cite{eberwine2014promise}.  

These different perspectives lead to the different mathematical formulations of the problem of {\em inferring a cancer progression model} from genomic data, and a need for versatile  computational tools to analyze data reproducibly -- two intertwined issues examined at length in this paper \cite{beerenwinkel2015cancer}. Indeed, such models and tools can  be  focused either on  characteristics of a population, i.e.  {\em ensemble-level}, or on multiple clonality in a {\em single-patient}. In general, both problems deal with understanding the {\em temporal ordering of somatic alterations} accumulating during cancer evolution, but use orthogonal perspectives and different input data -- see  Figure \ref{fig:probIem} for a comparison. \added{This paper proposes a new computational approach to efficiently deal with various aspects of the problem at a patient population level, relegating the other aspects to future publications.}


\paragraph{Ensemble-level cancer evolution.} It is thus desirable to extract a \emph{probabilistic graphical model}   explaining the statistical trend of accumulation of somatic alterations in a population of $n$  cross-sectional samples collected from patients diagnosed with a specific cancer. To normalize against the experimental conditions in which tumors are sampled, we only consider the {\em list of alterations detected per sample} -- thus, as 0/1 Bernoulli random variables. 

Much of the difficulty lies in estimating the true and unknown trends of \emph{selective advantage} among genomic alterations in the data,   from such observations. This hurdle is  not unsurmountable, if we constrain the scope to only those alterations that are \emph{persistent across tumor evolution in all sub-clonal populations}, since it yields a consistent model of a temporal ordering of mutations. Therefore, epigenetic and trascriptomic states, such as hyper and hypo-methylations or over and under expression, could only be used, provided that they are persistent through tumor development \cite{ramchandani1999dna}.

Historically,  the linear model of colorectal \addedthree{tumor} progression by Vogelstein is an instance of an early solution to the cancer progression  problem \cite{vogelstein1988genetic}. That approach was later generalized to accommodate  {\em tree-models of branched evolution} \cite{desper_1999,desper_2000,szabo,beerenwinkel_2005} and later, further generalized  to the inference of {\em directed acyclic graph} models, with several distinct strategies\cite{beerenwinkel_2007,gerstung2009quantifying,attolini2010mathematical,misra2014inferring}. We  contributed to this research program with the \algname{Cancer Progression Extraction with Single Edges} (\algname{CAPRESE})  and the \algname{Cancer Progression Inference} (\algname{CAPRI}) algorithms, which are currently implemented in  \algname{TRONCO},  an open source  \algname{R} package for \algname{Translational Oncology} available in standard repositories \cite{Loohuis:2014im,Ramazzotti15092015,TRONCO}. Both techniques rely on Suppes' theory of probabilistic causation to define estimators of selective advantage \cite{suppes1970probabilistic}, are robust to the presence of noise in the data and perform well even  with limited sample sizes.
 The former algorithm exploits  shrinkage-like statistics to extract a tree model of progression, the latter combines  bootstrap and maximum likelihood estimation with regularization to extract general directed acyclic graphs  that capture branched, independent and  confluent evolution. Both algorithms represent the  current state-of-the-art  approach to this problem, as they outperform others in speed, scale and predictive accuracy.
 
\paragraph{Clonal architecture in individual patients.} 
\added{A closely related problem addresses  the detection of clonal signatures and their prevalence in individual tumors, a problem complicated by {\em intra-tumor} heterogeneity.}

\added{Even though this phylogenetic version of the progression inference problem naturally relies on data produced from {\em single-cell sequencing} assays \cite{navin2014cancer,wang2014clonal},  the majority of approaches still make use of {\em bulk sequencing}  data, usually from multiple biopsies of the same tumors \cite{gerlinger2012intratumor, gerlinger2014genomic}. 
Indeed, several approaches try to extract the clonal signature of single tumors from {\em allelic imbalance proportions}, a problem made difficult as sequenced samples usually contain a large number of cells belonging to a collection of sub-clones resulting from the complex evolutionary history of the tumor \cite{oesper2013theta,oesper2014quantifying,miller2014sciclone, roth2014pyclone,jiao2014inferring,fischer2014high,zare2014inferring,garvin2015interactive,malikic2015clonality,el2015reconstruction}.}

 \added{We keep the current work   focused on the inference of progression models at the ensemble level,  and plan to return to  this variant to the problem in another publication.}

\section{The {\bpipe{}} pipeline}

\begin{figure*}[p]
 \centerline{\includegraphics[width=.95\textwidth]{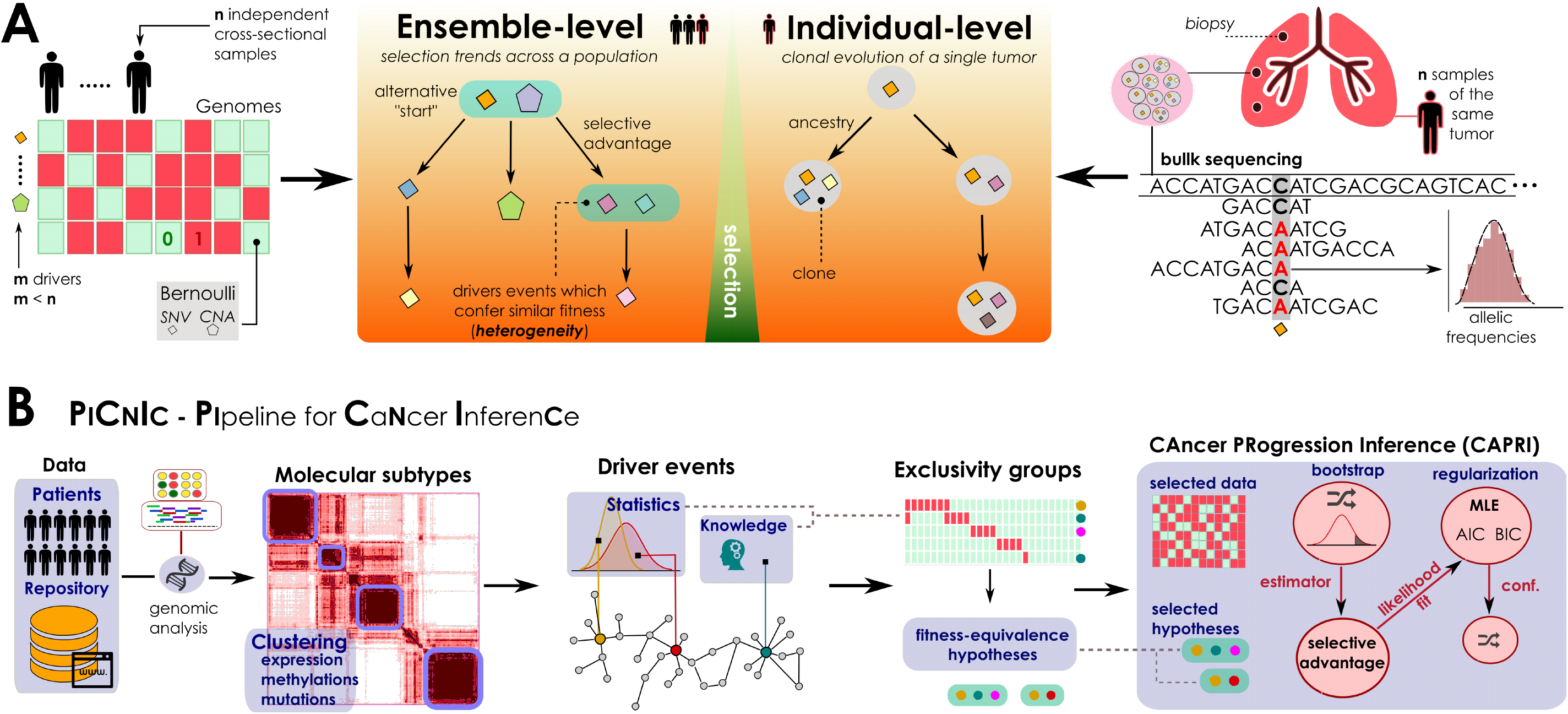}}
\caption[Problem statement and overview of the \algname{PicNic}  pipeline]{{\bf A.} Problem statement. (left) Inference of ensemble-level cancer progression models  from a cohort of $n$  independent patients (cross-sectional). By examining a list of somatic mutations or CNAs per patient (0/1 variables) we infer a probabilistic graphical model of the temporal ordering of fixation and accumulation of such alterations in the input cohort. Sample size and tumor heterogeneity  complicate the problem of extracting population-level trends, as this requires accounting for patients' specificities such as multiple starting events. (right)  For an individual tumor, its clonal phylogeny and prevalence is usually inferred from multiple biopsies  or single-cell sequencing data. Phylogeny-tree reconstruction from an underlying statistical model of reads coverage or depths  estimates  alterations' prevalence in each clone, as well as ancestry relations.  This problem is mostly worsened by the high intra-tumor heterogeneity and sequencing issues. {\bf B.}  The PiCnIc  pipeline for ensemble-level  inference includes several sequential steps to reduce tumor heterogeneity, before applying the CAPRI \cite{Ramazzotti15092015} algorithm.
 Available mutation, expression or methylation data  are first used to stratify patients into distinct tumor molecular subtypes, usually  by exploiting clustering tools.  
Then, subtype-specific alterations driving cancer initiation and progression are identified with  statistical tools and on the basis of prior knowledge. 
Next  is the identification of the fitness-equivalent groups of  mutually exclusive alterations across the input population,  again done with computational tools or biological priors. Finally, CAPRI processes a set of relevant alterations within such groups. Via bootstrap and hypothesis-testing,  CAPRI 
extracts a set of  ``selective advantage relations'' among them, which is eventually narrowed down via maximum likelihood estimation with regularization (with various scores). The ensemble-level progression model is obtained by combining such relations in a graph, and its confidence is assessed via various bootstrap and cross-validation techniques.}
\label{fig:probIem}
\end{figure*}

We report on the design, development and evaluation of  \added{the {\acronym{}} (\pipe{})}   to extract ensemble-level cancer progression models from cross-sectional data (Figure \ref{fig:probIem}).  \pipe{} is versatile, modular and \added{customizable}; it exploits state-of-the-art \added{data processing and machine learning} tools   to:
\added{ 
\begin{enumerate}
\item identify  {\em tumor subtypes} and then in each subtype;
\item select  (epi)genomic events  {\em relevant} to the progression;
\item identify  groups of events that are likely to be observed as {\em mutually exclusive};
\item  infer  {\em progression models} from groups and related data,  and annotate them with associated statistical confidence.
\end{enumerate}
 All these steps are necessary to minimize the confounding effects of  inter-tumor heterogeneity,  which are likely to lead to wrong results when data is not appropriately pre-processed\footnote{\added{The genuine selectivity  relationship sought to be inferred are subject to the vagaries of Simpson's paradox;  it can change, or worst  reverse, when we try to infer them from data not  suitably pre-processed. This effect (due to such paradox) manifests as data are sampled from a highly heterogenous mixture of populations of cells\cite{Ramazzotti15092015}. PiCnIc uses various mechanisms to avoid these pitfalls. In this context, it should be pointed out that  input bulk sequencing data  suffers also from  intra-tumor heterogeneity issues, which  are unfortunately intrinsic to the technology.}}.}

\added{In each stage of \pipe{} different techniques can be employed, alternatively or jointly, according to} specific research goals, input data, and cancer type. \added{Prior knowledge can be easily accommodated into our pipeline, as well as  the  computational tools discussed in the next subsections and summarized in Figure \ref{tab:pipeline}. The rationale   is similar  in spirit to workflows implemented by consortia such as \algname{TCGA} to analyze huge populations of cancer samples \cite{cancer2012comprehensive,cancer2013genomic}. One of the main novelties of our approach,  is the  exploitation of groups of exclusive alterations as a proxy to detect fitness-equivalent trajectories of cancer progression. This strategy is only feasible  by the hypothesis-testing features of the recently developed \algname{CAPRI} algorithm, an algorithm uniquely addressing this crucial aspect of the ensemble-level progression inference problem\cite{Ramazzotti15092015}.}

\added{
In the Results  section, we study in details a specific use-case for the pipeline, processing colorectal cancer data from \algname{TCGA}, where it is able to  re-discover much of the existing body of knowledge about colorectal cancer progression. Based on the output of this pipeline, we also propose novel experimentally-verifiable hypotheses.}

\subsection{Reducing inter-tumor heterogeneity by cohort subtyping}

 In general, for each of  $n$ tumors (patients) we assume relevant (epi)genetic data to be available. We do not put constraints on data gathering and selection, leaving the user to decide the appropriate ``resolution'' of the input   data. For instance, one might decide whether  somatic mutations  should be classified by type or by location, or aggregated. Or, one might decide to lift  focal  CNAs to the \deleted{wider} \addedthree{lower} resolution of cytobands or full arms \added{(e.g., in a kidney cancer cohort where    very long CNAs  are more common than focal events  \cite{cancer2013comprehensive})}.
These choices depend on data and on the overall understanding of such alterations and their functional effects for the cancer under study,
  and no single all-encompassing rationale may be provided. 

With these data at hand, we might wish to identify cancer subtypes in the  {\em heterogeneous mixture} of input samples. In some cases the classification can benefit from clinical biomarkers, such as evidences of certain cell types \cite{bennett1976proposals}, but  in most cases we will have to  rely on multiple {\em clustering} techniques at once, see, e.g., \cite{cancer2012comprehensive,cancer2013genomic}. Many common approaches  cluster expression profiles \cite{lu2005microrna}, often relying  on non-negative matrix factorization techniques \cite{gao2005improving} or earlier approaches such as $ k$-means, Gaussians mixtures or hierarchical/spectral clustering - see the review in  \cite{de2008clustering}.  For  glioblastoma and breast cancer, for instance, mRNA expression subtypes provides good correlation with clinical phenotypes  \cite{cancer2011integrated,konstantinopoulos2010gene,reis2011gene}. However, this is not always the case as, e.g., in colorectal cancer such clusters mismatch with  survival and chemotherapy response \cite{cancer2011integrated}. Clustering of  {\em full exome} mutation profiles or smaller panels of genes might be an alternative as it was shown for ovarian, uterine and lung cancers  \cite{hofree2013network,NBS-critique}. 

\added{Using pipelines such as \pipe{}, we expect that the resulting subtypes will be routinely investigated, eventually leading to distinct progression models, which shall be characteristic of the population-level trends of cancer initiation and progression.} 
 
 \subsection{Selection of driver events} 
In subtypes detection,   it becomes easier  to find similarities across input samples when more alterations are available, as features selection gains precision. In  progression inference, instead, one wishes  to focus  on $m \ll n$ {\em driver} alterations, which ensure also an appropriate statistical ratio between sample size ($n$, here the subtype size) and problem dimension ($m$). 

Multiple tools   filter  out driver from  passenger mutations. \algname{MutSigCV}  identifies drivers mutated more frequently than background mutation rate \cite{lawrence2013mutational}.   \algname{OncodriveFM} avoids such estimation but  looks for functional mutations \cite{gonzalez2012functional}. \algname{OncodriveCLUST} scans mutations  clustering in small regions of the protein sequence  \cite{tamborero2013oncodriveclust}. \algname{MuSiC} uses  multiple types of clinical data to establish correlations among mutation sites,  genes and pathways \cite{dees2012music}. Some other tools search for  driver  CNAs that affect protein expression \cite{tamborero2013oncodrive}.  All these approaches use different statistical measures to estimate signs of positive selection, and we suggest  using them in an orchestrated way, as done by  platforms such as \algname{Intogen} \cite{gundem2010intogen}. 

\added{We anticipate that such tools will run independently on each subtype, as  driver genes will likely differ across them, mimicking the different molecular properties of each group of samples; also, lists of genes produced by these tools might be augmented with  prior knowledge about tumor suppressors or oncogenes.}

\subsection{Fitness equivalence of exclusive alterations}
When working at the ensemble-level, identification of  ``groups of mutually exclusive'' alterations is crucial to derive a correct inference. \added{This step of \pipe{} is another  attempt to resolve part of the inter-tumor heterogeneity, as such alterations {\em could} lead to the same phenotype (i.e., hence resulting ``equivalent'' in terms of progression), despite being genotypically  ``alternative'', i.e., exclusive, across the input cohort. This information shall be used to detect alternative routes to cancer progression which capture the specificities of individual patients.}

 A plethora of recent tools can be used to detect groups of fitness equivalent alterations, according to the data available for each subtype;    greedy approaches   \cite{yeang2008combinatorial,miller2011discovering} or their optimizations, such as \algname{MEMO}, which constrain search-space with 
network priors    \cite{ciriello2012mutual}. This  strategy is further improved in \algname{MUTEX}, which  scans  mutations and focal CNAs for genes  with a common downstream effect in a curated signalling network, and selects only those genes that significantly contributes to the  exclusivity pattern \cite{babur2015systematic}. 
Other  tools such as \algname{Dendrix},  \algname{MDPFinder}, \algname{Multi-Dendrix},  \algname{CoMEt}, \algname{MEGSA} or \algname{ME}, employ advanced statistics  or generative approaches   without  priors\cite{vandin2012novo,zhao2012efficient,leiserson2013simultaneous,leiserson2015comet,MEGSA,szczurek2014modeling}. 

In such groups,  we distinguish between {\em hard} and {\em soft} forms of exclusivity, the former assuming strict exclusivity among alterations, with random errors accounting for possible overlaps (i.e., the majority of samples do not share alterations from such groups), the latter admitting co-occurrences (i.e., some samples might have common alterations, within a group) \cite{babur2015systematic}.  

 \algname{CAPRI} is currently the only algorithm which incorporates  this type of information, in inferring a model. Each of these groups are in fact   associated with a {\em ``testable hypothesis''} written in the well-known language of  {\em propositional  Boolean formulas}\footnote{There,  logical connectives such as $ \oplus$ (the logical  ``xor'') act as  a proxy for hard-exclusivity, and  $ \vee$ (the logical ``disjunction'') for  soft one. Besides from exclusivity groups, other connectives such as logical conjunction can be used.}. \added{Consider the following example: we might be informed that \g{apc} and \g{ctnnb1} {\em mutations} show a trend of  soft-exclusivity in our cohort -- i.e., some samples harbor both mutations, but the majority just one of the  two mutated genes.  Since such mutations lead to $\beta$-catenin deregulation (the phenotype), we might wonder whether such state of affairs could be responsible for progression initiation in the tumors under study. An affermative response would equate, in terms of progression, the two mutations. To {\em test this hypothesis}, one may  spell out formula \g{apc} $\vee$ \g{ctnnb1} to  \algname{CAPRI}, which means that we are suggesting to the inference engine that, besides the possible evolutionary trajectories that might be inferred by looking at the two mutations as {\em independent},  trajectories involving  such a ``composite'' event, shall be considered as well. It is then up to \algname{CAPRI} to decide which, of all such 
trajectories, is significant, in a {\em statistical sense}.}

\added{In general, formulas allow users to test general  hypotheses about complex model structures involving multiple genes and alterations. These  are useful  in many cases: for instance, where we are processing samples which harbour  homozygous losses or inactivating mutations in certain genes (i.e., equally disruptive genomic events), or when we know in advance that certain genes are controlling the same pathway, and we might speculate that a single hit in one of those decreases  the selection pressure on the others. We note that,  with no hypothesis, a model with such    alternative trajectories {\em cannot be analyzed,  due to various computational}   limitations inherent to   the inferential algorithms (see \cite{Ramazzotti15092015}).}

\added{From a practical point of view, \algname{CAPRI}'s  formulas/hypotheses-testing features ``help'' the inference process, but do not ``force'' it to select a specific model, i.e., the {\em inference is not biased}. In this sense, the trajectories inferred by examining these composite model structures  (i.e., the formulas) {\em are not given any statistical advantage} for inclusion in the final model. However, in spite of a natural temptation  to generate as many  hypotheses as possible, it is prudent to always limit the number of hypotheses according to the number of samples and alterations.  Note that this approach can also be 
extended to accommodate, for instance, co-occurrent alterations in 
significantly mutated subnetworks\cite{leiserson2015pan,vandin2011algorithms}.}

\subsection{Progression inference and confidence estimation}

We  use \algname{CAPRI} to reconstruct cancer progression models of each identified molecular subtype, provided that there exist a reasonable  list of driver events and the groups of fitness-equivalent exclusive alterations. \added{Since currently \algname{CAPRI}  represents the state of the art,  and supports complex formulas for groups  of alterations detected in the earlier \pipe{} step, it was well-suited for the task.}

\algname{CAPRI}'s input is a binary $ n \times (m+k)$ matrix  $\mathbf{M}$ with $ n$ samples (a subtype size), $m$ driver alteration events (0/1 Bernoulli random variables) and $ k$ testable formulas. \added{Each sample in $\mathbf{M}$ is described by a binary sequence: the 1's denote the presence of alterations.}
 \algname{CAPRI} first  performs a computationally fast {\em scan} of $ \mathbf{M}$ to identify a set \deleted{of} $\cal S$ \addedthree{of} plausible selective advantage relations among the driver alterations and the formulas;  then, it reduces $\cal S$ to  the  most  relevant ones, $\hat{\cal S} \subset\cal S$
\added{Each relation is represented as an edge connecting drivers/formulas in a Graphical Model -- which shall be termed  Suppes-Bayes Causal Network. This network represents the {\em joint probability distribution}\footnote{Technically, for a set of $m$ alterations modeled by  variables $\xx_1, \ldots, \xx_m$, such a network is a Graphical Model representing the factorization of the joint distribution -- $\Probab{\xx_1, \ldots, \xx_m}$ -- of observing any of the alterations in a  genome (i.e.,  $\xx_i=1$). This factorization is made compact as the model encodes the statistical dependencies in its structure   via  
\[
\Probab{\xx_1, \ldots, \xx_m} = \prod_{i=1}^m  \Probab{\xx_i \mid \pi_i}
\]
 where $\pi_i = \{ \xx_j \mid \xx_j \to \xx_i \in \hat{\cal S}\}$ are the ``parents''  of the $i$-th node. These are those from which 
 the presence of the $i$-th alteration is predicted. In our approach these edges are the pictorial representation of the  selective advantage relations where the alterations in $\pi_i$ select for $\xx_i$.}  of observing a set of driver alterations in a cancer genome, subject to constraints imposed by Suppes' {\em probabilistic causation} formalism\cite{suppes1970probabilistic}.}


\addedtwo{Set $\cal S$ is built by a statistical procedure. Among any pair of input drivers/formulas $x$ and $y$, \algname{CAPRI} postulates that $x\to y \in {\cal S}$ could be a selective advantage relation with ``$x$  selecting for $y$''   if   it estimates  that two conditions hold
\begin{enumerate}
\item  ``$x$ is earlier than $y$'';
\item ``$x$'s presence increases the probability of observing $y$''.
\end{enumerate}
Such claims, grounded in Suppes' theory of probabilistic causation, are expressed as inequalities over {\em marginal} and {\em conditional} distributions of $x$ and $y$. These are assessed via a standard Mann-Withney U test after the distributions are estimated from a reasonable number (e.g., 100) of {\em non-parametric bootstrap resamples} of $\mathbf{M}$ (see Supplementary Material). \algname{CAPRI}'s increased performance over existing methods can be motivated by the reduction of the state space within which models are searched, via ${\cal S}$.}

\added{Optimization of $\cal S$  is central to our tolerance to {\em false positives} and {\em negatives}  in $\hat{\cal S}$.  We would like to select only the minimum number of relations  which are true and statistically supported, and build  our model from those.  \algname{CAPRI}'s implementation in \algname{TRONCO} \cite{TRONCO}  selects a subset by optimizing  a {\em score function} which assigns to a model a real number equal to its  {\em log-likelihood} (probability of generating data for the model) minus a {\em penalty term} for model complexity -- a regularization term increasing with  $\hat{\cal S}$'s size, and hence penalizing overly complex models.} It is a standard approach to avoid overfitting, and usually relies on  the  \algname{Akaike} or the \algname{Bayesian Information Criterion}  (\algname{AIC} or \algname{BIC}) as regularizers.  Both scores  are  approximately correct; \algname{AIC} is more prone to overfitting but likely to provide also good  predictions from data and is better  when  false negatives are   more misleading than positive ones. \algname{BIC} is more prone to underfitting errors, thus   more  parsimonious and better in opposite direction. As often done, we suggest approaches that to combine but  distinguish which relations are selected by \algname{BIC} versus \algname{AIC}. Details on the algorithm are provided as see Supplementary Material.

\paragraph{Statistical confidence of a model.} \added{In-vitro and in-vivo experiments  provide the most convincing validation for the newly suggested selective advantage relations and hypotheses, yet this is out of reach in some cases.} 


\addedtwo{Nonetheless, statistical validation approaches can be used almost universally to assess the  confidence of edges, parent sets and  whole models, either via  {\em hypothesis-testing} or  {\em bootstrap} and {\em cross-validation} scores for Graphical Models. We briefly discuss  approaches that are implemented in \algname{TRONCO}, and refer to the Supplementary Materials for additional details.} 

\addedtwo{First, \algname{CAPRI}   builds ${\cal S}$ by computing  two p-values per edge,  for the  confidence in  condition $(1)$ and $(2)$. In addition, for each edge $x \to y$, it computes a third p-value via hypergeometric testing against the hypothesis that the co-occurrence of  $x$ and $y$ is due to chance. These p-values measure confidence in the  direction of each edge and the amount of statistical dependence among $x$ and $y$.}

\addedtwo{Second, for each model inferred with  \algname{CAPRI} we can estimate (\emph{a posteriori}) how frequently our edges would be retrieved if we resample from our data ({\em non-parametric} bootstrap), or from the model itself, assuming its correctness ({\em parametric} bootstrap) \cite{efron1994introduction}.  Also, we can measure the bias in \algname{CAPRI}'s construction of ${\cal S}$ due to the random procedure which estimates the  distributions in condition $(1)$ and $(2)$ ({\em statistical} bootstrap).} 

\added{Third, scores can be computed to quantify the consistency for the model against bias in the data and models.  For instance, {\em non-exhaustive} $k$-fold  cross-validation   can be used to compute the  {\em entropy loss} for the whole model, and the {\em prediction} and {\em posterior classification errors}  for each edge or parent set \cite{bayesian_learning}.}

\section{Results}

\subsection{Evolution in a population of MSI/MSS colorectal tumors.} \label{sec:msi-mss}

It is common knowledge that {\em colorectal cancer} (CRC) is a heterogeneous disease comprising different molecular entities. Indeed, it is currently accepted that colon tumors can be classified according to their global genomic status into two main types: {\em microsatellite \deleted{instable} \addedthree{unstable} tumors} (MSI), further classified as high or low,  and {\em microsatellite stable} (MSS) tumors (also known as tumors with {\em chromosomal instability}).  This taxonomy plays a significant role in determining pathologic, clinical and biological characteristics of CRC tumors \cite{ogino2008molecular}.  Regarding molecular progression, it is also well established that each subtype arises from a distinctive molecular mechanism. While MSS tumors generally follow the classical adenoma-to-carcinoma progression described in the seminal work by Vogelstein and Fearon \cite{fearon1990genetic}, MSI tumors result from the inactivation of DNA mismatch repair genes like \g{mlh}-1 \cite{vilar2010microsatellite}.

\added{With the aid of the \algname{TRONCO} package, we instantiated  \pipe{}   to process colorectal tumors freely available  through \algname{TCGA} project \algname{COADREAD} \cite{cancer2012comprehensive} (see Supplementary Figure S1), and inferred  models for  the MSS and MSI-HIGH tumor subtypes  (shortly denoted MSI) annotated by the consortium. In doing so, we used a combination of  background knowledge produced by \algname{TCGA} and new computational predictions;  to a different degree,  some knowledge comes from manual curation of data and other from    tools mentioned in  \pipe{}'s description (see Figure \ref{tab:pipeline}). Data and exclusivity groups for MSI tumors are shown in  Figure \ref{fig:MT-trdataset-MSI}, the analogous  for MSS tumors is provided as Supplementary Material.} 

\added{For the models inferred, which are shown in Figures \ref{fig:MT-model-MSS} and \ref{fig:MT-model-MSI}, we evaluated various forms of statistical confidence measured as  p-values, bootstrap scores (in what follows, \algname{npb} denotes non-parametric bootstrap and the closer to 100 the better), and cross-validation statistics reported in the Supplementary Material.  Many of the  postulated selective advantage relations  (i.e., model edges) have very strong statistical support for \algname{COADREAD} samples,   although events with similar marginal frequency may lead to ambiguous imputed  temporal ordering (i.e., the edge direction). In general, we observed that  overall the estimates are slightly better in the MSS cohort (entropy  loss $< 1\%$ versus $3.8\%$), which is expected given the difference in sample size of the two datasets (152 versus 27 samples), see Material and Methods for details.}

\paragraph{Interpretation of the models.} 

Our models  capture the well-known features distinguishing MSS and MSI tumors: for the former \g{apc}, \g{kras} and \g{tp53}  mutations as primary events  together with chromosomal aberrations, for the latter \g{braf} mutations and  lack of chromosomal alterations. Of all 33 driver genes, 15 are common to both models - e.g., \g{apc}, \g{braf}, \g{kras}, \g{nras}, \g{tp53} and \g{fam123b} among others (mapped to pathways like \g{wnt}, \g{mapk}, apoptosis or activation of T-cell lymphocites),  although in different relationships (position in the model), whereas new (previously un-implicated) genes stood out from our analysis and deserve further research. 

\begin{description}
\item {\em MSS (Microsatellite Stable).} In agreement with the known literature, in addition to \g{kras}, \g{tp53} and \g{apc} as primary events, we identify \g{pten} as a late event in the carcinogenesis, as well as \g{nras} and \g{kras} converging in \g{igf2} amplification, the former being ``selected by''   \g{tp53} mutations \added{(\algname{npb} 49\%)}, the latter ``selecting for''  \g{pik3ca} mutations \added{(\algname{npb} 81\%)}.  The leftmost portion of the model links many \g{wnt} genes, in agreement with the observation that multiple concurrent lesions affecting such pathway confer selective advantage. In this respect, our model predicts  multiple routes for the selection  of alterations in \g{sox9} gene, a transcription factor known to be active in colon mucosa \cite{abdel2011minisox9}. Its mutations are directly selected  by \g{apc}/\g{ctnnb1}  alterations \added{(though with low \algname{npb} score)}, by \g{arid1a}  \added{(\algname{npb} 34\%)} or  by   \g{fbxw7} mutations \added{(\algname{npb} 49\%)},  an early mutated gene that both directly, and in a redundant way via \g{ctnnb1}, relates to  \g{sox9}. The \g{sox} family of transcription factors have emerged as modulators of canonical \g{wnt}/$\beta$-catenin signaling in many disease contexts \cite{kormish2010interactions}. Also interestingly, \g{fbxw7} has been previously reported to be involved in the malignant transformation from adenoma to carcinoma \cite{li2014sequential}. The rightmost part of the model involves genes from various pathways, and outlines the relation between \g{kras} and the \g{pi3k} pathway. We indeed  find selection of \g{pik3ca} mutations by \g{kras} ones, as well as selection of  the  whole \algname{MEMO} module \added{(\algname{npb} 64\%)},   which is responsible for  the activation of the \g{pi3k} pathway \cite{cancer2012comprehensive}. \g{smad4} proteins relate either to \g{kras} \added{(\algname{npb} 34\%)}, and \g{fam123b} (through \g{atm}) and \g{tcf7l2}  converge in \g{dkk2} or \g{dkk4} \added{(\algname{npb} 81, 17 and 34\%)}.

\item {\em MSI-HIGH (Microsatellite \deleted{Instable} \addedthree{Unstable}).}  
In agreement with the current literature, \g{braf} is the most commonly mutated gene in MSI tumors  \cite{kim2014molecular}. \algname{CAPRI}  predicted convergent evolution of  tumors harbouring \g{fbxw7} or \g{apc} mutations towards deletions/mutations of \g{nras} gene \added{(\algname{npb} 21, 28 and 54\%)}, as well as selection of \g{smad2} or \g{smad4} mutations by \g{fam123b} mutations \added{(\algname{npb} 23 and 46\%)}, for these tumors. Relevant to all MSI tumors seems again the role of the \g{pi3k} pathway. Indeed, a   relation among  \g{apc} and \g{pik3ca} mutations was inferred \added{(\algname{npb} 66\%)}, consistent with recent experimental evidences pointing at a synergistic role of these mutations, which co-occurr in the majority of human colorectal cancers \cite{deming2014pik3ca}. Similarly, we find consistently a selection trend among \g{apc} and the whole \algname{MEMO} module \added{(\algname{npb} 48\%)}. Interestingly, both mutations in \g{apc} and \g{erbb3} select for \g{kras} mutations \added{(\algname{npb} 51 and 27\%)}, which might point to interesting therapeutic implications. 
In contrast, mutations in \g{braf} mostly select for mutations in \g{acvr1b} \added{(\algname{npb} 36\%)}, a receptor that once activated phosphorylates \g{smad} proteins. It forms receptor complex with \g{acvr2a}, a gene   mutated in these tumors that selects for \g{tcf7l2} mutations \added{(\algname{npb} 34\%)}. Tumors harbouring \g{tp53} mutations are those selected by \deleted{exhibit} mutations in \g{axin2} \added{(\algname{npb} 32\%)}, a gene implicated in \g{wnt} signalling pathway, and related to \deleted{instable} \addedthree{unstable} gastric cancer development \cite{kim2009frameshift}. Inactivating mutations in this gene are important, as it  provides serrated adenomas with a mutator phenotype in the MSI tumorigenic pathway \cite{muto2014dna}. Thus, our results reinforce its putative role as driver gene in these tumors.
\end{description}

By comparing these models we can find  similarity in the prediction of a potential new early event for CRC formation, \g{fbxw7}, as other authors have recently described \cite{li2014sequential}. This tumor suppressor  is frequently inactivated in human cancers, yet the molecular mechanism by which it exerts its anti-tumor activity remains unexplained \cite{zhan2015fbxw7}, and our models provide a new hypothesis in this respect. 

\section{Discussion}

This paper represents our continued \deleted{scientific} exploration of the nature of somatic evolution in cancer, and its translational exploitation through models of cancer progression, models of drug resistance (and efficacy), left- and right-censoring, sample stratification, and therapy design. Thus this paper emphasizes the engineering and dissemination of production-quality computational tools as well as validation of its applicability via use-cases carried out in collaboration with translational collaborators: e.g., colorectal cancer, analyzed jointly with epidemiologists currently studying the disease actively. As anticipated, we reasserted that the proposed model of somatic evolution in cancer not only supports the heterogeneity  seen in tumor population, but also suggests a selectivity/causality relation that can be used in analyzing (epi)genomic data and exploited in therapy design \added{-- which we  introduced in our earlier works  \cite{Loohuis:2014im,Ramazzotti15092015}}. \added{In this paper,   we have introduced an open-source  pipeline, \pipe{},  which minimizes the confounding effects arising from inter-tumor heterogeneity, and we have shown that \pipe{} can be effective in extracting ensemble-level evolutionary trajectories of cancer progression.}

When applied to a highly-heterogeneous  cancer such as  colorectal, \added{\pipe{}} was able to infer the role of  many known events in colorectal cancer progression (e.g., \g{apc}, \g{kras} or \g{tp53} in MSS tumors, and \g{braf} in MSI ones), confirming the validity of our approach\footnote{\added{As a further investigation for CRC, we leave as future work to check whether the inferred progression are also  representative of other  subtyping strategies for colorectal cancer, with particular reference to recent works which show marked interconnectivity between different independent classification systems coalescing into four consensus molecular subtypes  \cite{guinney2015consensus}.} 
}. Interestingly, new players in CRC progression stand out from this analysis such as \g{fbxw7} or \g{axin2}, which deserve further investigation. In colon carcinogenesis, although each model identifies  characteristic early mutations suggesting different initiation events,  both models appear to converge in common pathways and functions  such as \g{wnt} or \g{mapk}.  

However, both models have some clear distinctive features. \deleted{Private} \addedthree{Specific} events in MSS include mutations in intracellular genes like \g{ctnnb1} or in \g{pten}, a well-known tumor suppressor gene.  On the contrary, \deleted{private} \addedthree{specific} mutations in MSI tumors appear in membrane receptors such as  \g{acvr1b}, \g{acvr2a}, \g{erbb3}, \g{lrp5}, \g{tgfbr1} and  \g{tgfbr2}, as well as in secreted proteins like \g{igf2}, possibly suggesting that such tumors need to disturb cell-cell and/or cell-microenvironment communication to grow. At the pathway level,  genes exclusively appearing in the MSI progression model accumulate in \deleted{private} \addedthree{specific} pathways such as cytokine-cytokine receptor, endocytosis and \g{tgf}-$\beta$ signaling pathway. On the other hand, genes in MSS progression model are implicated in \g{p53}, \g{mTOR}, sodium transport or inositol phosphate metabolism. 

\added{Our study also highlighted the translational relevance of the models that we can produce with \pipe{} (see Supplementary Figure S12). The evolutionary trajectories depicted by our models can, for instance, suggest previously-uncharacterized phenotypes, help in finding biomarker molecules predicting cancer progression and therapy response, explain drug resistant phenotypes and predict metastatic outcomes. The  logical structure of the formulas describing alterations with equivalent fitness (i.e., the exclusivity group) can also point to novel targets of therapeutic interventions.  In fact,  exclusivity groups that are found to have a role in the progression can be screened for  {\em synthetic lethality} among such genes -- thus explaining why we do not observe  phenotypes where such alterations co-occur. In this sense, our models  describe also such clonal signatures which, though theoretically possible, are not   selected. We call such conspicuously absent phenotypes
 \emph{anti-hallmarks\/} \cite{loohuis2014cancer}.}

\added{Our models   have other applications to both computational and cancer research. Our models, as encoded by Suppes-Bayes Causal Networks could be used as informative {\em generative models} for the genomic  profiles for the cancer patients. 
In fact, as known in machine learning, such generative models are extremely useful in creating better representation of  data in terms of, e.g.,  discriminative kernels, such as Fisher  \cite{budilya2016}. In practice, this change of representations would allow framing common classification problems  in the domain of our generative  structures, i.e., the models, rather than the data. As a consequence, it is possible to create a new class of  more robust classification and prediction systems.}

\added{One may think of these representations as those bringing us closer to phenotypic (and causal) representation of the patient's tumor, replacing its genotypic (and mutational) representation. We suspect that such representations will improve the accuracy of measurement of the biological clocks, dysregulated in cancer and critically needed to be measured in order to predict survival time, time to metastasis, time to evolution of drug resistance, etc. We believe that these ``phenotypic clocks'' can be used immediately to direct the therapeutic intervention.}

\added{Clearly, applicability and reliability of techniques such as \pipe{} is very much dependent  on the background of data available. At the time of this writing,  the quality, quantity and reliability of (epi)genomic data available, e.g., in public databases, is related to the ever increasing computational and technological improvements characterizing the wide area of cancer genomics. } \added{Of similar importance is the availability of wet-lab technologies for models validation.} \added{Our recent work on SubOptical Mapping technology, for instance,  points to the ability to cheaply and accurately characterize translocation, indels and epigenomic modifications at the single molecule and single cell level \cite{reed2012identifying,sundstrom2012image}. This technology also provides the ability to directly validate (or refute) the hypotheses generated by \pipe{} via gene-correction and single cell perturbation approaches.}

\added{To conclude,  the precision of any statistical inference technique, including \pipe{},  is influenced by the quality, availability and idiosyncrasies of the input data --  the goodness of the outcomes improving along with the expected advancement in the field. Nevertheless, the strength of the proposed approach lies in the efficacy in managing possibly \deleted{unreliable} \addedthree{noisy/ biased} or insufficient data, and in proposing refutable hypotheses for experimental validation.}

\section{Materials and methods}

\paragraph{\bf Processing COADREAD samples with PiCnIc.} 

We instantiated  \spipe{}  to process clinically annotated high MSI-HIGH  and MSS  colorectal tumors collected from   The Cancer Genome Atlas  project {``Human Colon and Rectal Cancer''} (COADREAD) \cite{cancer2012comprehensive} -- see Supplementary Figure S1. Details on the implementation and the  source code to replicate this study are available as Supplementary Material.  COADREAD has enough samples, \added{especially for MSS tumors,} to implement a   consistent and significant statistical validation of our findings -- see  Supplementary Table S1.

 In brief, we split  subtypes by the microsatellite status of each tumor as annotated by the consortium \added{(so, step I of \spipe{} is done by exploiting background knowledge rather than computational predictors). It should be expected that if this step is skipped or this classification is incorrect, the resulting models would noticeably differ. Once split into groups, the input COADREAD data is processed to maintain only samples for which both high-quality curated mutation and CNA  data are available; for CNAs we use focal high-level  amplifications and homozygous deletions.}

Then, for each sample we select only alterations (mutations/CNAs) from a list of  33 driver  genes manually annotated to 5  pathways in \cite{cancer2012comprehensive} - \sg{wnt}, \sg{raf}, \sg{tgf}-$\scriptstyle\beta$, \sg{pi3k} and \sg{p53} (Supplementary Figures S2 and S3). \added{This list of drivers, step II of \spipe{}, is produced by TCGA, as a result of manual curation and running MutSigCV.}

In the next module of the pipeline, we fetch groups of exclusive alterations. We scanned these groups by using the  MUTEX  tool (Supplementary Table S2), and merged its results with  the group  that TCGA detected by  using the MEMO tool, which involves mainly genes from the \g{pi3k} pathway. Knowledge on the potential exclusivity among genes in the \sg{wnt} (\sg{apc},\sg{ctnnb1})  and \sg{raf} (\sg{kras},\sg{nras},\sg{braf}) pathways was exploited as well.
Groups were then used to create CAPRI's formulas; \added{we also included hypotheses for genes which  harbour mutations and homozygous deletions across different samples,  see Supplementary Table S3.} Data and exclusivity groups for MSS tumors are shown  in Supplementary Figure S4 and S5.

CAPRI was run, as the last step of \spipe{}, on each subtype, by  selecting recurrent alterations from the pool of  33 pathway genes  and using both AIC/BIC regularizer. \added{Timings to run the relevant steps of the pipeline are reported in the Supplementary Material.} In the   models of  Figures \ref{fig:MT-model-MSS} and Figure \ref{fig:MT-model-MSI}  each edge  mirrors  selective advantage among the upstream and downstream nodes, as estimated by CAPRI; \addedtwo{Mann-Withney U test  is carried out with statistical  significance  0.05, after 100 non-parametric bootstrap iterations.}

\added{The significance of the reconstructed models and the input data is assessed by computing all the  statistics/tests discussed in the Main text (temporal priority, probability raising and hypergeometric testing p-values, bootstrap  and cross-validation scores). Motivation and background on each of these measures is available in the   Supplementary Materials. A table with their values for  edges with  highest  non-parametric bootstrap scores is in  Supplementary Figure S8.} 

\added{For the MSS cohort all the p-values are strongly significant ($\scriptstyle p\ll0.01$) except for the temporal priority of the edges connecting mutations in  \sg{fam123b} and \sg{atm}, and \sg{erbb2} alterations (mutations and amplifications), which leads us to conclude that, even if these pairs of genes seem to undergo selective  advantage, the temporal ordering of their occurrence is ambiguous and failed to be imputed correctly from the datasets, analyzed here. The same situation occurs in MSI-HIGH tumors,  for the relation between \sg{kras} and \sg{erbb3}. Non-parametric and statistical bootstrap estimations are used to assess the strength of all the findings (Supplementary Figures S6 and S7). Moreover, any bias in the data is finally evaluated by cross-validation (Supplementary Figures S8-S11) and common statistics such as entropy loss, posterior classification and prediction errors. In general, most of the selective advantage relations depicted by the inferred models present a strong statistical support, with the MSS cohort presenting the most reliable results.}

\begin{figure*}[p]
\centerline{\includegraphics[width=1\textwidth]{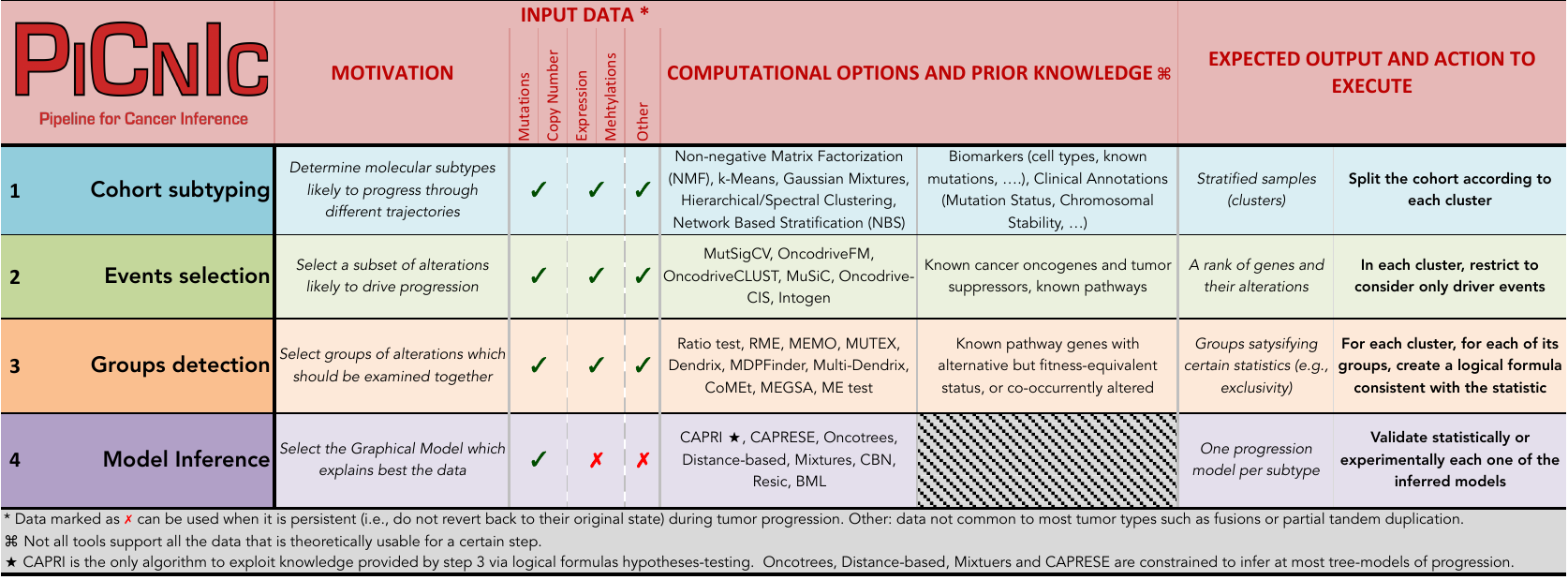}}
\caption[The  \algname{PicNic} pipeline]{\added{The  PiCnIc pipeline. We do not provide a unique all-encompassing rationale to instantiate PiCnIc as all  steps refer to  research area currently development, where the optimal approach is often dependent on  the type of data available and prior knowledge about the cancer under study. References are provided for each tool that can be used to instantiate PiCnIc: NMF \cite{gao2005improving}, k-Means, Gaussian Mixtures, Hierarchical/Spectral Clustering \cite{de2008clustering}, NBS \cite{hofree2013network}, MutSigCV\cite{lawrence2013mutational}, OncodriveFM \cite{gonzalez2012functional},  OncodriveCLUST\cite{tamborero2013oncodriveclust}, MuSiC\cite{dees2012music} Oncodrive-CIS \cite{tamborero2013oncodrive} Intogen\cite{gundem2010intogen}, Ratio  \cite{yeang2008combinatorial}, RME \cite{miller2011discovering},  MEMO   \cite{ciriello2012mutual}, MUTEX \cite{babur2015systematic}, Dendrix \cite{vandin2012novo},  MDPFinder \cite{zhao2012efficient}, Multi-Dendrix \cite{leiserson2013simultaneous}, CoMEt \cite{leiserson2015comet}, MEGSA \cite{MEGSA}, ME \cite{szczurek2014modeling},  CAPRI \cite{Ramazzotti15092015}, CAPRESE \cite{Loohuis:2014im}, Oncotrees \cite{desper_1999,szabo}, Distance-based \cite{desper_2000}, Mixtures \cite{beerenwinkel_2005}, CBN \cite{beerenwinkel_2007,gerstung2009quantifying}, Resic \cite{attolini2010mathematical} and BML \cite{misra2014inferring}.}}
\label{tab:pipeline}
\end{figure*}

\begin{figure*}[p]
\centerline{\includegraphics[width=0.9\textwidth]{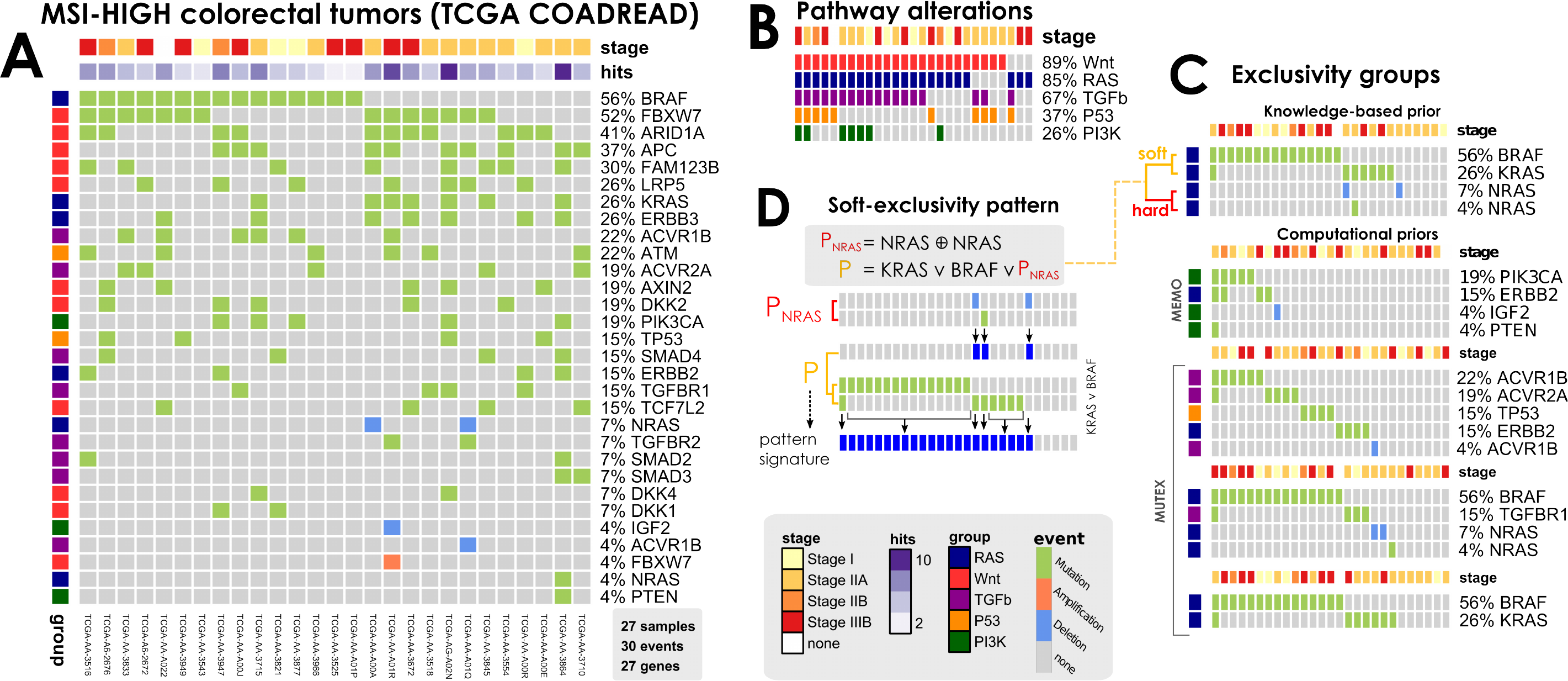}}
\caption[Data processed for MSI-HIGH tumors]{{\bf A.}  MSI-HIGH colorectal tumors  from the TCGA COADREAD project \cite{cancer2012comprehensive}, restricted to 27 samples with both somatic mutations and high-resolution CNA data available and a selection out of 33 driver genes annotated to   \g{wnt}, \g{ras}, \g{pi3k}, \g{tgf-$\beta$} and \g{p53} pathways. This dataset is used to infer the  model  in Figure \ref{fig:MT-model-MSI}. {\bf B.} Mutations and CNAs in MSI-HIGH tumors mapped to pathways confirm heterogeneity even at the pathway-level.  {\bf C.} Groups of mutually exclusive alterations   were obtained from  \cite{cancer2012comprehensive} - which run the  MEMO\cite{ciriello2012mutual} tool - and by  MUTEX\cite{babur2015systematic} tool. In addition, previous knowledge about exclusivity among genes in the \g{ras} pathway was exploited. 
{\bf D.} A Boolean formula input to CAPRI tests the hypothesis that alterations in  the \g{ras} genes \g{kras}, \g{nras} and \g{braf} confer equivalent selective advantage.  The formula accounts for hard exclusivity of alterations in \g{nras} mutations and deletions, jointly with soft exclusivity with  \g{kras} and \g{nras} alterations.}
\label{fig:MT-trdataset-MSI}
\end{figure*}

\added{Summary implementation for COADREAD (\spipe{} steps,  Figure \ref{tab:pipeline}): (1) TCGA clinical classification, (2) MutSigCV and  TCGA manual curation, (3) MEMO, MUTEX and knowledge of \sg{wnt} and \sg{raf} pathways and (4) CAPRI.}

\paragraph{Implement your own case study with PiCnIc/TRONCO.} 
\added{TRONCO started as a project before \spipe{}, and is our effort at collecting, in a free R package,  algorithms to infer progression models from genomic data. In its current version it offers the implementation of the CAPRI and CAPRESE algorithms, as well as a set of routines to pre-process genomic data. With the invention of \spipe{}, it started accommodating software routines to easily interface CAPRI and CAPRESE to some of the tools that we mention  in Figure \ref{tab:pipeline}.  In particular, in its current 2.0 version it supports input/output for the Matlab Network Based Stratification tool (NBS) and the Java MUTEX tool, as well as the possibility to fetch data available from the   cBioPortal for Cancer Genomics (\url{http://cbioportal.org}{http://cbioportal.org}), which provides a Web resource for exploring, visualizing, and analyzing multidimensional cancer genomics data.}

\added{We plan to extend TRONCO in the future to support other similar tools and become an integral part of daily laboratory routines, thus facilitating  application of PiCnIc to additional use cases. }

\vspace{2.0cm}

\paragraph{Authors contributions}
This work follows up on our earlier project initiated by BM and carried out by Milan-Bicocca and the Catalan Institute of Oncology, based on a framework discussed at the 2014 School on Cancer, Systems and Complexity (CSAC).   \algname{PicNic} was designed and constructed by MA's Bioinformatics lab at University of Milan-Bicocca, within a project led and supervised by GC. GC, AG and DR designed the pipeline, and GC, DR and LDS coded and executed it. Data gathering and model interpretation was done by GC, LDS, DR, AG together with BM, VM and RSP. GM, MA, VM and BM provided overall organizational guidance and discussion. GC, AG, RSP and BM wrote the original draft of the paper, which all authors reviewed and revised in the final form. BM and MA are co-senior authors.

\paragraph{Acknowledgments}
MA, GM, GC, AG, DR acknowledge the SysBioNet project, a MIUR initiative for the Italian Roadmap of European Strategy Forum on Research Infrastructures (ESFRI) and Regione Lombardia (Italy) for the research projects RetroNet through the ASTIL Program [12-4-5148000-40]; U.A 053 and Network Enabled Drug Design project [ID14546A Rif SAL-7], Fondo Accordi Istituzionali 2009. BM acknowledges founding  by the NSF grants CCF-0836649, CCF-0926166 and a NCI-PSOC grant.  VM and RSP acknowledge the Instituto de Salud Carlos III supported by The European Regional Development Fund (ERDF) grants PI11-01439, PIE13/00022, the Spanish Association Against Cancer (AECC) Scientific Foundation, and the Catalan Government DURSI, grant 2014SGR647.

We wish to thank the anonymous reviewers for their help in improving the quality and rigor of the presentation.

\bibliographystyle{plain}

\newpage
\FloatBarrier

\newpage
\FloatBarrier

\FloatBarrier

\FloatBarrier

\begin{figure*}[p]
\centerline{\includegraphics[width=0.95\textwidth]{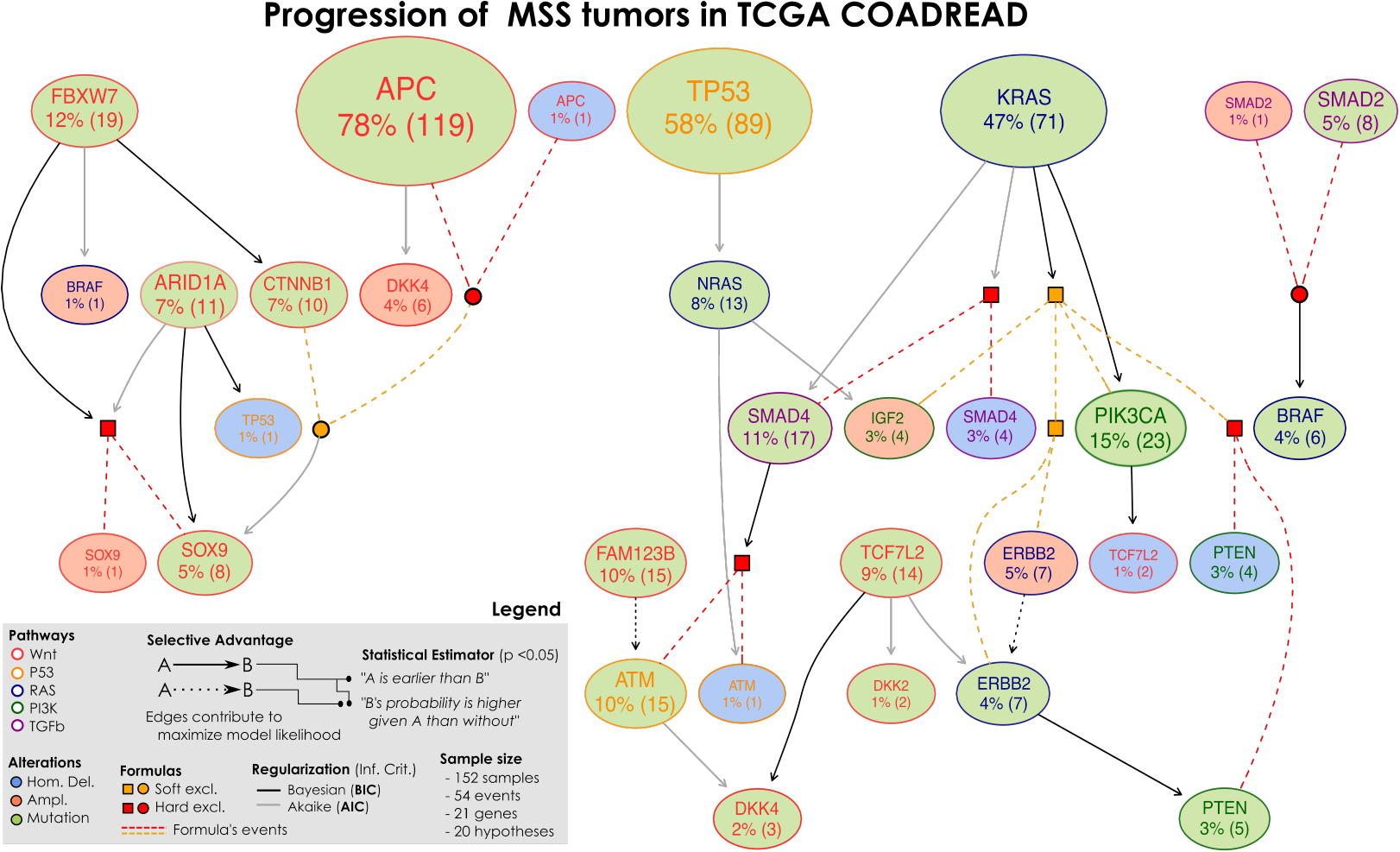}}
\caption[Progression of MSS tumors]{Selective advantage relations inferred by CAPRI constitute MSS  progression; input dataset  in Supplementary Figure S3 and S4.  \addedtwo{Formulas  written on groups of exclusive  alterations, e.g., \g{sox9} amplifications and mutations,  are  displayed in expanded form; their events  are connected by dashed lines with  colors representing the type of  exclusivity (red for hard, orange for soft), logical connectives are squared  when the formula is selected, and circular when the formula selects for a downstream node.}
 \added{For this model of  MSS tumors in COADREAD, we find strong statistical support for many edges (p-values, bootstrap scores and cross-validation statistics shown as Supplementary Material), as well as the overall model. This model  captures both current knowledge about CRC progression -- e.g, selection of alterations in \g{pi3k} genes by the \g{kras} mutations (directed or via the MEMO group, with BIC) -- as well as novel interesting testable hypotheses -- e.g., selection of \g{sox9} alterations by \g{fbxw7} mutations (with BIC).}} 
\label{fig:MT-model-MSS}
\end{figure*}

\FloatBarrier

\begin{figure*}[p]
\centerline{\includegraphics[width=0.8\textwidth]{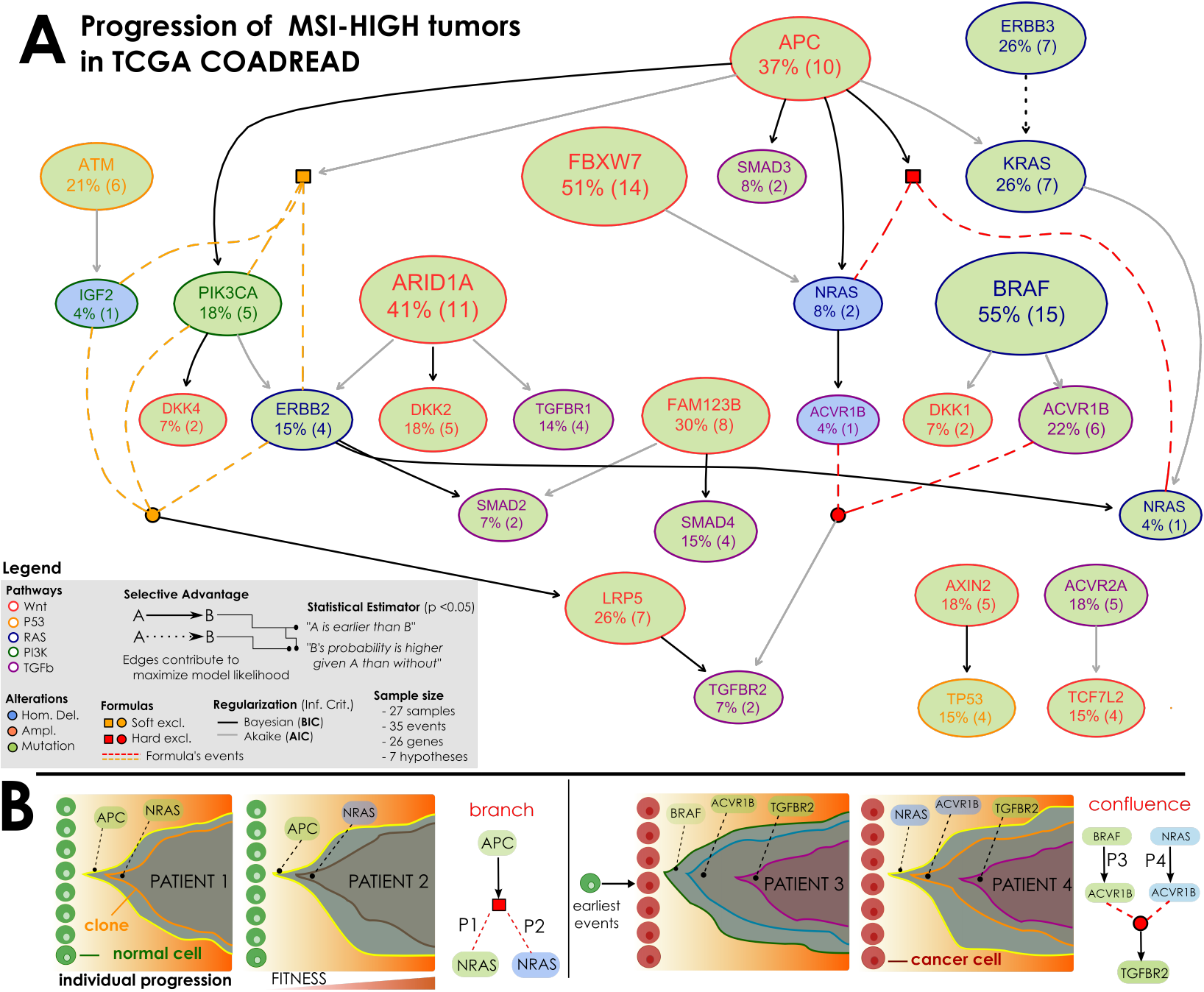}}
\caption[Progression of MSI-HIGH tumors]{{\bf A.} Selective advantage relations inferred by CAPRI constitute MSI-HIGH  progression; input dataset  in Figure \ref{fig:MT-trdataset-MSI}.  Formulas written on groups of exclusive  alterations  are  expanded  as in Figure \ref{fig:MT-model-MSS}. \added{For each relation,  confidence is estimated as for MSS tumors and reported as Supplementary Material. In general, this model is supported by weaker statistics than MSS tumors -- possibly because of this small sample size ($\scriptstyle n=27$). Still, we can find interesting relations involving} \g{apc} mutations which select for \g{pik3ca} ones (via BIC) as well as  selection of the MEMO group (\g{erbb2}/\g{pik3ca} mutations or \g{igf2} deletions) predicted by AIC. Similarly, we find a strong selection trend among mutations in \g{erbb2} and \g{kras}, \added{despite in this case the temporal precedence among those mutations is not disentangled as the two events have the same marginal frequencies ($\scriptstyle 26\%$)}.
{\bf B.} Evolutionary trajectories of clonal expansion predicted  from  two  selective advantage relations in the model.  \g{apc}-mutated clones shall enjoy  expansion, up to acquisition of  further selective advantage via  mutations  or  homozygous deletions in  \g{nras}. These cases should be representative of different individuals in the population, and the ensemble-level interpretation should be that ``\g{apc} mutations select for  \g{nras} alterations, in hard exclusivity'' as no sample harbour both alterations. A similar argument can show that the clones of patients harbouring distinct alterations in \g{acvr1b} -- and different upstream events -- will enjoy further selective advantage from mutation in the \g{tgfbr2} gene.}
\label{fig:MT-model-MSI}
\end{figure*}

\FloatBarrier

\renewcommand\thefigure{S\arabic{figure}}    
\renewcommand\thetable{S\arabic{table}}    
\appendix

\newpage

\section{Reproducing this study}

\definecolor{mycolor}{rgb}{0.242, 0.242, 0.242}

\begin{tcolorbox}[
colframe=mycolor,
colback=gray!10,
coltitle=black!20!white,  
title= \bf Code availability]
The implementation of PiCnIc  shown in the Main Text was performed by using, as core, the {\sc r} language, and other external Java tools which we reference in this document. In  {\sc r}, much of the data processing and inference is done by exploiting the current version of the open-source
\begin{center}
 {\bf TRanslational ONCOlogy}  ({\tt TRONCO}, \cite{TRONCO}, version 2.3) 
 \end{center}
 package which implements up-to-date statistical algorithms to estimate cancer progression models from a list of genomic lesions (e.g., somatic mutations, copy number variations or persistent epigenetic states) in a population of independent tumors, or in a single patient.\\

{\tt TRONCO}'s official webpage is reachable from the Software section of our group's webpage
\begin{center}
\url{http://bimib.disco.unimib.it/}
\end{center}
By navigating to the Case Studies section of {\tt TRONCO}'s official webpage one can find the source code to replicate this study (i.e., the \pipe{}'s implementation) along with the  documentation detailing all the implementation, \added{as well as the data that we used}. This should allow easy implementation of similar studies in different contexts.

\vspace{1.0cm}

\begin{center}
\includegraphics[width = 4cm]{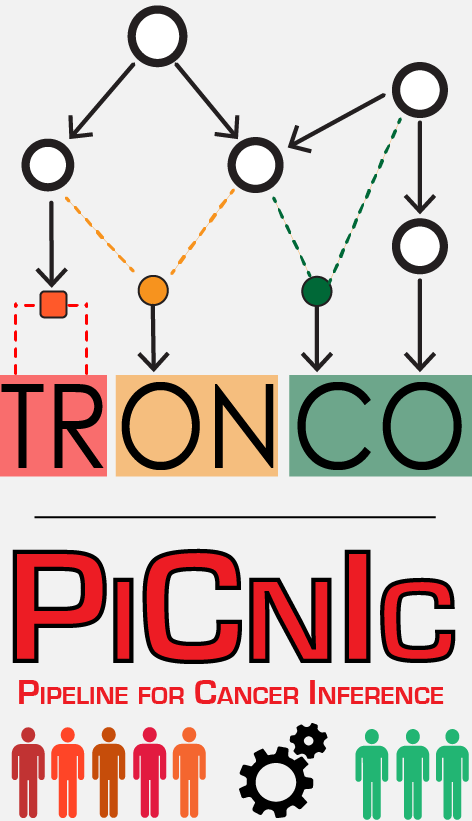}
 \end{center}

\vspace{1.0cm}

$\quad\\$

\end{tcolorbox}

\newpage

\section{\added{Glossary}}

\renewcommand{\underline}[1]{\textbf{\textcolor{YellowOrange}{#1}}}

\added{This glossary of terms shall be of help to readers not familiar with  the concepts mentioned in the Main Text. 
For clarity, terms are separate in two categories according to the fact that they are   common to the statistics or the cancer biology communities. Each term which is included in this glossary appears in color. }\\

\noindent
\renewcommand{\arraystretch}{1.4}


\begin{tabular}{p{2.3cm} | p{13cm}}
\multicolumn{2}{c}{\bf Terms common to the statistics community}\\
\textbf{Term}                                          & \textbf{Meaning}                                                                                                                                                                                                                                                                                                                                                                                                                   \\ 
\underline{\small Boolean formula} & \small{In CAPRI, a formula written with standard logical operators which capture a relation among a group of alterations. In \pipe{}, these are used to detect alternative routes to \underline{selective advantage} from \underline{mutually exclusive alterations}.  See: \underline{Fitness equivalence}}.\\ \hline

\underline{\small Ensemble level progression inference}               & \small{Detection of the relations of \underline{selective advantage} across the permanent  \underline{alterations}  in a cohort of independent tumors (\underline{cross-sectional} data). When aggregated in a \underline{graphical model}, these shall  picture the most common evolutionary trajectories in the population/cancer under study. See also:  
\underline{inter-tumor heterogeneity}}. \\ \hline 

\underline{\small Graphical models} & \small{In this context, a direct acyclic graph with nodes (\underline{alterations}) and edges (\underline{selective advantage relations}), as a shorthand to represent the joint probability of observing a set of alterations in a sample (i.e., a cancer genotype). See also: \underline{Suppes-Bayes Causal Network}, \underline{Model selection}.}\\ \hline

\underline{\small Individual level progression inference} & \small{Detection of \underline{clonal} signatures and their prevalence in individual tumors by scanning  \underline{multi-region} or \underline{single-cell} sequencing data; clones are then  displayed in a \underline{phylogenetic tree structure}. {See also}: \underline{intra-tumor heterogeneity}.} \\ \hline

\underline{\small Model selection} & \small{The process of selecting a model which fits data, according to some criterion. In CAPRI, this is done by balancing model likelihood (a measure of to which degree data can be explained by the model) and model complexity (the size of the \underline{graphical model}). See: \underline{regularization}.}\\ \hline

\underline{\small Phylogenetic tree} & \small{In this context, rooted tree where each node is a \underline{clone}, and edges represent ancestry relations among clones. }\\ \hline

\underline{\small Regularization} & \small{Common approach to avoid overfitting (false-positives) during \underline{model selection} -- in CAPRI this is achieved by using the standard AIC/BIC which penalize with different severity \underline{graphical models} which contain many \underline{selective advantage relations}. }\\ \hline

\underline{\small Simposon's paradox} & \small{A paradox in statistics, in which a trend appears in different groups of data but disappears or reverses when these groups are combined. In this context, this shall refer to genuine \underline{selective advantage relations} which are not inferred unless data coming from different populations is separated before doing inference. See: \underline{heterogeneity}, \underline{subtypes}, \underline{formulas
}.}\\ \hline

\underline{\small Suppes-Bayes Causal Networks} & \small{A specific type of \underline{graphical model} returned by CAPRI algorithm, where each edge satisfies Suppes's conditions of probabilistic causation subsuming  temporal ordering and  positive statistical dependence -- the statistical approach to estimate \underline{selective advantage} among the alterations. } \\ \hline

\end{tabular}

\begin{tabular}{p{2.1cm} | p{14cm}}
\multicolumn{2}{c}{\bf Terms common to the cancer biology community}\\
\textbf{Term}                                          & \textbf{Meaning}                                                                                                                                                                                                                                                                                                                                                                                                                   \\ \hline
\underline{\small Alterations} &  \small{Somatic mutations: A change in the genome of a cell that is not inherited from a parent, but is acquired. CNVs: Structural variation of large regions of DNA segments, including deletions, insertions, duplications and complex multi-site variants.}\\ \hline
\underline{\small Bulk sequencing} & \small{Genome sequencing from single tumor samples, each containing  a large number of cells. The resulting genomic profiles are derived from a mixture of cells with potentially distinct evolutionary histories.}\\ \hline
\underline{\small Clones; Clonal expansion} & \small{Clone: group of cells sharing an identical genome and that derive from a common ancestor. Clonal expansion: the production of descendent cells all arising originally from a single cell. In the scenario of cancer development, tumors develop through a series of clonal expansions, in which the most favorable clonal population survives and proliferate.}\\ \hline
\underline{\small Cross-sectional data} & \small{Unique snapshots of data derived from samples that are collected at unknown time points. Usually derived from bulk sequencing technologies.}\\ \hline
\underline{\small Driver; Passenger}
& \small{Driver: (epi)genetic  \underline{alteration} that provides a \underline{selective advantage} to a \underline{cancer clone}. Passenger: alteration of a cancer cell that does not increase its \underline{fitness}.
}\\ \hline

\underline{\small Exclusivity of alterations}  & \small{Group of alterations which manifest few or no  co-occurences  in a cohort of different samples, and might be  \underline{fitness-equivalent} for tumor progression. Hard exclusivity: when co-occurrences shall be considered the result of random errors. Soft exclusivity: when few co-occurrences shall be possible.
See:  \underline{formulas}.}\\ \hline

\underline{\small Fitness} & \small{A cell's ability of surviving, proliferating and adapting to environmental changes, usually within an environment with limited and depleting resources (e.g., oxygen or nutrients). }\\ \hline

\underline{\small Fitness equivalence} 
 & \small{Groups of driver \underline{alterations}, functional to  the same pathway or equally dysruptive, that can independently confer a \underline{selective advantage} to a cancer cell. Multiple co-occurrence of such alterations  to provide no further advantage, hence leading to \underline{mutually exclusive} alteration patterns across distinct samples.}\\ \hline
\underline{\small Hallmark of cancer} & \small{Common traits or phenotypic properties that are supposed to drive the transformation of normal cells to cancer cells. Anti-hallmark: \underline{clonal  profiles} that are usually not observed, yet being theoretically possible.} \\ \hline

\underline{\small Inter-tumor heterogeneity}
 & \small{The phenomenon according to which different patients with the same cancer type usually display a few common alterations. This is the major problem of \underline{inferring ensemble-level cancer progression models}.
 }\\ \hline

\underline{\small Intra-tumor heterogeneity} 
 & \small{Intra-tumor heterogeneity is related to possible coexistence of different cancer clones, with different evolutionary histories and different mutational profiles, within the same tumor.  This is the major problem of \underline{inferring individual-level cancer progression models}.
 }\\ \hline

\underline{\small Multiregion sequencing} 
 & \small{Collection of genomic data obtained by processing multiple spatially separated biopsy samples from the same individual tumor.}\\ \hline

\underline{\small Next Generation Sequencing (NGS)} 
 & \small{New  technologies for sequencing genomes at high speed and low cost, including, e.g., full-genome/exome sequencing, genome resequencing, transcriptome profiling (RNA-Seq), DNA-protein interactions (ChIP-Seq), and epigenome characterization.}\\ \hline

\underline{\small  Selective advantage relation}
 & \small{In  successive waves of \underline{clonal expansions} one or more cells of the same clone can (progressively) increase their \underline{fitness} through the acquisition of additional \underline{driver alterations}, leading to the emergence and development of a fitter clone.  In this case a relation of selective avantage connects the earlier to the succeeding \underline{alterations}. 
}\\ \hline

\underline{\small Single-cell sequencing}
 & \small{Recent technology based on the retrieval and analysis of genomic information from individual cells, rather than from mixtures of cells.}\\ \hline

\underline{\small Synthetic lethality}
 & \small{The phenomenon according to which two otherwise non-lethal \underline{alterations} lead the cell  death when they co-occur within the same cell. See: \underline{Anti-hallmark}}\\ \hline

\end{tabular}

\section{\pipe{}'s implementation for COADREAD samples}

Here we detail all the steps implemented to use \pipe{} for CRC progression inference. 

\subsection{TCGA COADREAD project data}

COADREAD provides genome-scale analysis of  samples with exome sequence, DNA copy number, promoter methylation, messenger RNA and microRNA expression data, which we used to define      \added{the input} dataset.   In particular,   only samples with both mutations and CNAs profiles were used in the analysis. Supplementary Table \ref{fig:SM-table-datasets}  details  the dataset.

\paragraph{Dataset \added{used to infer models presented in the Main Text}.} Samples published in  \cite{cancer2012comprehensive} were used   \added{as, to the best of our knowledge,  these represent the highest--quality data made available by COADREAD as of today};  for these samples TCGA provides somatic mutation profiles and high-resolution focal CNAs via GISTIC. These are obtained from TCGA data freeze  as of 2 February 2012,  downloaded on 12 March 2015, from  repository:
\begin{center}
 \url{https://tcga-data.nci.nih.gov/docs/publications/coadread\_2012/}
\end{center}
The following files were processed to produce the   data:
\begin{itemize}
\item \texttt{TCGA\_CRC\_Suppl\_Table2\_Mutations\_20120719.xlsx} \\
{\em Somatic mutations} profiles obtained via whole-exome sequencing of 224 colorectal tumors by TCGA. Data available consists of
15995 mutations  in 228 samples, provided in the  {\em Manual Annotation Format} (MAF). Samples were selected to univocally match  the 224 patients as of the TCGA guidelines for aliquote disambiguation, see \url{https://wiki.nci.nih.gov/display/TCGA/TCGA+barcode}.
\added{
All the mutations annotated by TCGA  --  truncating ({\tt De\_novo\_Start\_OutOfFrame}, {\tt Frame\_Shift\_Del},  {\tt Frame\_Shift\_In}, {\tt Nonsense\_Mutation}, {\tt Splice\_Site}, {\tt Frame\_Shift\_Ins}, {\tt In\_Frame\_Del}),  silent ({\tt Silent}) and missense ({\tt Missense\_Mutation}) -- were considered for  analysis; notice that the majority of them are missense, see  Figures \ref{fig:SM-mutrate} and \ref{fig:SM-lolliplot}.}

\item \texttt{crc\_gistic.txt.zip}\\
Focal {\em Copy Number Alterations} (CNAs) for  564 patients derived from   whole-genome sequencing using the Illumina HiSeq platform.  {\em High-level gains} and {\em homozygous deletions} were considered for  analysis by selecting entries with GISTIC scores  $\pm$ 2;

 \item \texttt{crc\_clinical\_sheet.txt}\\
{\em Clinical data} summary with patient stage and {\em Micro Satellite Stabe/\deleted{Instable}\addedthree{Unstable}} (MSS/MSI) status being any of: MSS, MSI-high and MSI-low.
\end{itemize}

The list of patients used was first reduced to those having  {\em both} CNAs and somatic mutation data, and then was split into two groups: MSI-HIGH and MSS. The training cohort has 152 MSS  and 27 MSI-HIGH samples; samples flagged as low MSI were excluded from the study as they have not been shown to differ in their clinicopathologic features or in most molecular features from MSS tumors \cite{pawlik2004colorectal}. 
%
%

\subsection{Driver events selection}

\added{In the TCGA COADREAD study\cite{cancer2012comprehensive} integrated analysis of mutations, copy number and mRNA expression changes in 195 tumours with complete data was performed. Part of the analysis was carried out by using the MutSig tool \cite{lawrence2013mutational}, as well as manual curation. Samples  were grouped by  mutation status, and   recurrent alterations in key CRC pathways   were identified in \cite{cancer2012comprehensive} (Fig. 4, Supplementary Fig. 6 and Supplementary Table 1) as a result, we can use the consortium's  list of 33 driver genes annotated to 5  pathways and use these to extract our progression models.}
These are well-known cancer genes, frequently reported as relevant to colorectal progression and  to the major pathways involved in CRC. Driver events are alterations in:
\begin{itemize}
\item \g{wnt} genes (14):  \g{apc}, \g{dkk-4}, \g{tcf7l2}, \g{ctnnb1}, \g{lrp5}, \g{fbxw7}, \g{dkk-1}, \g{fzd10}, \g{arid1a},  \g{dkk-2}, \g{fam123b}, \g{sox9}, \g{dkk-3} and  \g{axin2};
\item \g{rtk}/\g{ras}  genes (5):  \g{erbb2}, \g{erbb3}, \g{nras},  \g{kras} and  \g{braf};
\item \g{tgf-$\beta$} genes (5):   \g{tgfbr1},  \g{smad3},   \g{tgfbr2},   \g{smad4},  \g{acvr1b},  \g{acvr2a} and \g{smad2};
\item \g{igf2}/\g{pi3k}   genes (5):   \g{igf2}, \g{irs2}, \g{pik3ca}, \g{pik3r1} and \g{pten};
\item \g{p53} genes (2):   \g{tp53}  and \g{atm}.
\end{itemize}
In the Main Text, \g{rtk}/\g{ras}  and  \g{igf2}/\g{pi3k}  pathways are shortly  denoted as  \g{ras} and  \g{pi3k}.

\added{The distinct types of  mutations detected in these genes are shown in Figure \ref{fig:SM-mutrate}, as well as the overall rate of COADREAD mutations. The spatial distribution (per gene) of such mutations is shown in Figure \ref{fig:SM-lolliplot}.}

\subsection{Mutual exclusivity groups of alterations.} 

Groups of alterations showing a trend of mutual exclusivity  were scanned with   MUTEX  and  mutations and CNA hitting any of the  33 selected genes as input. MUTEX was  run independently on MSS and MSI-HIGH groups (Supplementary Table \ref{fig:SM-table-mutex}, running times: approximately 6 and 3.5 hours, respectively, on a standard Desktop machine).

We selected only groups with score $< 0.2$, where the  score is derived from {\em p-values corrected for  false discovery rate}.  3 groups are found for MSI-HIGH tumors and 6 for MSS. For MSI-HIGH tumors, the three predicted groups consists of genes \g{acvr1b}, \g{acvr2a},  \g{tp53} and  \g{erbb2}, of genes \g{braf},  \g{nras} and  \g{tgfbr2}, and of genes \g{kras} and \g{braf}. 

Further groups of exclusive alterations were considered consistent with results reported in \cite{cancer2012comprehensive}. These include groups derived by  consolidated  knowledge of colorectal progression: the well-known  \g{wnt} alterations in   \g{apc}/\g{ctnnb1}   \cite{gerstein2002apc}, as well as \g{ras} alterations in  \g{kras}, \g{nras} and \g{braf} genes   \cite{yeang2008combinatorial}. Similarly, we used also a group collected by scanning non-hypermutated tumors with the  MEMO tool in  \cite{cancer2012comprehensive} - this group includes \g{pik3ca}, \g{pten}, \g{erbb2} and \g{igf2} genes.  These groups were restricted to account only for genes actually altered in a certain subtype, e.g., MSI-HIGH tumors lack  \g{ctnnb1} mutation, making the Wnt group irrelevant. Groups for MSS tumors are shown as Supplementary Figure \ref{fig:SM-exclusivity-MSS}, groups for MSI-HIGH tumors are in the Main Text.

\subsection{CAPRI's execution} \label{sec:CAPRI}
 
 \paragraph{Background on the algorithm.} \addedtwo{CAPRI algorithm can be executed in two different modes, originally dubbed as ``supervised'' when formulas are given in input for testing, and ``unsupervised'' when this is not the case. This paper deals with the former; see the Main Text for  the interpretation of formulas in this context and \cite{Ramazzotti15092015} for a derivation of the algorithm. CAPRI is a three-steps procedure, which we briefly recall here.} 

\begin{enumerate}
\item \addedtwo{CAPRI starts by creating a ``lifted'' representation of the input data $\mathbf{M}$ which includes the input formulas; each formula -- which is written in propositional logic -- is so evaluated to yield a new column in the dataset. This is the input processed by the algorithm, which starts  by selecting a set of candidate model edges, which are then used to constrain a score-based Bayesian model-selection problem \cite{Ramazzotti15092015}.}

\item  \added{The initial set of selective advantage relations $\cal S$, which determines  the model edges $x \to y$, is computed  by evaluating the following inequalities
 \begin{align*} 
\text{\tt(Temporal priority)} & \quad p(x) > p(y) \\
\text{\tt(Probability raising)} & \quad p(y\mid x) > p(y\mid \neg x)\, ,
\end{align*}}
where  $ p(\cdot)$ is a  marginal probability, $ p(\cdot \mid \cdot)$ is a conditional, $ \neg x$ is the negation of $x$ and either $x$ or $y$ is not a formula. \addedtwo{It is interesting to observe that   {\tt probability raising}  implies that $x$ and $y$  are {\em positively statistically dependent}, thus imposing a minimum threshold on their association  \cite{Loohuis:2014im}.}

\addedtwo{In each of CAPRI's executions, the distribution of the observed marginals and  conditionals are estimated by $K$ non-parametric bootstrap resamples; practically, this means that we create a bootstrapped approximation for each of the four populations  $\hat{p}(x)$, $\hat{p}(y)$, $\hat{p}(y\mid x)$ and $\hat{p}(y\mid \neg x)$. Then, we use a single-tail non-parametric Mann-Withney U test of the difference in mean to test the hypothesis that one of the two populations is more probable than the other, e.g., $\hat{p}(x) > \hat{p}(y)$. With this test, we can  compute two p-values, one for each condition. An edge is included in $\cal S$ if at least the p-value for  {\tt probability raising} is below  a  significance  threshold $p_\ast$ and an edge is said not to be orientable if its p-value for  {\tt temporal priority}  is above the same threshold. Cycles $x_1 \to x_2 \to \ldots x_k \to x_1$ that might appear in $\cal S$ are broken by deleting the edge with minimum p-value;  both $K$ and $p_\ast$ are custom parameters.}

\item Optimization of $ \cal S$, namely detection of the subset $ {\cal S}^\ast$ of $ {\cal S}$ with the edges that we include in the final progression model is done by optimizing, via {\em hill climbing} or {\em tabu search},  the  {\em score with regularization}   
\begin{align}\tiny  
{\cal S}^\ast \triangleq \arg\min_{\hat{\cal S}\subset\cal S}  \left\{ -2 \log[{\cal L}(\hat{\cal S} \mid \mathbf{M})] + \theta |\hat{\cal S}|  \right\}\, ,
\end{align}
where $ {\cal L}(\cdot)$ is the {\em model likelihood} and $\mathbf{M}$ is the input data; the estimated optimal solution is ${\cal S}^\ast$, \addedtwo{which is displayed as a Suppes-Bayes Causal Network}. The different regularization strategies mentioned in the Main Text, BIC and AIC, are obtained by the following parametrization:
 \begin{align*} 
\text{\tt(Bayesian Information Criterion)} & \quad \theta \triangleq \log(n) \\
\text{\tt(Akaike Information Criterion)} & \quad \theta \triangleq 2.
\end{align*}
\addedtwo{Besides edges, a model has a set of parameters $\boldsymbol{\theta}$ which define the {\em conditional probability table} of each edge and should be fit from data; these are necessary if one wishes to use a model as a  ``generator'' of  further data.
For discrete-valued graphical models,  for each parent set  $\pi_x\to y$,  parameter $\boldsymbol{\theta}({y})=p(y \mid \pi_x)$ can be  taken either as the {\em maximum likelihood estimate} from the lifted input, or by using a Bayesian interpretation  \cite{bayesian_learning}.}

\end{enumerate}

\paragraph{Usage in this context.}
CAPRI was run, on each group of tumors, by  selecting alterations from the pool of  33 pathway genes; every alteration on a gene $x$ is included if {\em any} of these apply:
\begin{itemize}
\item the  alteration frequency of $x$ - sum of mutation and CNA  frequency - is greater than 5\%;
\item  $x$ it is part of an exclusivity group.
\end{itemize}
The set of selected events for  MSI-HIGH training tumors is shown in the Main Text,  the analogous  set for MSS tumors is shown as Supplementary Figure \ref{fig:SM-trdataset-MSS}.

CAPRI was executed in its {supervised mode} by writing formula over groups and genes with multiple alterations associated, as explained in the Main Text. For instance, for MSI-HIGH tumors with alterations in \g{ras} pathway we grouped hard exclusivity of \g{nras} mutations and deletions, with soft exclusivity of  \g{kras} and \g{braf} mutations. Our aim was to account for a small subset of samples with concurrent \g{kras} and \g{nras} alterations (see Figure 2, Main Text). The list of all Boolean formulas written over groups is in Table \ref{fig:SM-table-formulas}; 
this approach was adopted also when a gene harbors multiple alterations in a subtype, e.g., \g{erbb2} in MSS training samples which shows a trend of soft exclusivity between mutations and amplifications. \addedtwo{We used both AIC and BIC scores to regularize inference after 100 non-parametric bootstrap iterations for estimation of the preliminary selective advantage relations --  Mann-Whitney U test was performed with a minimum threshold $p_\ast=0.05$.} In most cases p-values are  orders of magnitude below $p_\ast$ - exact values reported as Dataset File S1.  CAPRI's models with such p-values and non-parametric bootstrap confidence are shown in Figures \ref{fig:SM-MSS-Bootstrap} and  \ref{fig:SM-MSI-Bootstrap}, \added{statistical validation of the models is discussed in the next section}.

\section{\added{Statistical validation of the models}}

\addedtwo{P-values from hypothesis testing, as well as  scores from $k$-fold {\em cross-validation} and various {\em bootstrapping} techniques can be used to measure the statistical consistency of models and data, each one capturing different potential errors in the inference process.  Approaches such as cross-validation and bootstrap  are sometimes also used in the (\emph{ex novo}) generation and inference of models from data (see, e.g., \emph{bootstrap consensus models} \cite{felsenstein2004inferring}), but we only use them here for the \emph{a posteriori}  evaluation of a model's confidence, and we interpret them as a quantitative measure for the \emph{relative} assessment of each model's relation. }

\added{All the p-values and the scores that we present here are computed within {\tt TRONCO}.}

\subsection{\addedtwo{Edge p-values}}

\addedtwo{As explained in \S \ref{sec:CAPRI}, for each model edge $x\to y$ we get two p-values by assessing temporal priority and probability raising via Mann-Whitney U testing.}  

\addedtwo{For each edge, a p-value for the {\em hypergeometric test} of overlap between  alteration profiles $x$ and $y$ can be computed. More precisely, we test if there is a difference between the number of samples containing {\em both}  $x$ and $y$  versus the total population of samples with $x$, $y$, or both. We would like  the overlap to be significant as those samples -- that determine the joint probability of  $x$ and $y$ -- are those  supporting the presence of a selection trend among $x$ and $y$.}

\addedtwo{An edge is fully supported if all three p-values are below a custom significance threshold, e.g., $0.05$ or even better $0.01$. Some edges might have the p-value for temporal priority above the threshold. If so, the selection trend might be still significant, but the temporal order of $x$ and $y$ -- i.e., the  {\em direction of selection} --  is not  supported by the data.}

\subsection{\added{Bootstrap}}

We used non-parametric and statistical bootstrap techniques to measure the {\em goodness-of-fit}, \addedtwo{as originally proposed in \cite{szabo2002estimating}.}
In this case, we distinguish two type of errors that one could make in the inference process, estimating the presence or absence of edges in the model:
\begin{itemize}
\item {\em Type I errors:} incorrect rejection of a true $H_0$ (null-hypthesis), i.e., a {\em ``false positive''} edge that we wrongly include. 
\item {\em Type II errors:} incorrect acceptance (failure to reject) of a false $H_0$, i.e., a {\em ``false negative''} edge that we miss.
\end{itemize}

{\em Non-parametric bootstrap} \cite{efron1994introduction} computes scores to be interpreted here as follows.
If a model contains an  edge $x \to y$   that is a true positive, we expect its  score  to be high. In other words, when we sample with repetition subsets of the original data and re-run the inference process we  expect to often find models which contain the edge $x \to y$. Conversely, for a node $y$ without incoming edges, or equivalently for any  edge $x \to y$ which is correctly excluded from a model -- a  true negative -- we would expect its  bootstrap score to be low. However, this reasoning can be also generalized to whole models, where we count how many times we re-infer exactly the same model.  \addedtwo{Clearly, such scores will depend also on the empirical probabilities of the nodes in our data, and their deviation from the true probabilities of the phenomenon. So, one might expect  rare events to be less frequently bootstrapped, which results in a  lower estimate; however, such counts can be anyway interpreted as measures of repeatability  of our findings\footnote{\addedtwo{Bootstrapping techniques have been widely used to gauge uncertainty in estimates, but also subjects of philosophical debate about their precise interpretation, especially when coupled with various
 significance thresholds -- unlike the situation with a p-value for a null hypothesis. An exhaustive review on the topic is provided by Soltis in \cite{soltis2003applying}. We follow the  ideas originally developed in the area of   phylogenetic analysis \cite{felsenstein1985confidence}, suggesting that the scores  can be alternatively interpreted as a measure of accuracy of the method or of the robustness of the data\cite{efron1996bootstrap}.}}.}


\added{The above bootstrap approach depends on two  random number generators: one to  shuffle  data (the a posteriori bootstrap), and one to    evaluate  CAPRI's inequalities via hypothesis testing (the internal CAPRI's bootstrap, see  \ref{sec:CAPRI}). Thus, to ensure that no bias is introduced by the random number generators, we performed a {\em statistical bootstrap} by holding data fixed, and re-estimating CAPRI's inequalities with generators initialized with different seeds. We evaluated the robustness of our scores for all edges imputed to be genuine and hence, high-scoring.
}

\addedtwo{Notice that, in principle, even  {\em parametric bootstrap}  scores could be computed if we  used the model to generate bootstrapped data \cite{efron1994introduction}.  However,  as the support of the distribution subsumed by a model with $n$ nodes consists of  $2^n$ possible outcomes,  sampling uniformly from large models might be computationally hard. For this reason, and because such scores are  overestimates of the non-parametric ones, we did not include them in our computation.}

\added{The MSS and MSI progression models are annotated with the non-parametric bootstrap scores in Figures \ref{fig:SM-MSS-Bootstrap} and \ref{fig:SM-MSI-Bootstrap}. Non-parametric and statistical bootstrap scores for a set of selected edges are shown and commented in  Figure \ref{tab:table-stat}. For the same set of edges, we also report the p-values for CAPRI's inequalities assessed in the MSS and MSI progression models (temporal priority and probability raising, as a measure of the selectivity among the alterations, and hypergeometric, as a measure of the randomness in the overlap of two alteration profiles). Additional comments are in the  caption.
}

\subsection{\added{Cross-validation}} \added{Next we study the sufficiency of the data sizes for model inference and its ability to characterize  the underlying  progression ({\em goodness-of-data}). Thus, we focus on the Type III errors,  which occur when  the sample size is inadequate or the sample is a poor descriptor for the reference phenomenon\footnote{Consider the case where the samples are from two different unknown and heterogeneous groups, with random chance making low values to be sampled from a group that actually has a majority of  high values, and vice versa. In this case, the samples will not be the best descriptor of the two groups.}, and thus failing to represent the  progression. For this purpose, we used cross-validation with the data used for the models built (see the Main Text), and followed the best practices developed by the  Bayesian Networks community \cite{bayesian_learning}.
}

\addedtwo{ {\tt TRONCO} exploits  the cross-validation routines implemented in  the {\tt bnlearn} package \cite{bnlearn}. The approach that we adopt is a  $k$-fold non-exhaustive cross-validation, which we   repeat $10$ times to average its results. Exhaustive strategies might be used for datasets of small sample size. Each run of cross-validation  consists in  computing a loss function for a model; its steps are the followings:
\begin{itemize}
\item split randomly the data in $k=10$ groups, and then repeat the following two steps,  for each group in turn:
\begin{enumerate}
\item set one of the groups to be the ``training'' $\mathbf{M}_{tr}$;
\item merge  the others $k-1$  to be the ``test'' $\mathbf{M}_{te}$;
\item by  holding fixed the model structure (i.e., the edges in ${\cal S}^\ast$),  fit the model parameters $\boldsymbol{\theta}$ over the training data via {\em maximum likelihood estimates},  compute a score over the test $\mathbf{M}_{te}$ (see below) and the corresponding loss;
\end{enumerate}
\item combine the  $k$ loss estimates to give an overall loss for data.
\end{itemize}
}

Let $\boldsymbol{\theta}_{tr}$ be the parameters fit from the training set $\mathbf{M}_{tr}$, and ${\cal S}^\ast$ the edges in the model,  three scores are computed with cross-validation:
\begin{enumerate}
\item the  {\em negative entropy} of a  model -- i.e., the negated expected log-likelihood of the test set for the Bayesian network fitted from the training set\addedtwo{, that is
\[
{\tt eloss}({\cal S}^\ast, \boldsymbol{\theta}_{tr}) = - \mathbb{E}[ {\cal L}({\cal S}^\ast, \boldsymbol{\theta}_{tr}  \mid \mathbf{M}_{te}) ]\, .
\]}

\item the {\em prediction error} for a single node $x$ and its parents set $X$, i.e.,   we measure how precisely we can predict the values of $x$ by  using only the information present in its local distribution \addedtwo{$\boldsymbol{\theta}_{tr}(x)$. This parameter corresponds to computing the misclassification rates from ${p}_{te}(x)$,  the  empirical marginal probability of $x$ estimated from the test.}

\item the  {\em posterior classification error} for a single node $x$ and one of its parent node  $y \in X$ --  i.e., the values of $x$  are predicted using only the information present in $y$ by likelihood weighting and Bayesian posterior estimates. 
\end{enumerate}
\addedtwo{See   \cite{murphy2012machine} for a discussion on these loss functions.} \added{The first statistics measures the {\em log-likelihood loss} when we ``forget'' some of the samples used to infer a model (indeed, the test samples); see Figure \ref{tab:eloss}.  Roughly, 
we are measuring how the model's predictive power changes as we look at the data from different viewpoints. This is a score for a whole model, it  has no scale -- so cannot be used to say how good the models/data are, in any absolute sense. Nonetheless, it can be used to evaluate how ``stable'' a model is, for a certain dataset.}

\added{The second and third statistics measure the accuracy of the parent-set, $X$, for a child $x$; the second statistics dealing with the whole parent set as predictor, and the third, the individual contribution of each of the parents. For these two statistics, we desire the prediction error to be low, as a measure of goodness. These are shown in Figure \ref{tab:table-stat} (selected edges), \ref{tab:prederr-all-mss} and \ref{tab:prederr-all-msi} (all edges, prediction error).
}

\FloatBarrier
\section{Supplementary Tables and Figures}

\begin{table}[h]
\begin{center} \small
\begin{tabular}{l | lll | lll}
 \multicolumn{7}{c}{{\bf Datasets} \added{(CNAs and mutations provided by TCGA)}} \\
& \multicolumn{3}{c}{\bf statistics} & \multicolumn{3}{c}{\bf alteration type} \\
{\bf cancer$^\dagger$}& {\small $n$}& {\small $m$}& {\small $|G|$}& {\small \em mutations} &   {\small \em amplifications} & {\small \em deletions} \\ \hline
{MSI-HIGH} &27&16100&13798  &  11556 &  2888  & 1656  \\ \hline
{MSS} &152 & 21317& 16371 &  12417 &  6925  & 1975  \\ \hline 
\multicolumn{7}{p{10cm}}{\footnotesize$^\dagger$ Samples were classified as MSI-HIGH/LOW and MSS by  TCGA; see flag {\tt MSI\_status} in clinical data available for the COADREAD project.} \\ \hline
\end{tabular}
\end{center}
\caption[COADREAD Data]{{\bf COADREAD Data.} Data used in this study, derived from the TCGA COADREAD project \cite{cancer2012comprehensive}.} 
\label{fig:SM-table-datasets}
\end{table}

\begin{table}[h]
\begin{center} 
\centerline{
\begin{tabular}{llp{8cm}}
\multicolumn{3}{c}{\bf MUTEX parameters}\\\hline
{\bf Parameter} & {\bf Value}  & {\bf Description} \\ \hline
{\tt \footnotesize signalling-network} & - & {\em MUTEX network$^\dagger$}\\
{\tt \footnotesize max-group-size} &  {\tt 5} & {\em maximum size of a result  group} \\
{\tt \footnotesize  first-level-random-iteration} &  {\tt 10000} & {\em number of randomisation to estimate null distribution of member p-values in  groups }\\
{\tt \footnotesize  second-level-random-iteration} &  {\tt 100} & {\em number of runs to estimate the null distribution of final scores }\\
{\tt \footnotesize  fdr-cutoff} & {-} &  {\em false-discovery-rate cutoff  maximising the expected value of true positives - false positives is estimated from data} \\
{\tt \footnotesize search-on-signaling-network} & {\tt TRUE} & {\em  reduce the search space using the signalling network } \\\hline
\multicolumn{3}{p{15cm}}{\footnotesize $^\dagger$ Manually curated from Pathway Commons, SPIKE and SignaLink databases. Provided with the tool; available for download at {\tt https://code.google.com/p/mutex/}.}\\\hline
\end{tabular}}

\vspace{.5cm}
\begin{tabular}{r | lll}
\multicolumn{4}{c}{\bf MUTEX groups with score $<.2$ }\\\hline
&{\bf MSI-HIGH Groups} & {\em score}  & {\em $q$-value} \\ \hline
\bf 1&\g{kras}, \g{braf}, & 0.095 & {0.48}\\
\bf 2&\g{nras}, \g{braf}, \g{tgfbr1}& 0.1677	&0.45	\\
\bf 3&\g{erbb2}, \g{tp53}, \g{acvr1b}, \g{acvr2a}& 0.1703	&0.355	\\
\\
&{\bf MSS Groups} & {\em score}  & {\em $q$-value} \\ \hline
\bf 1&\g{tp53}, \g{atm}, & 0.051 & {0.34}\\
\bf 2&\g{arid1a}, \g{tp53} & 0.075	&0.193	\\
\bf 3&\g{kras}, \g{nras}, \g{braf}, & 0.0864 & {0.1975}\\
\bf 4&\g{ctnnb1}, \g{apc}, \g{dkk2}, & 0.098 & {0.144}\\
\bf 5&\g{dkk1}, \g{tp53}, \g{atm}, \g{dkk2} & 0.1387 & {0.176}\\
\bf  6&\g{pik3ca}, \g{tp53}, \g{atm} & 0.164 & {0.207}
\end{tabular}
\end{center}
\caption[MUTEX: parameters and results]{{\bf MUTEX: parameters and results.}  Top: Parameters used to run MUTEX  on the   \added{original TCGA}  MSS/MSI-HIGH datasets with input CNA and somatic mutations in the pathway genes described in text.  Bottom: MUTEX identified 3 and 6 groups of alterations showing a trend of mutual exclusivity in these groups with score below the suggested cutoff of $0.2$.} 
\label{fig:SM-table-mutex}
\end{table}

\begin{table}[h]
\centerline{
\begin{tabular}{r | r  | r | }
\multicolumn{3}{c}{\bf  Formulas input for testing to CAPRI$^\dagger$}\\\hline
&{\bf MSI-HIGH tumors} & {\em description} \\ \hline
\bf 1& (\g{nras}:m $\oplus$ \g{nras}:d) $\vee$ \g{kras}:m $\vee$ \g{braf}:m & \g{raf} exclusivity  \\
\bf 2& \g{pik3ca}:m $\vee$ \g{erbb2}:m $\vee$ \g{pten}:m $\vee$ \g{igf2}:d & MEMO group \\
\bf 3& (\g{acvr1b}:m $\oplus$ \g{acvr1b}:d) $\vee$ \g{acvr2a}:m $\vee$ \g{tp53}:m $\vee$ \g{erbb2}:m & MUTEX group \\
\bf 4& (\g{nras}:m $\oplus$ \g{nras}:d) $\vee$ \g{tgfbr1}:m $\vee$ \g{braf}:m & MUTEX group  \\
\bf 5& \g{kras}:m $\vee$ \g{braf}:m & MUTEX group  \\
\bf 6& \g{acvr1b}:m $\oplus$ \g{acvr1b}:a & multiple  alterations \\
\bf 7& \g{nras}:m $\oplus$ \g{nras}:a & multiple  alterations \\
\bf 8& \g{fbxw7}:m $\vee$ \g{fbxw7}:a & multiple  alterations $^\ddagger$ \\\hline
&{\bf MSS tumors} & {\em description} \\ \hline
\bf 1& (\g{apc}:m $\oplus$ \g{apc}:d) $\vee$ \g{ctnnb1}:m & \g{wnt} exclusivity  \\
\bf 2 &  (\g{kras}:m $\vee$ \g{kras}:a)  $\vee$ (\g{nras}:m $\oplus$ \g{nras}:a) $\vee$ (\g{braf}:m $\oplus$ \g{braf}:a)  & \g{raf} exclusivity  and MEMO group \\
\bf 3& \g{pik3ca}:m $\vee$ (\g{erbb2}:m $\vee$ \g{erbb2}:a) $\vee$ (\g{pten}:m $\oplus$ \g{pten}:d)  $\vee$ \g{igf2}:a & MEMO group \\
\bf 4&   (\g{tp53}:m $\oplus$ \g{tp53}:d) $\vee$ (\g{atm}:m $\oplus$ \g{atm}:d)   & MUTEX group \\
\bf 5&   (\g{tp53}:m $\oplus$ \g{tp53}:d) $\vee$  \g{arid1a}:m    & MUTEX group \\
\bf 6&   (\g{tp53}:m $\oplus$ \g{tp53}:d) $\vee$  \g{arid1a}:m    & MUTEX group \\
\bf 7& (\g{apc}:m $\oplus$ \g{apc}:d) $\vee$ \g{ctnnb1}:m $\vee$ \g{dkk2}:m & MUTEX group  \\
\bf 8&   (\g{tp53}:m $\oplus$ \g{tp53}:d) $\vee$ (\g{atm}:m $\oplus$ \g{atm}:d)  $\vee$ \g{dkk2}:m $\vee$ \g{dkk1}:m & MUTEX group \\
\bf 9&   (\g{tp53}:m $\oplus$ \g{tp53}:d) $\vee$ (\g{atm}:m $\oplus$ \g{atm}:d)  $\vee$ \g{pik3ca}:m  & MUTEX group \\
\bf 10& (\g{apc}:m $\oplus$ \g{apc}:d)  & multiple  alterations  \\
\bf 11& (\g{tp53}:m $\oplus$ \g{tp53}:d)  & multiple  alterations  \\
\bf 12& (\g{smad4}:m $\oplus$ \g{smad4}:d)  & multiple  alterations  \\
\bf 13& (\g{tcf7l2}:m $\oplus$ \g{tcf7l2}:d)  & multiple  alterations  \\
\bf 14& (\g{atm}:m $\oplus$ \g{atm}:d)  & multiple  alterations  \\
\bf 15& (\g{nras}:m $\oplus$ \g{nras}:d)  & multiple  alterations  \\
\bf 16& (\g{erbb2}:m $\vee$ \g{erbb2}:a)  & multiple  alterations  \\
\bf 17& (\g{pten}:m $\oplus$ \g{pten}:d)  & multiple  alterations  \\
\bf 18& (\g{smad2}:m $\oplus$ \g{smad2}:a)  & multiple  alterations  \\
\bf 19& (\g{dkk4}:m $\oplus$ \g{dkk4}:a)  & multiple  alterations  \\
\bf 20& (\g{sox9}:m $\oplus$ \g{sox9}:d)  & multiple  alterations  \\
\bf 21& (\g{braf}:m $\oplus$ \g{braf}:a)  & multiple  alterations  \\ \hline
\multicolumn{3}{p{15cm}}{\footnotesize $^\dagger$ Events type: mutation (m), deletion (d), amplification (a). Hard ($\oplus$) and soft ($\vee$) exclusivity.}\\
\multicolumn{3}{p{15cm}}{\footnotesize $^\ddagger$ Formula not included as it creates a duplicated signature in the dataset.}\\\hline
\end{tabular}
}
\caption[CAPRI's formulas from exclusivity groups]{{\bf CAPRI formulas  from exclusivity groups.} Formulas created for the groups, and input to CAPRI for testing. These are either derived from exclusivity groups or from genes involved in different types of alterations.
} 
\label{fig:SM-table-formulas}
\end{table}


\begin{figure}[h]
\centerline{\includegraphics[width=1\textwidth]{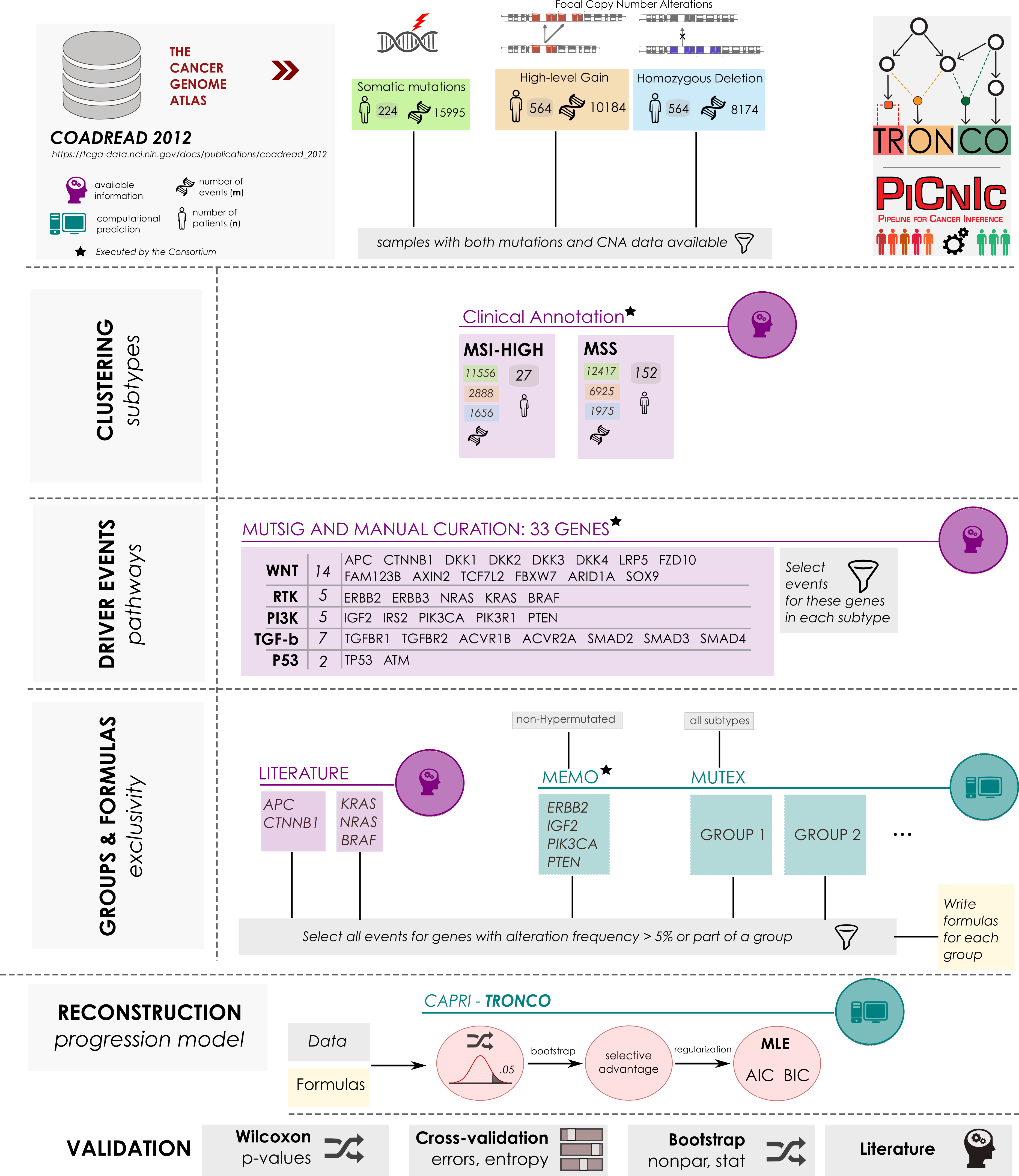}}
\caption[\pipe{} pipeline processing MSI-HIGH/MSS tumors]{{\bf \pipe{}  pipeline processing MSI-HIGH/MSS tumors.} We  process  with \pipe{}   Microsatellite Stable    and   highly  \deleted{instable} \addedthree{unstable} tumors collected from the  The Cancer Genome Atlas  project   ``Human Colon and Rectal Cancer'' (subtypes annotations provided  as clinical data). We implement a    study on selected somatic mutations and focal CNAs in 33 driver  genes manually annotated with 5  pathways in the COADREAD project. We scan groups of exclusive alterations with computational tools run by us and by TCGA,  and we exploit previous knowledge on CRC; we select which alterations we input to CAPRI. Next, inference is performed with various settings of regularization and confidence. Statistical confidence of the models is assessed with standard techniques from the literature (p-values from statistical testing, bootstrap scores and cross-validation statistics). }
\label{fig:SM-CRC-pipeline}
\end{figure}

\begin{figure}[h]
\centerline{\includegraphics[width=1.0\textwidth]{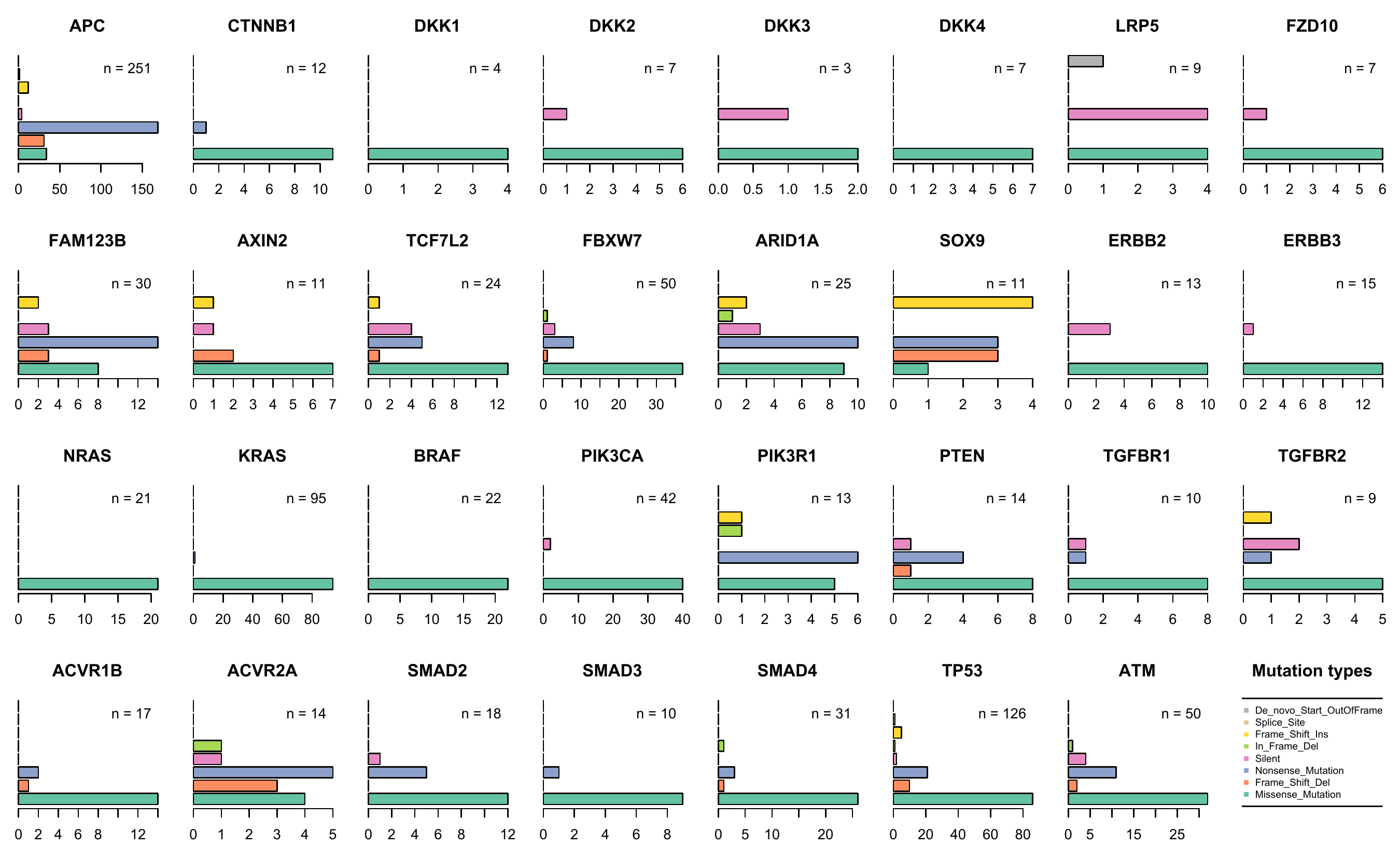}}
\center
\colorbox{white}{\includegraphics[width = 0.5\textwidth]{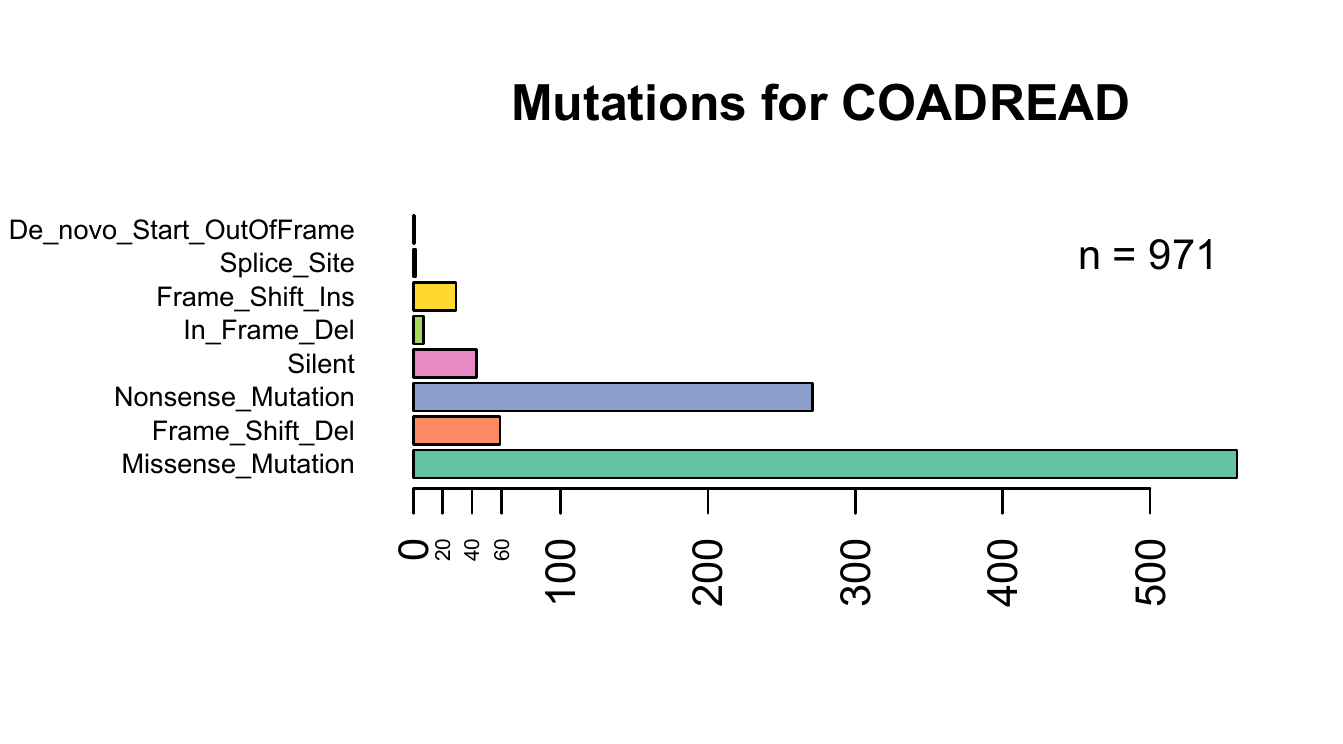}}
\caption[\added{Mutations annotated by TCGA for the driver genes}]{\added{{\bf Mutations annotated by TCGA for the driver genes.} Top: the majority of the mutations annotated by TCGA for the driver genes that we consider  are {\em missense} --  this in almost all genes and in all the cohort.
 Bottom: overall summation of the frequencies determine the mutations across all driver genes.}
} 
\label{fig:SM-mutrate}
\end{figure}

\begin{figure}[h]
\colorbox{white}{\includegraphics[width = 0.5\textwidth]{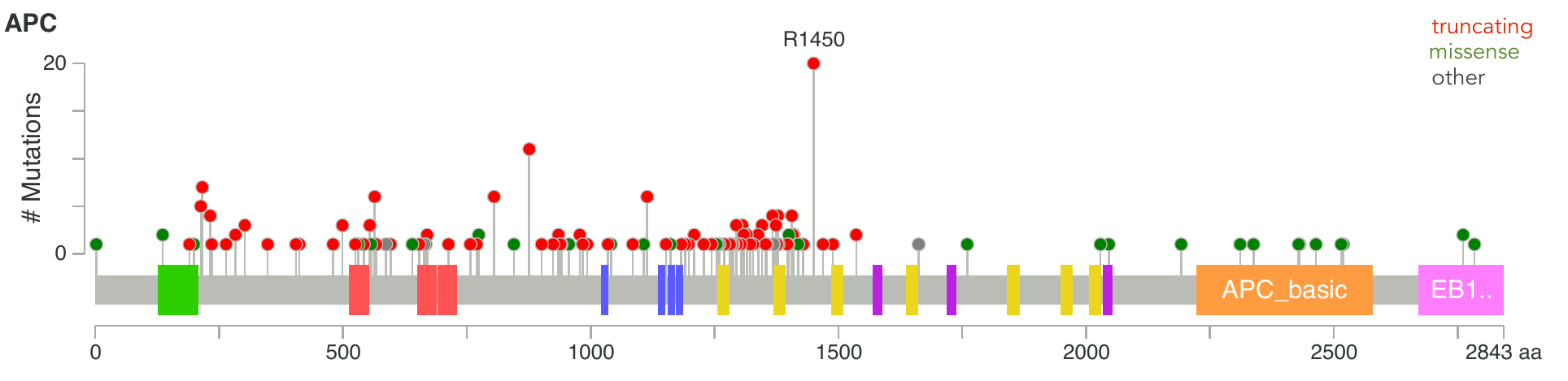}}
\colorbox{white}{\includegraphics[width = 0.5\textwidth]{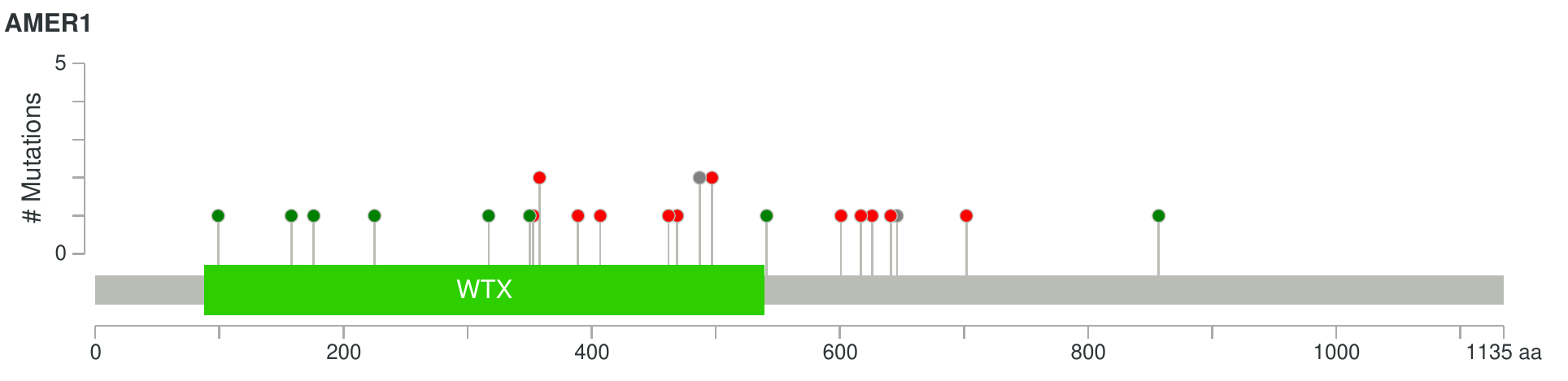}}
\colorbox{white}{\includegraphics[width = 0.5\textwidth]{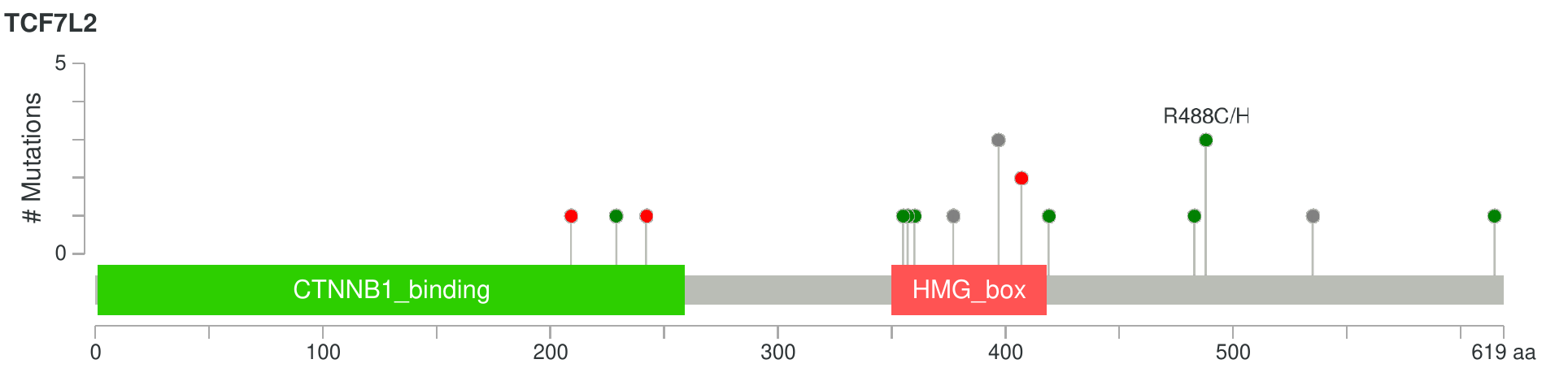}}
\colorbox{white}{\includegraphics[width = 0.5\textwidth]{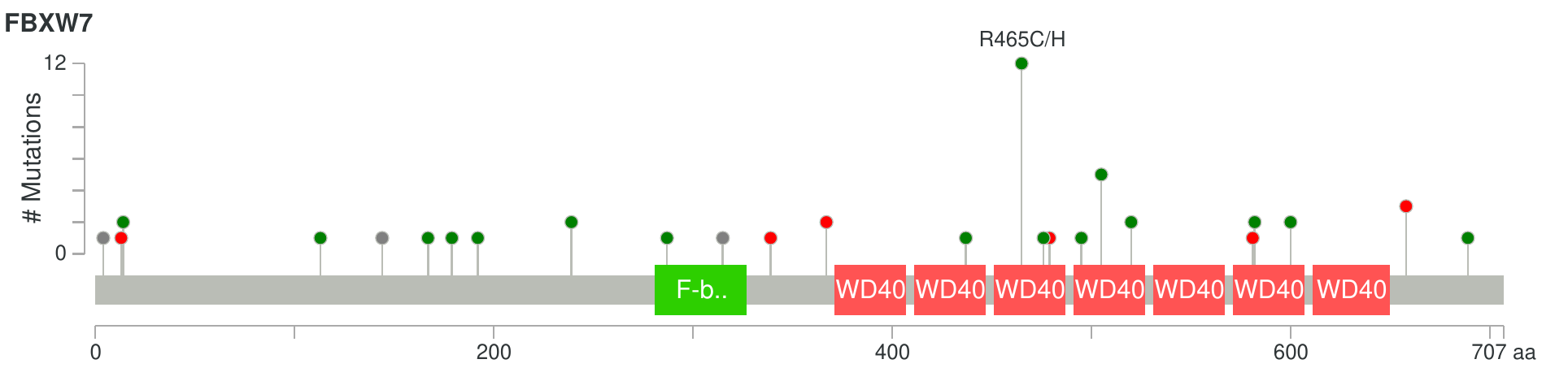}}
\colorbox{white}{\includegraphics[width = 0.5\textwidth]{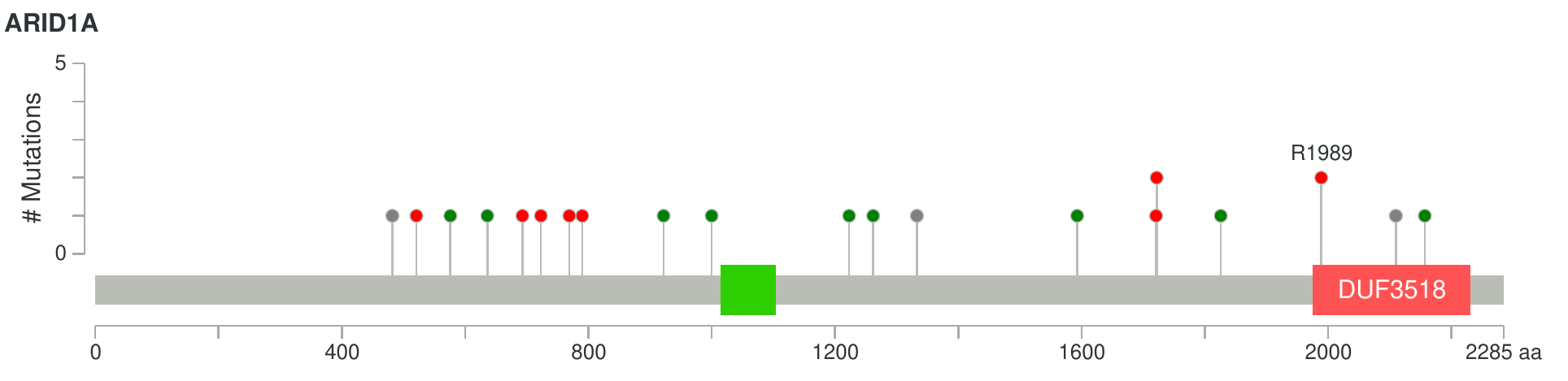}}
\colorbox{white}{\includegraphics[width = 0.5\textwidth]{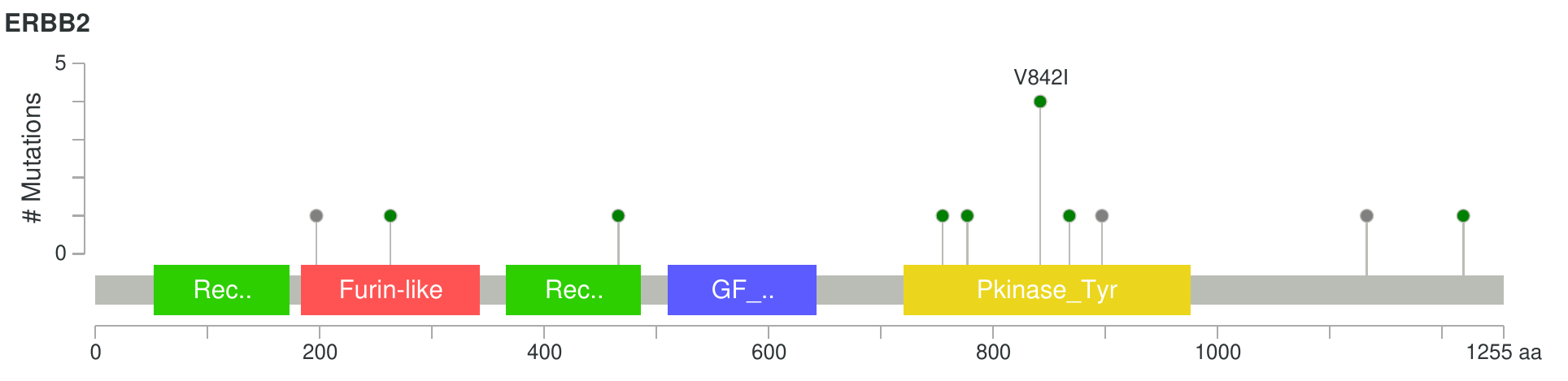}}
\colorbox{white}{\includegraphics[width = 0.5\textwidth]{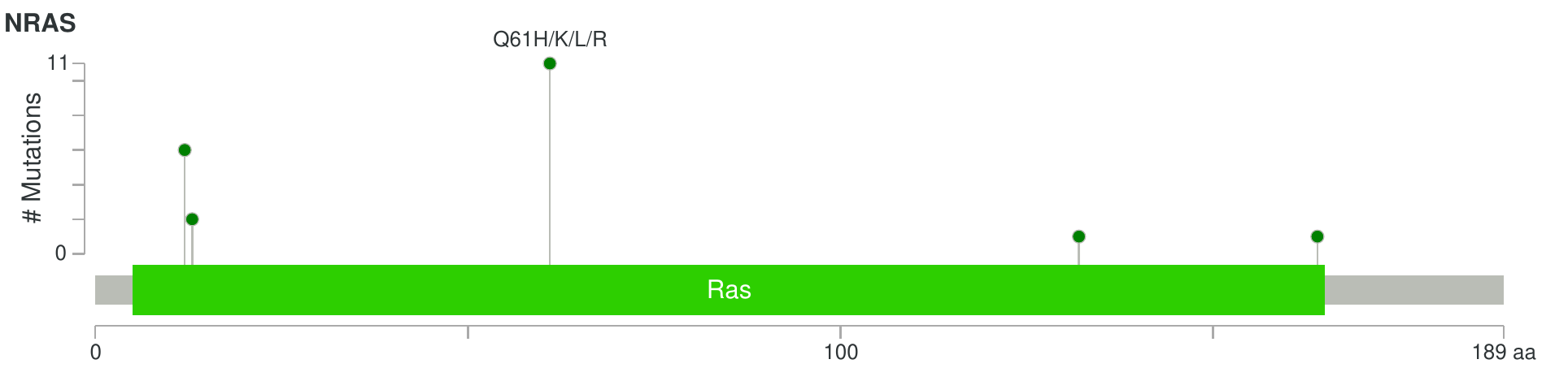}}
\colorbox{white}{\includegraphics[width = 0.5\textwidth]{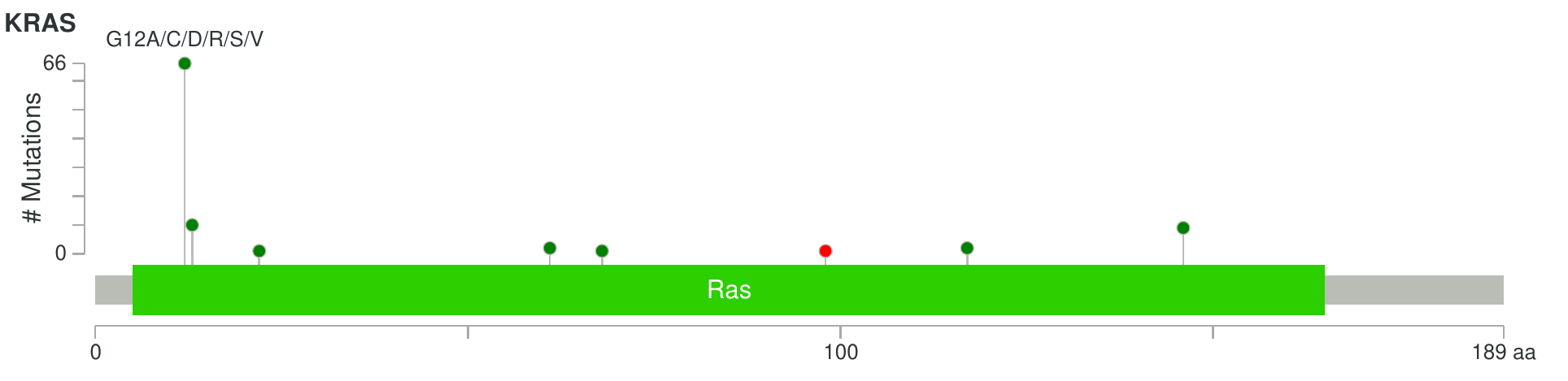}}
\colorbox{white}{\includegraphics[width = 0.5\textwidth]{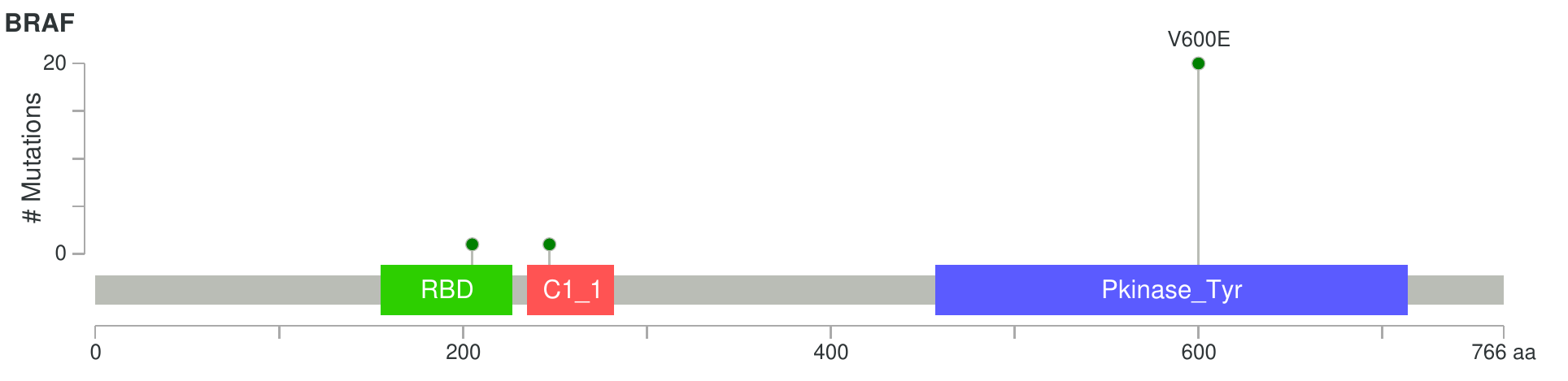}}
\colorbox{white}{\includegraphics[width = 0.5\textwidth]{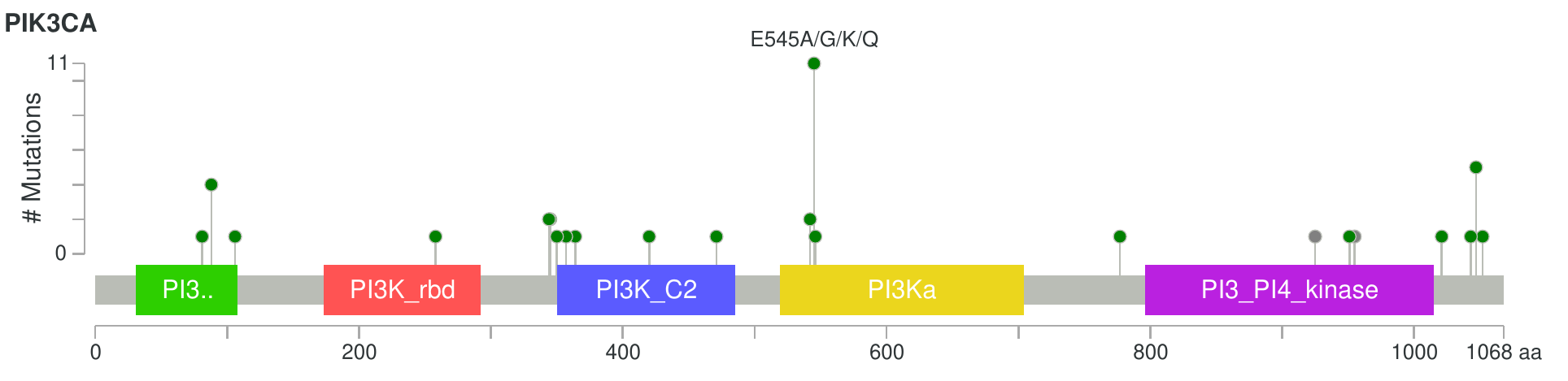}}
\colorbox{white}{\includegraphics[width = 0.5\textwidth]{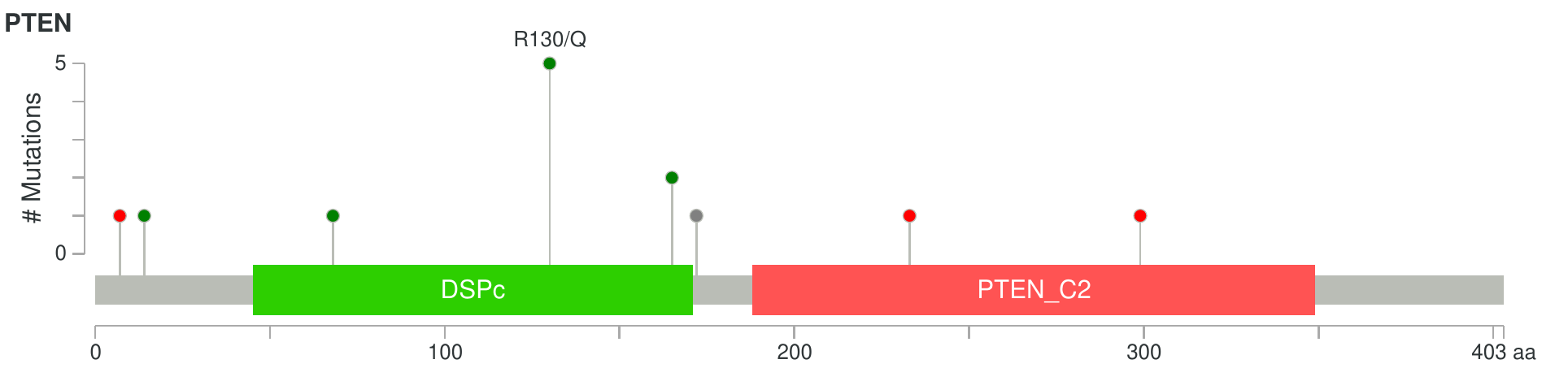}}
\colorbox{white}{\includegraphics[width = 0.5\textwidth]{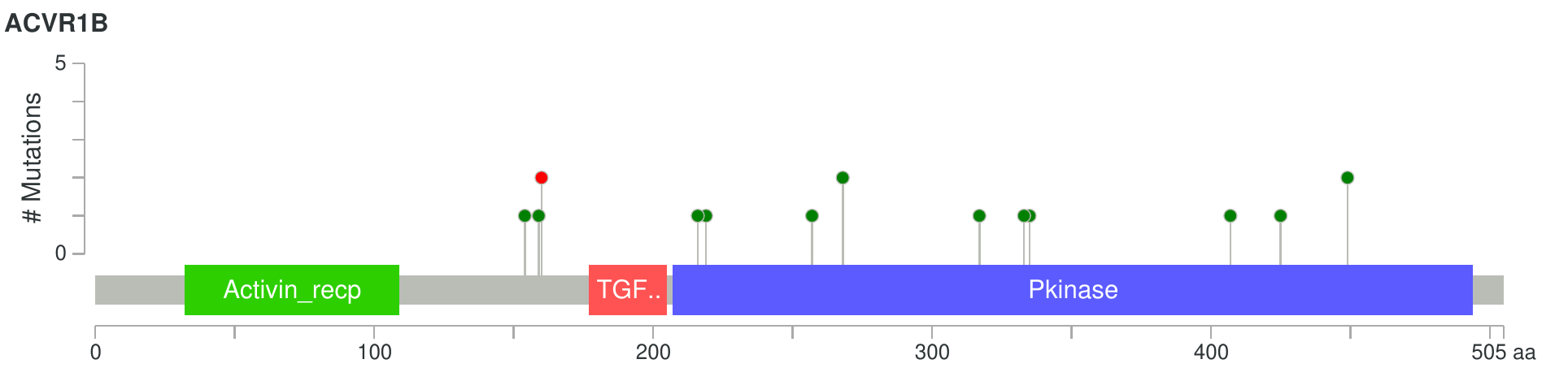}}
\colorbox{white}{\includegraphics[width = 0.5\textwidth]{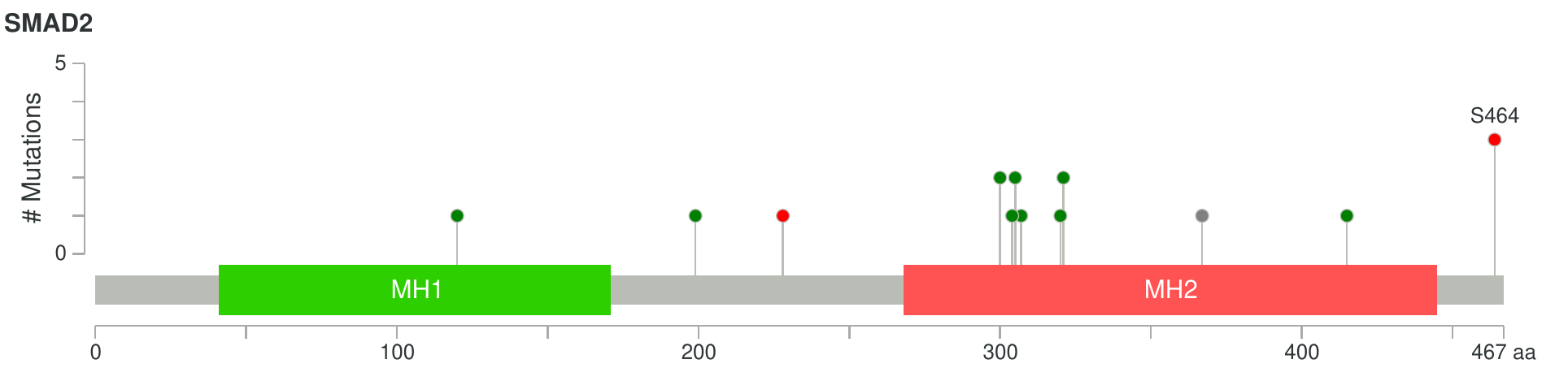}}
\colorbox{white}{\includegraphics[width = 0.5\textwidth]{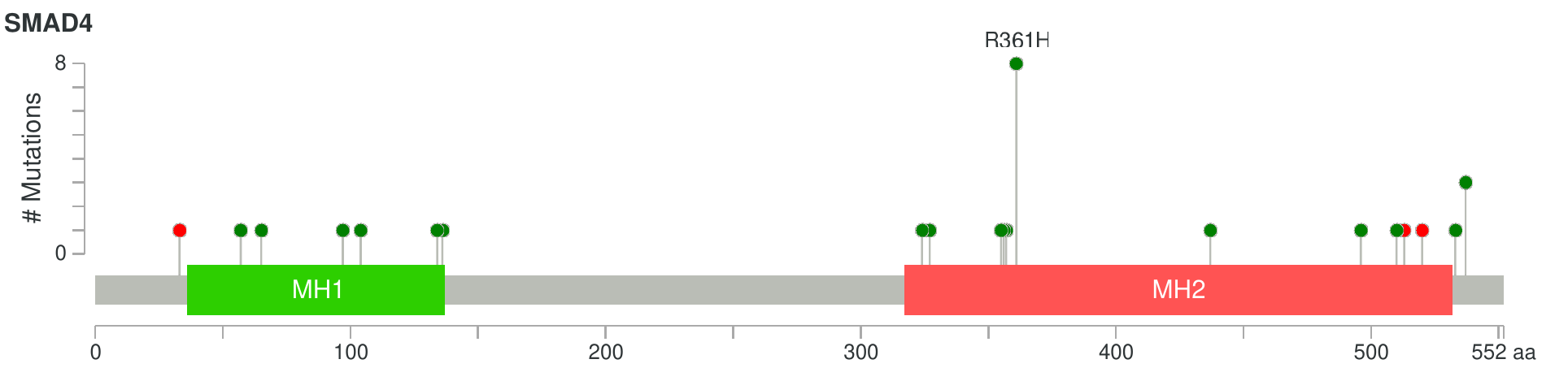}}
\colorbox{white}{\includegraphics[width = 0.5\textwidth]{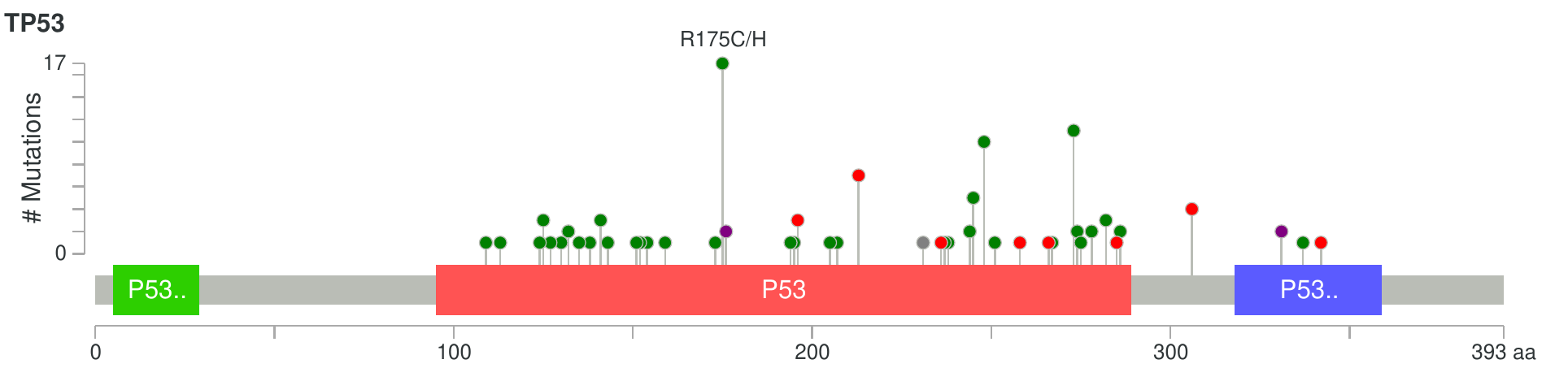}}
\colorbox{white}{\includegraphics[width = 0.5\textwidth]{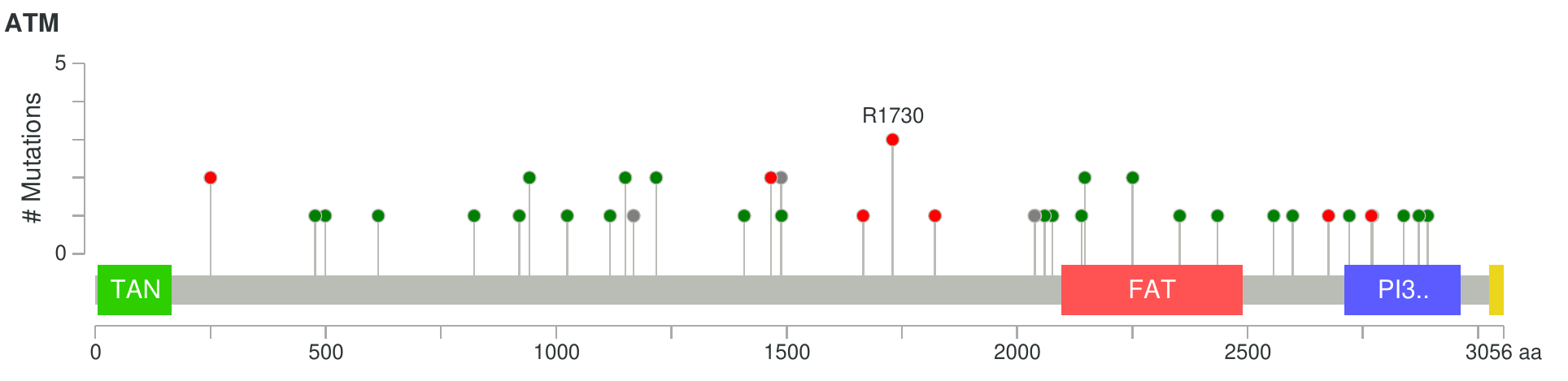}}
\center
\colorbox{white}{\includegraphics[width = 0.67\textwidth]{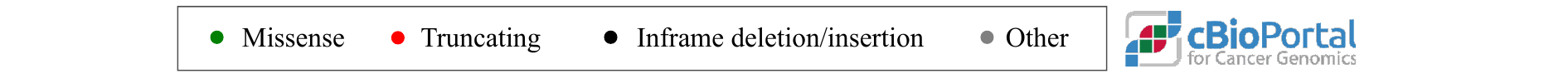}}
\caption[\added{Lolliplot diagrams of TCGA mutations}]{\added{{\bf Lolliplot diagrams of TCGA mutations.} Diagrams generated from the cBio portal  for the COADREAD project (see \url{http://www.cbioportal.org/}). These display the physical distribution of the annotated mutations for each gene. Here we shown only genes with a total mutation count greater than $15$; {\sc fam123b} is called with its synonym {\sc amer1}, as in the portal.}
} 
%
\label{fig:SM-lolliplot}
\end{figure}

%

\begin{figure}
\centerline{\includegraphics[width=1.0\textwidth]{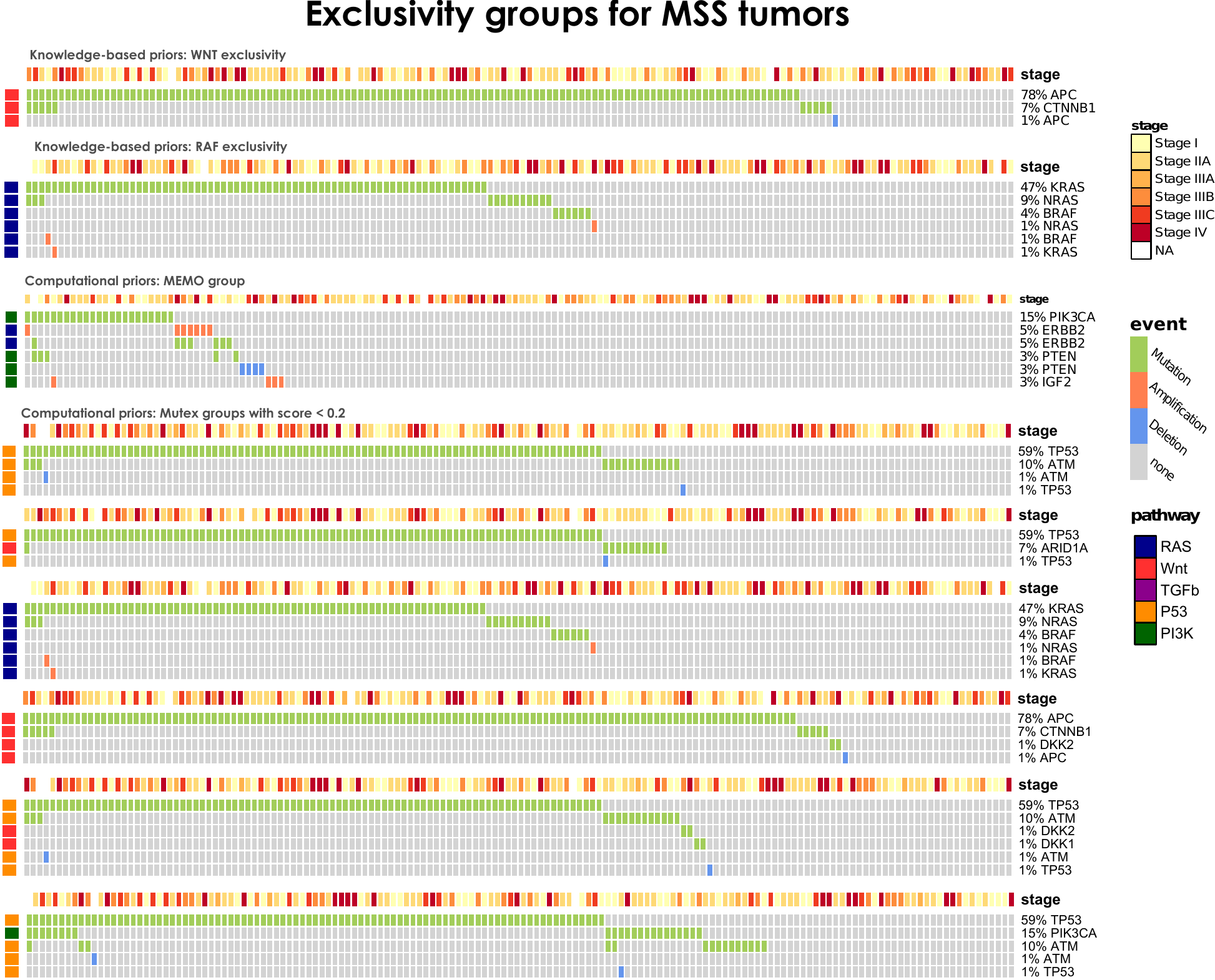}}
\caption[Groups of exclusive alterations for MSS tumors]{{\bf Groups of exclusive alterations for MSS tumors.} Knowledge-based groups of exclusive alterations consist of: 
\g{kras}, \g{nras} and \g{braf} genes (\g{raf} pathway) and  \g{apc} and \g{ctnnb1} genes (\g{wnt} pathway). The MEMO\cite{ciriello2012mutual} group identified in \cite{cancer2012comprehensive} in this cohort consists of genes \g{pik3ca}, \g{erbb2}, \g{igf2} and \g{pten}.  Finally, 6 groups are predicted by MUTEX \cite{babur2015systematic} with score below $.2$, one of these is equivalent to the known exclusive alterations in \g{raf} pathway.
} 
\label{fig:SM-exclusivity-MSS}
\end{figure}

\begin{figure}[p]
\centerline{\includegraphics[width=1.0\textwidth]{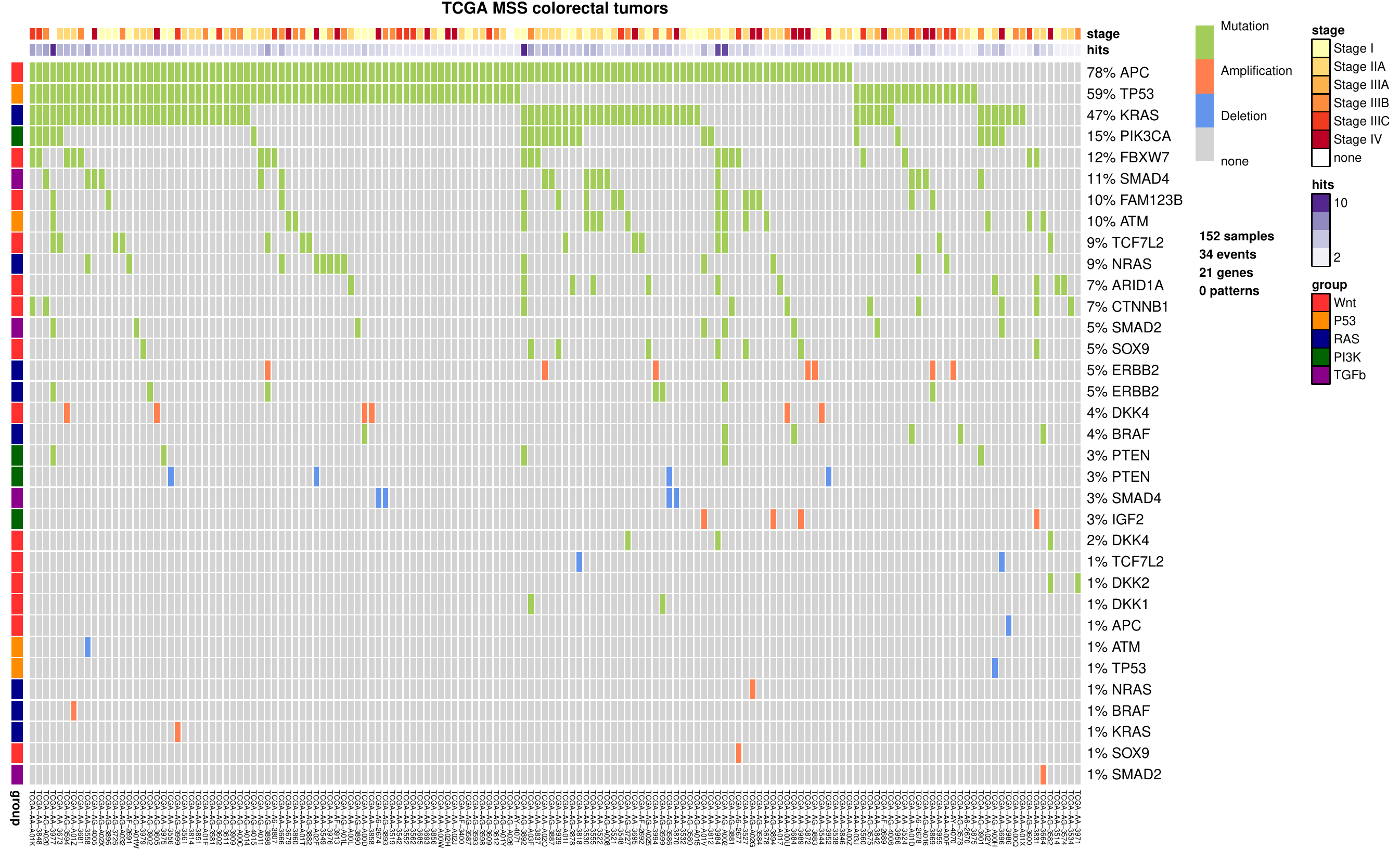}}
\caption[Selected data for MSS tumors]{{\bf Selected data for MSS tumors.} Colorectal tumors with Microsatellite Stable clinical status in the TCGA COADREAD project, restricted to 152 samples with both somatic mutations and CNA data available. 33 driver genes annotated with  5 pathways are selected from the list published in \cite{cancer2012comprehensive} to automatically detect groups of mutually exclusive alterations. Events selected for reconstruction are those involving genes altered in at least 5\% of the cases, or part of group of alterations showing an exclusivity trend  (see Figure \ref{fig:SM-exclusivity-MSS}). This dataset is used to infer the set of selective advantage relations which constitute the MSS progression model presented in the Main Text.
} 
\label{fig:SM-trdataset-MSS}
\end{figure}

\begin{figure}[p]
\centerline{\includegraphics[width=1.0\textwidth]{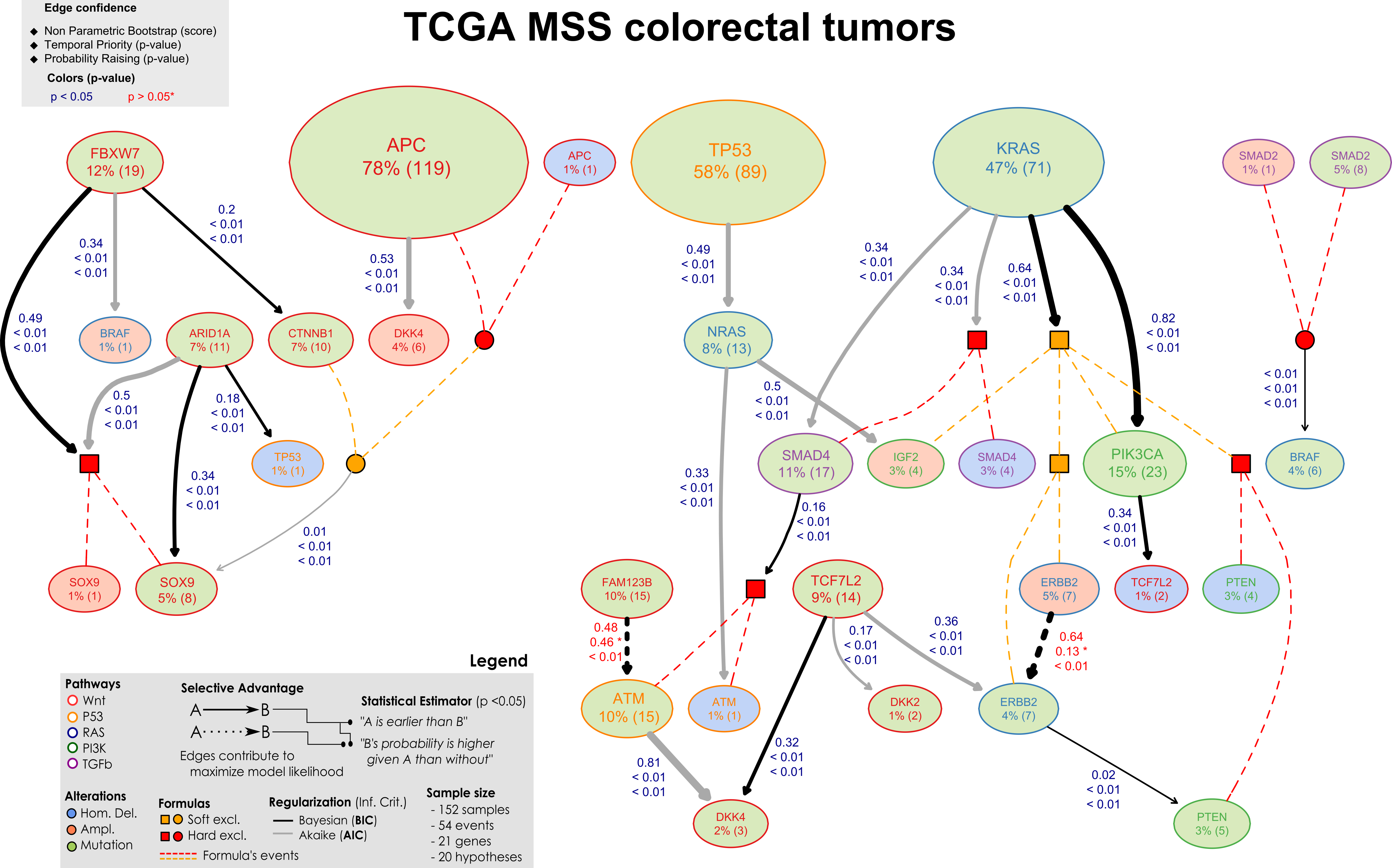}}
\caption[Non-parametric bootstrap scores for MSS progression]{{\bf Non-parametric bootstrap scores for MSS progression.} Progression model for MSS tumors with confidence shown as edge labels. The first label represents the relation confidence estimated with 100 non-parametric bootstrap iterations, the second and third are p-values for temporal priority and probability raising. Red p-values are above the minimum significane threshold of $.05$. \addedthree{See Figure 4 in the Main Text for an interpretation of this model.}
} 
\label{fig:SM-MSS-Bootstrap}
\end{figure}

\begin{figure}[p]
\centerline{\includegraphics[width=1.0\textwidth]{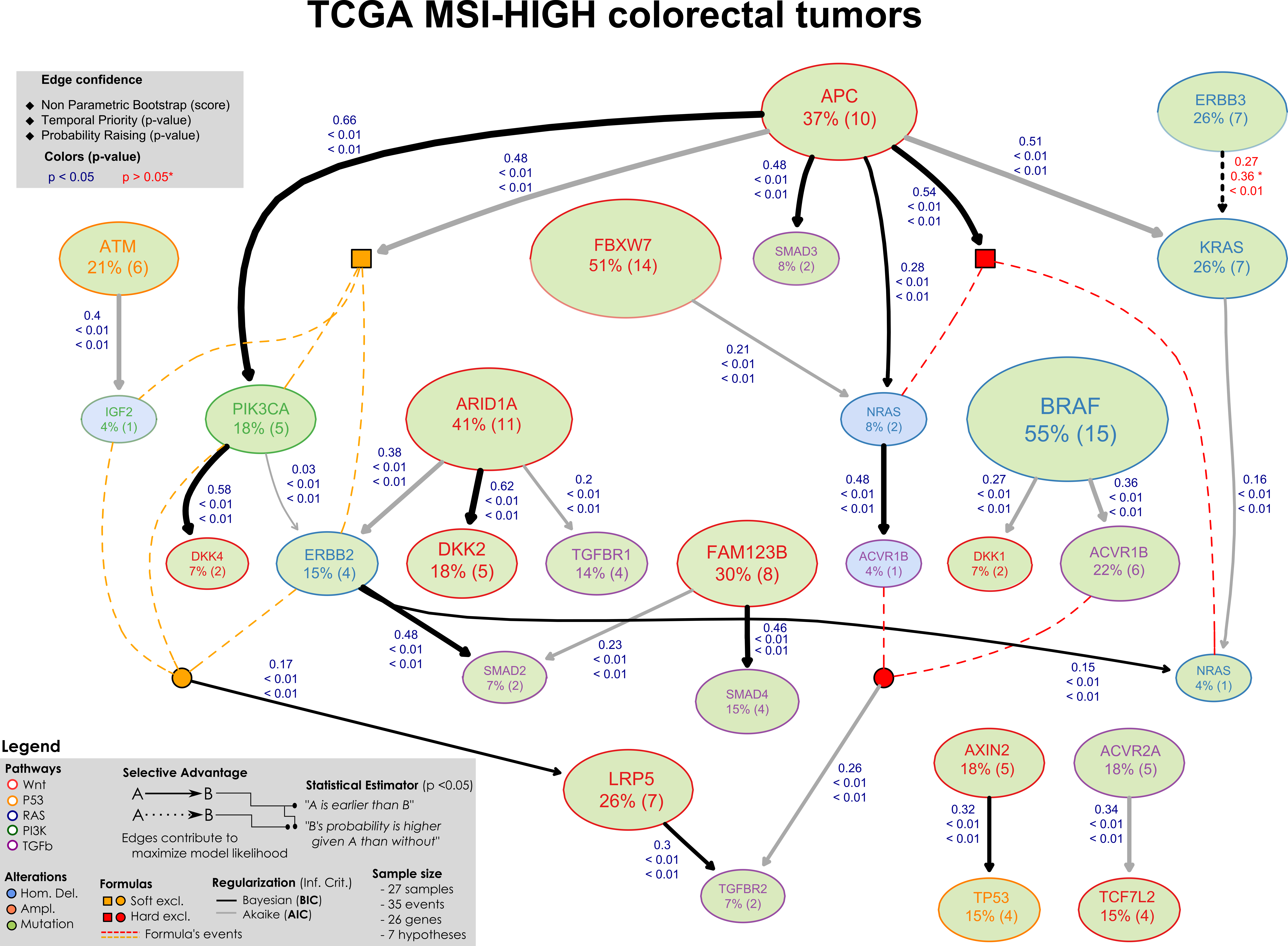}}
\caption[Non-parametric bootstrap scores for MSI-HIGH progression]{{\bf Non-parametric bootstrap scores for MSI-HIGH progression.} Progression model for MSI-HIGH tumors with confidence shown as edge labels. The first label represents the relation confidence estimated with 100 non-parametric bootstrap iterations, the second and third are p-values for temporal priority and probability raising. Red p-values are above the minimum significane threshold of $.05$. \addedthree{See Figure 5 in the Main Text for an interpretation of this model.}
} 
\label{fig:SM-MSI-Bootstrap}
\end{figure}

\begin{figure}
\center
\colorbox{white}{\includegraphics[width=17cm]{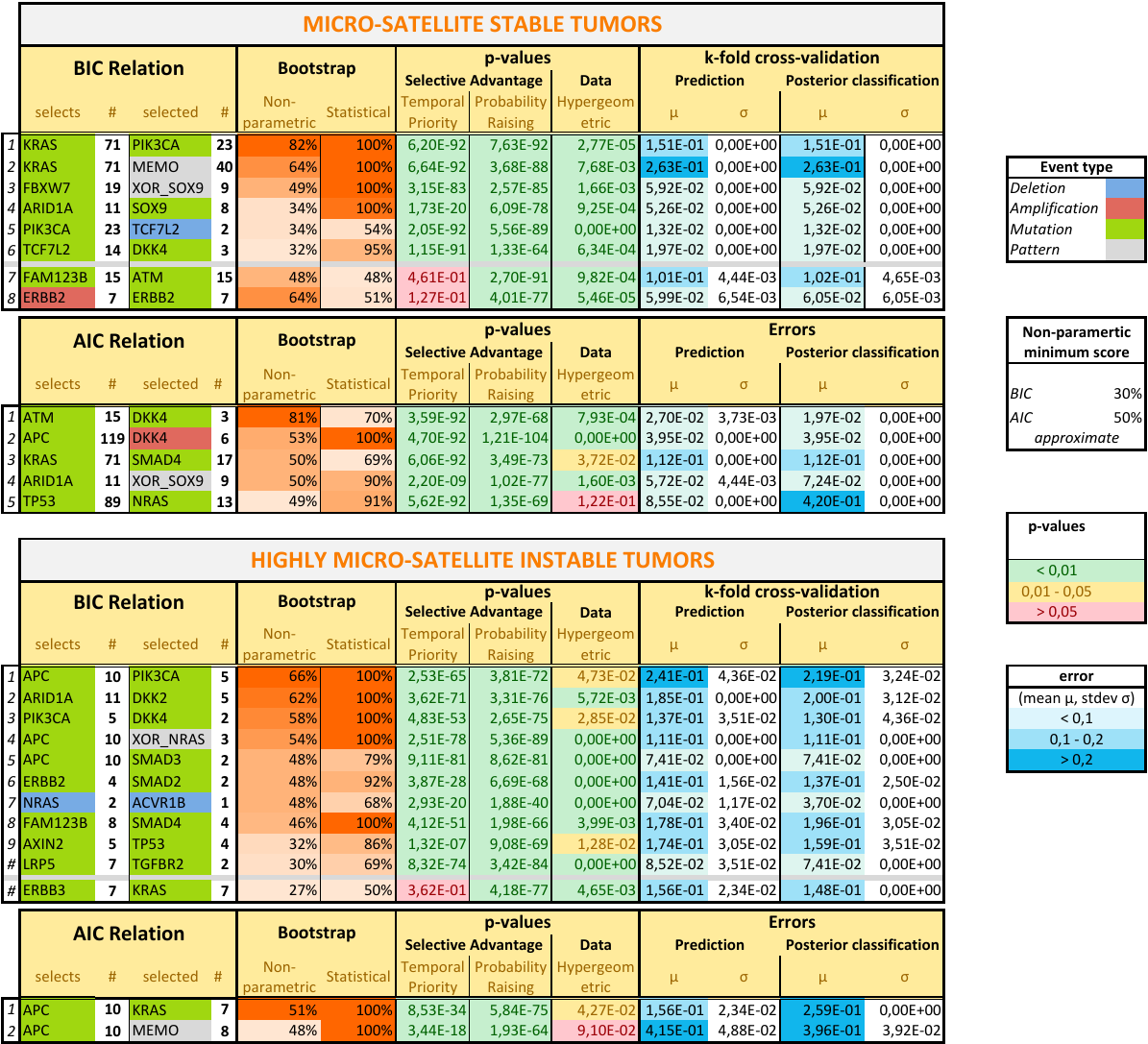}}
\caption[\added{COADREAD statistics for models confidence}]{\added{{\bf COADREAD statistics for models confidence.} 
For BIC models we show statistics for edges with non-parametric bootstrap score approximately greater than $30\%$, for AIC models those greater than $50\%$s. $i)$ {\em p-values} (100 repetition of non-parametric bootstrap, prior to Wilcoxon testing) for each edge statistics  of selective advantage ({\em direction and statistical dependence, and hypergeometric}). In general, the edges that we selected show very strong support ($p \ll 10^{-10}$), but for those edges connecting events with the same marginal frequencies, where we can not be confident in the edge direction ($p > 0.05$) but still we find strong statistical dependence. $(ii)$ {\em A posteriori} model confidence against Type I and II errors estimated with {\em non-parametric and statistical bootstraps} (100 repetitions) --   edges annotated in  Figures   \ref{fig:SM-MSS-Bootstrap} and \ref{fig:SM-MSI-Bootstrap}.   $(iii)$ Values of  {\em posterior classification and prediction errors} are estimated from 10 repetitions of 10-fold {\em cross-validation}.  The former reports how much error is due to predicting, for each set of edges $X=\{x_1,\ldots,x_n\} \to y$, the value of $y$ according to the value of each $x_i\in X$. The latter reports the same statistics when we predict $y$ from the whole set of parents $X$.}}
\label{tab:table-stat}
\end{figure}

\begin{figure}
\center
\colorbox{white}{\includegraphics[width=8cm]{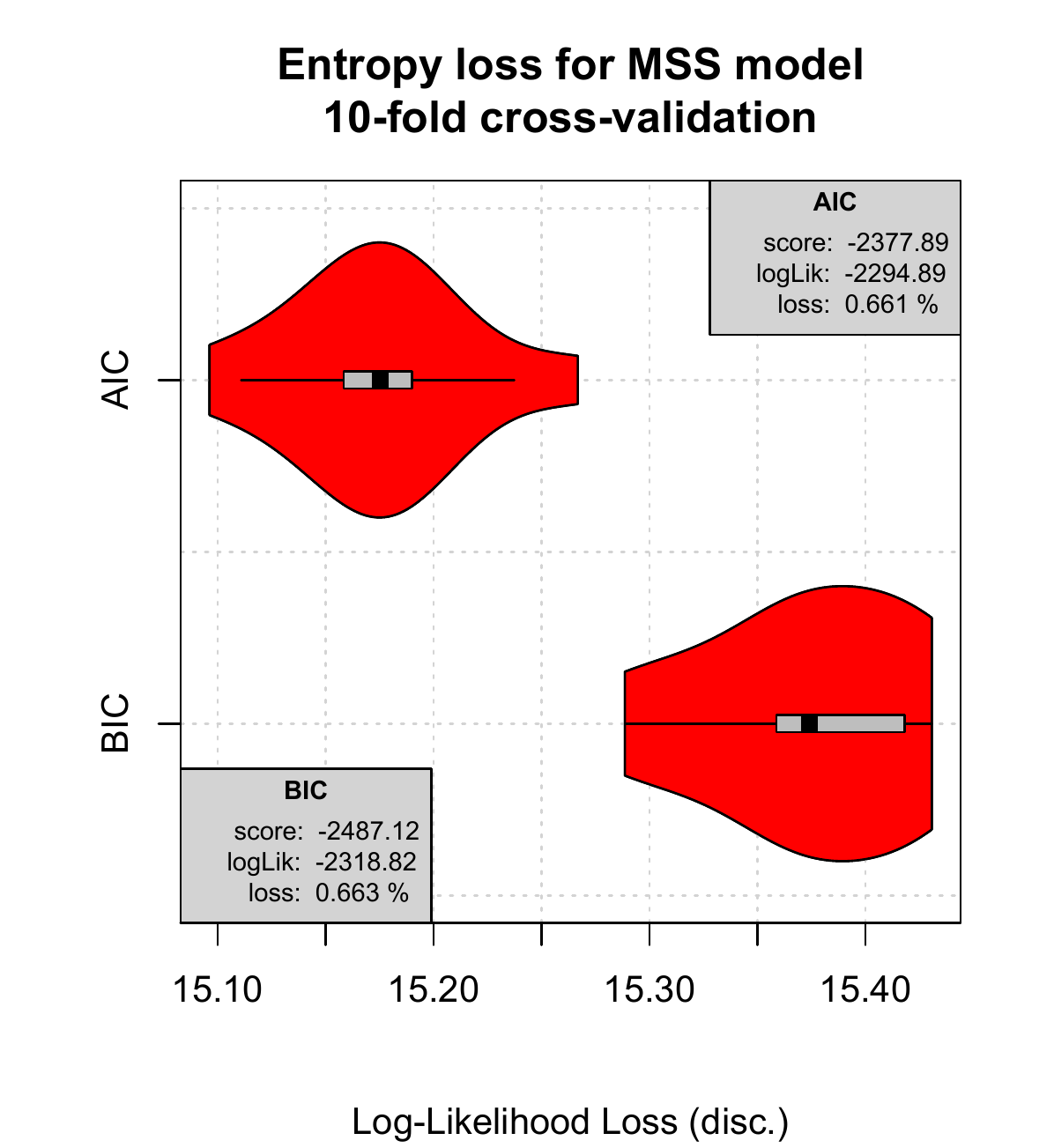}}
\colorbox{white}{\includegraphics[width=8cm]{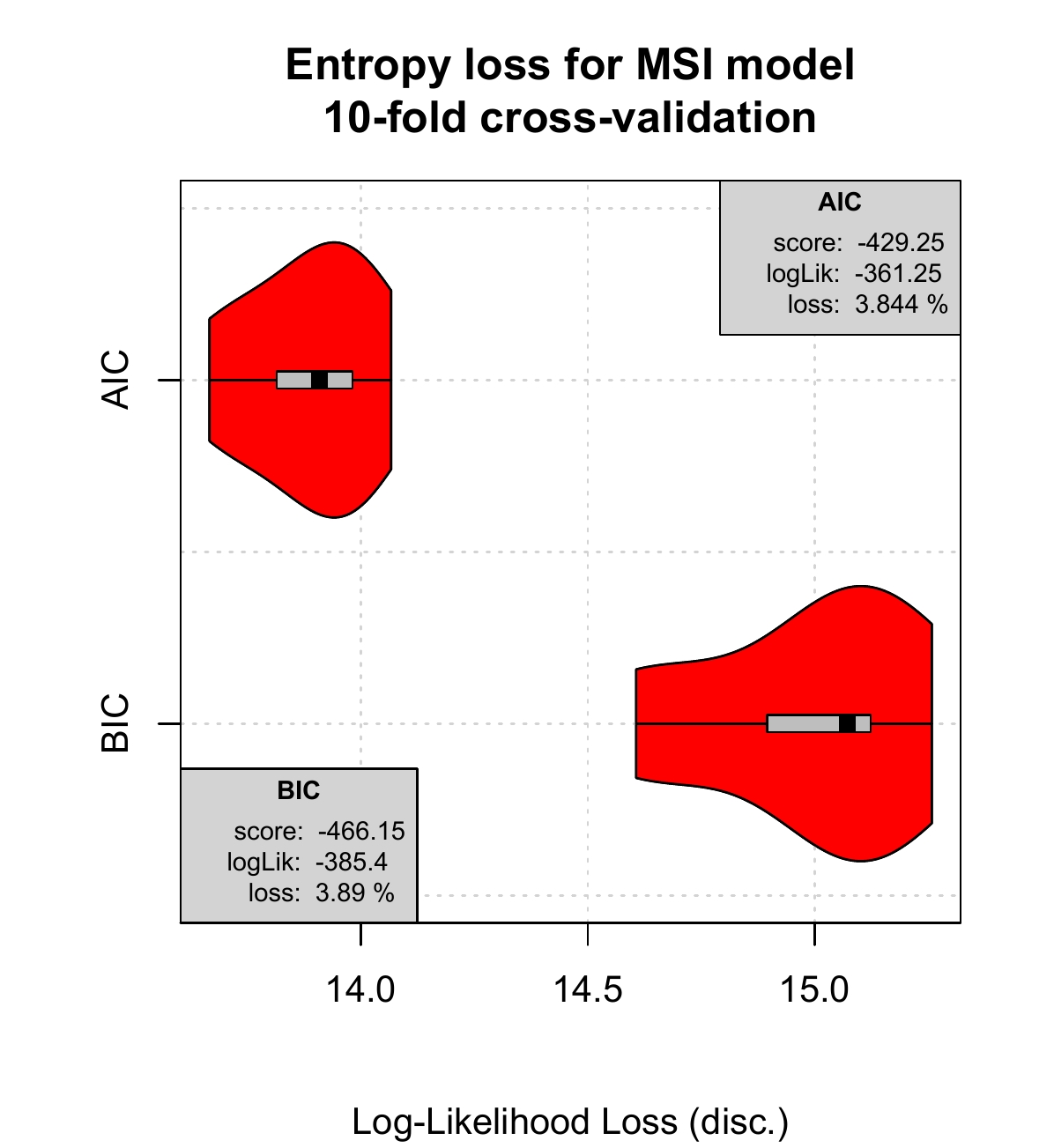}}
\colorbox{white}{\includegraphics[width=12cm]{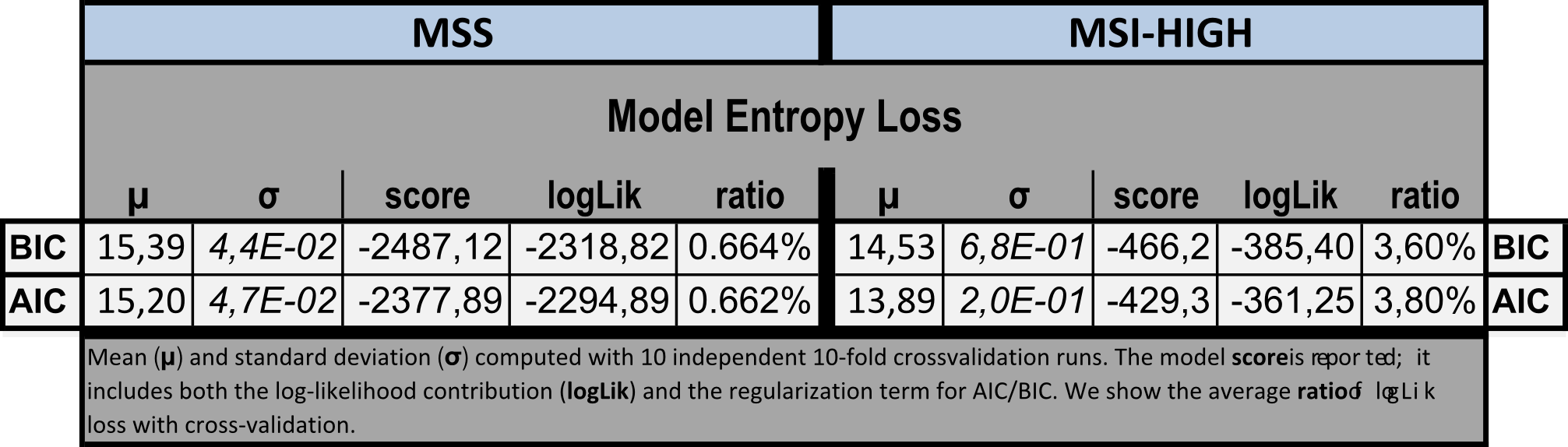}}
\caption[\added{Entropy loss for MSI-HIGH/MSS models}]{\added{{\bf Entropy loss for MSI-HIGH/MSS models.} 
Violin plot computed from $10$ runs of $k$-fold cross-validation with $k=10$, where we compute the  ``loss of log-likelihood'' at each fold. In the plot and in the table we report also the overall  log-likelihood, as well as the BIC and AIC scores for the models. We present the ratio of log-likelihood loss as a measure of stability of these models for these two datasets -- we can observe that the MSS models lose $<1\%$ of their likelihood, while the MSI  lose slightly more  (still, $<4\%$), possibly because of the smaller sample size. From a statistical point of view, the greater (despite small) loss of likelihood by the  models regularized via BIC confirms its tendency to underestimate the true model (i.e.,  the model should have false negatives, which could be  AIC's edges). 
}}
\label{tab:eloss}
\end{figure}


\begin{figure}
\center
\colorbox{white}{\includegraphics[width=17cm]{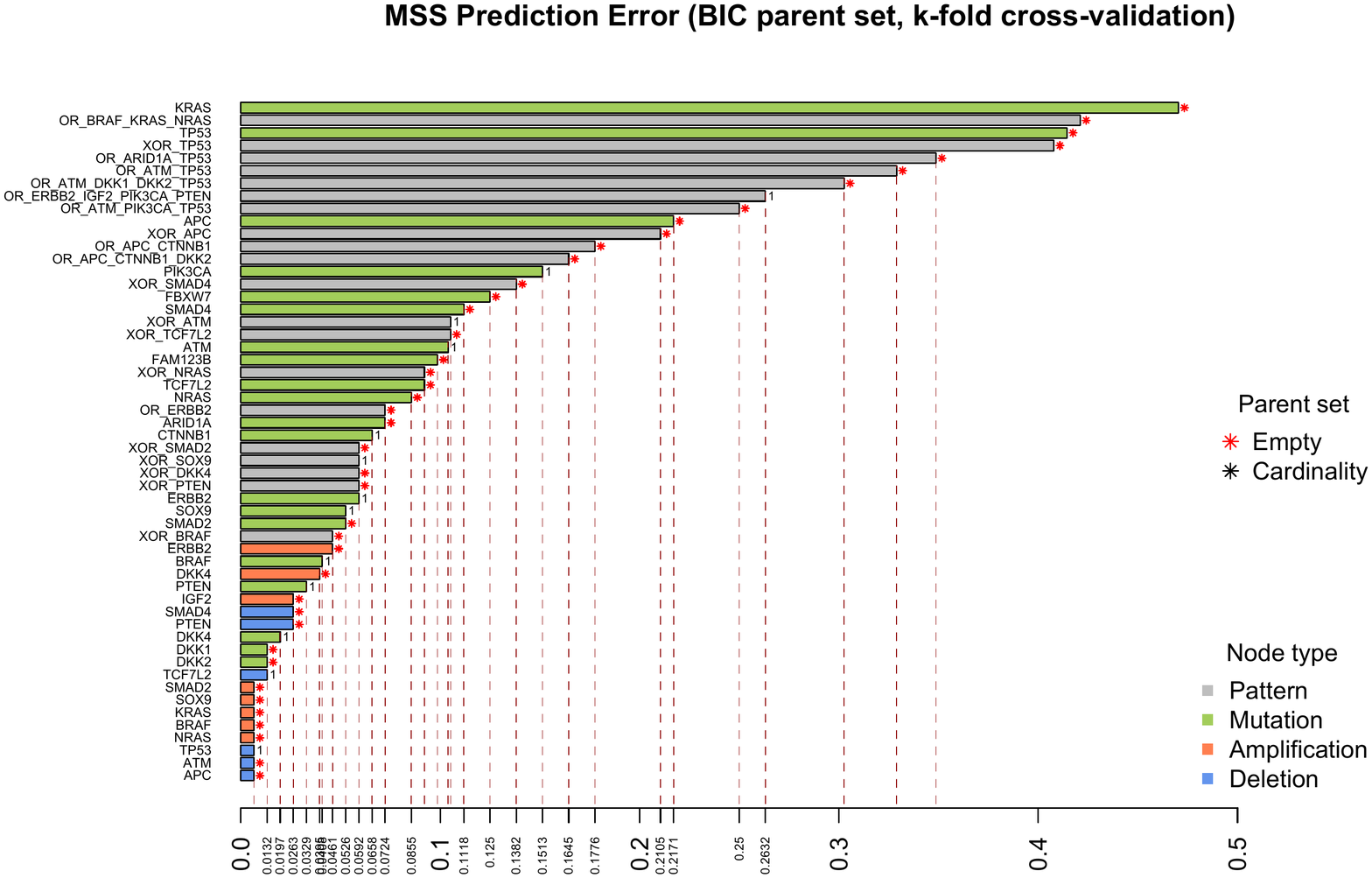}}
\colorbox{white}{\includegraphics[width=17cm]{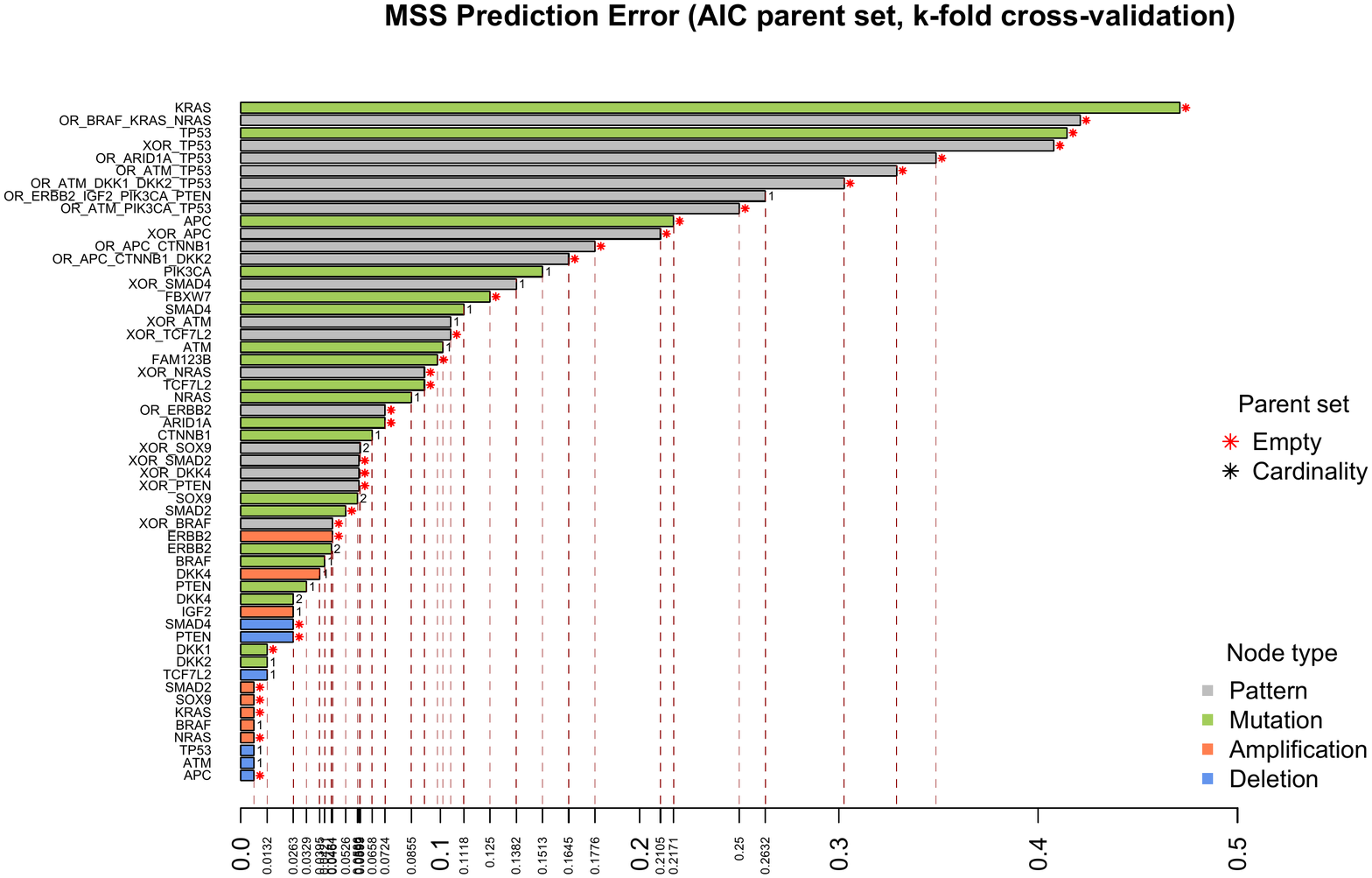}}
\caption[\added{Prediction error for each parent set of BIC and AIC models of MSS tumors}]{
\added{{\bf Prediction error for each parent set of BIC and AIC models of MSS tumors.} }}
\label{tab:prederr-all-mss}
\end{figure}

\begin{figure}
\center
\colorbox{white}{\includegraphics[width=19cm]{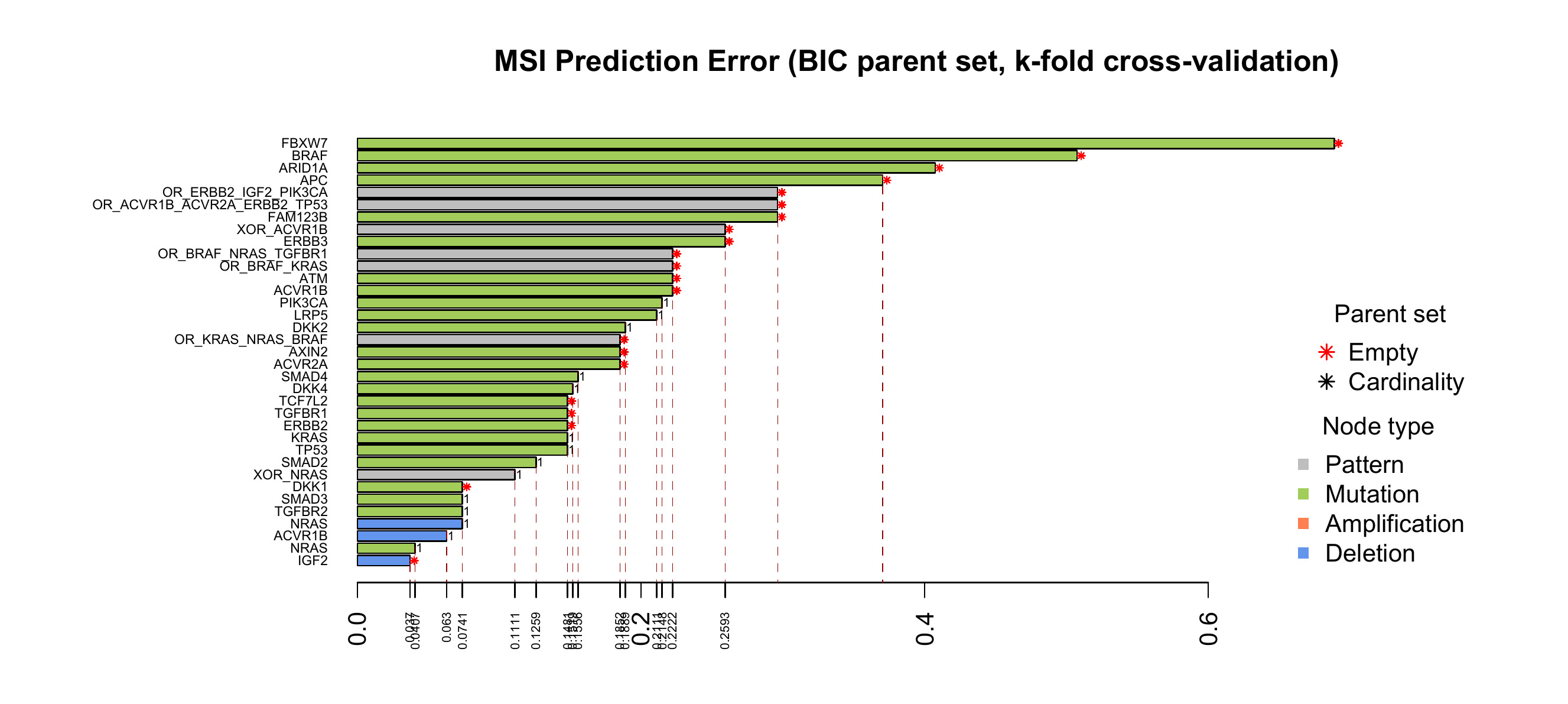}}
\colorbox{white}{\includegraphics[width=19cm]{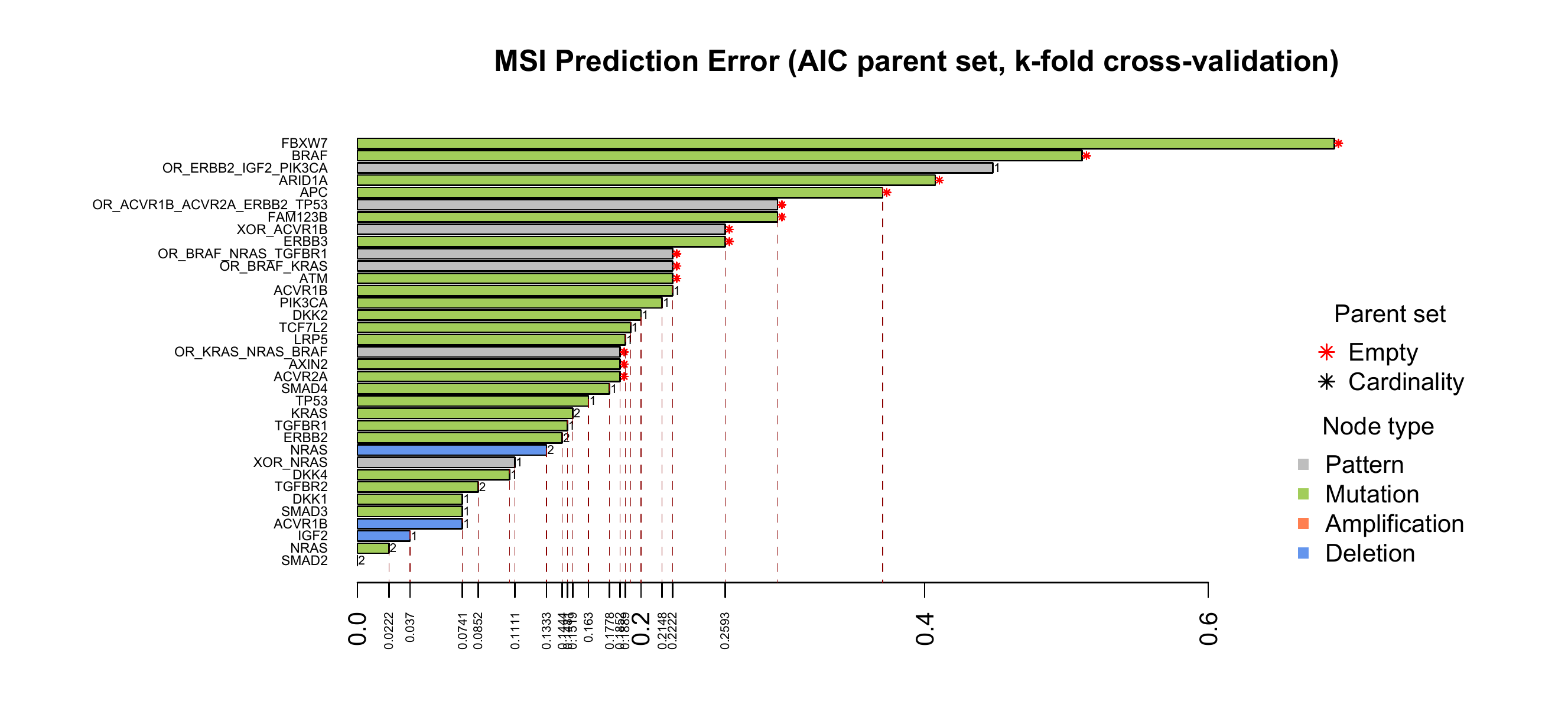}}
\caption[\added{Prediction error for each parent set of BIC and AIC models of MSI tumors}]{\added{{\bf Prediction error for each parent set of BIC and AIC models of MSI-HIGH tumors.} }}
\label{tab:prederr-all-msi}
\end{figure}

\begin{figure}
\center
\colorbox{white}{\includegraphics[width=14cm]{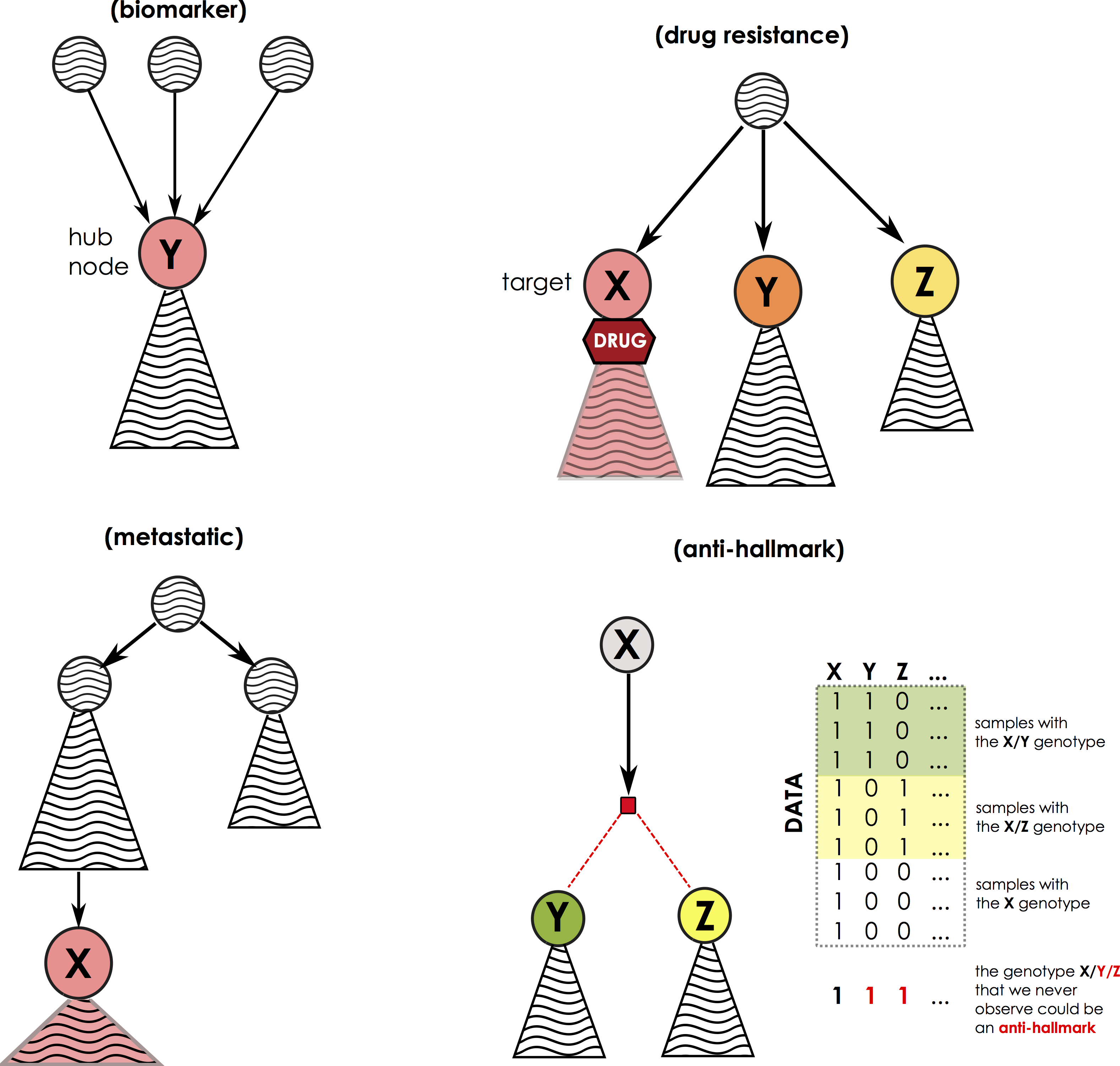}}
\caption[\added{Models and the phenotype that they might explain}]{\added{{\bf Models and the phenotype that they might explain.}}
{\bf (biomarker)} Independent evolutionary trajectories depicted by a model might share common routes through a certain alteration Y; that could point to a new biomarker  harbored by most of the tumors under study. {\bf (drug resistance)}  When a progression model branches in many independent sub-progressions, each one identified by alterations X, Y and Z,  if  a certain drug is known to target only a certain type of such clones  (e.g., those where biomarker X is present), we might get insights on which are the biomarkers which make the drug ineffective for certain patients (e.g., those were cancer evolves through Y and Z). {\bf (metastatic)} When a  model is extracted from data representative of various tumor stages,  we might discover which ``late events''  are those  conferring a metastatic phenotype to a tumor -- X in the figures. {\bf (anti-hallmarks)} Relation between anti-hallmarks and formulas. Exclusivity formulas  allow to capture fitness-equivalent events (Y and Z in the figure), and the presence of alternative routes -- here those identified by the genotypes X/Y or  X/Z. These could point us to genotype/phenotype that we do no observe in our cohort -- here the X/Y/Z -- which could be exploited for targeted therapy if a synthetic lethality is screened among Y and Z, the anti-hallmark.}
\label{tab:model-phenotype}
\end{figure}

\end{document}